\documentclass[aps,prb,reprint,superscriptaddress]{revtex4-2}
\usepackage{graphicx}
\usepackage{amsmath}
\usepackage{amssymb}
\usepackage{bm}
\usepackage{physics}
\usepackage{subfigure}
\usepackage[colorlinks=true,linkcolor=blue,urlcolor=blue,citecolor=blue]{hyperref}
\usepackage{times}
\usepackage{changes}
\usepackage{algorithm}
\usepackage{algpseudocode}
\usepackage{algorithmicx}
\begin{document}

\newcommand{\Hop}{\hat{H}}
\newcommand{\Himp}{\hat{H}_{\rm imp}}
\newcommand{\Hint}{\hat{H}_{\rm int}}
\newcommand{\Uimp}{\hat{U}_{\rm imp}}
\newcommand{\gHimp}{\mathcal{H}_{\rm imp}}
\newcommand{\gUimp}{\mathcal{U}_{\rm imp}}
\newcommand{\Hbath}{\hat{H}_{\rm bath}}
\newcommand{\Hhyb}{\hat{H}_{\rm hyb}}
\newcommand{\Hel}{\hat{H}_{\rm el}}
\newcommand{\Hph}{\hat{H}_{\rm ph}}
\newcommand{\Iop}{\hat{I}}

\newcommand{\aop}{\hat{a}}
\newcommand{\adop}{\hat{a}^{\dagger}}

\newcommand{\bop}{\hat{b}}
\newcommand{\bdop}{\hat{b}^{\dagger}}

\newcommand{\sgp}{\hat{\sigma}^+}
\newcommand{\sgx}{\hat{\sigma}^x}
\newcommand{\sgy}{\hat{\sigma}^y}
\newcommand{\sgz}{\hat{\sigma}^z}
\newcommand{\nop}{\hat{n}}

\newcommand{\cop}{\hat{c}}
\newcommand{\cdop}{\hat{c}^{\dagger}}
\newcommand{\hc}{{\rm H.c.}}
\newcommand{\rhotot}{\hat{\rho}_{\mathrm{tot}}}
\newcommand{\rhoop}{\hat{\rho}}
\newcommand{\rhoimp}{\hat{\rho}_{\mathrm{imp}}}
\newcommand{\rhobath}{\hat{\rho}_{\mathrm{bath}}}
\newcommand{\Zimp}{Z_{{\rm imp}}}
\newcommand{\mea}{\mathcal{D}}
\newcommand{\gK}{\mathcal{K}}
\newcommand{\gI}{\mathcal{I}}
\newcommand{\gIel}{\mathcal{I}_{\rm el}}
\newcommand{\gIph}{\mathcal{I}_{\rm ph}}
\newcommand{\gF}{\mathcal{F}}
\newcommand{\bolda}{\bm{a}}
\newcommand{\boldabar}{\bar{\bm{a}}}
\newcommand{\boldc}{\bm{c}}
\newcommand{\boldcbar}{\bar{\bm{c}}}
\newcommand{\abar}{\bar{a}}
\newcommand{\im}{{\rm i}}
\newcommand{\contour}{\mathcal{C}}
\newcommand{\gA}{\mathcal{A}}
\newcommand{\gB}{\mathcal{B}}
\newcommand{\gM}{\mathcal{M}}
\newcommand{\boldeta}{\bm{\eta}}
\newcommand{\boldetabar}{\bar{\bm{\eta}}}
\newcommand{\etabar}{\bar{\eta}}
\newcommand{\parity}{\mathcal{P}}
\newcommand{\current}{\mathcal{J}}
\newcommand{\pronyerror}{\varsigma_p}
\newcommand{\Dph}{D^{\rm ph}}
\newcommand{\Del}{D^{\rm el}}
\newcommand{\nbar}{\bar{n}}
\newcommand{\boldn}{\bm{n}}

\newcommand{\Zel}{Z_{\rm el}}
\newcommand{\Zph}{Z_{\rm ph}}

\newcommand{\mtrace}{{\rm Tr}}

\newcommand{\WI}{{\rm W}^I}
\newcommand{\WII}{{\rm W}^{II}}

\newcommand{\EqDef}{\stackrel{\mathrm{def}}{=}}

\newcommand{\tmmathbf}[1]{\ensuremath{\boldsymbol{#1}}}
\newcommand{\tmop}[1]{\ensuremath{\operatorname{#1}}}
\newcommand{\mathd}{\mathrm{d}}
\newcommand{\equallim}{\mathop{=}\limits}

\newcommand{\rem}[1]{{\color{red}{\sout{#1}}}}
\newcommand{\gcc}[1]{{\color{red}#1}}

\definechangesauthor[name=RF,color=green]{RF}

\title{Tensor network algorithm to solve polaron impurity problems}

\author{Ruofan Chen}
\affiliation{College of Physics and Electronic Engineering, Sichuan Normal University, Chengdu 610068, China}
\affiliation{Center for Computational Sciences, Sichuan Normal University, Chengdu 610068, China}

\author{Lei Gu}
\affiliation{College of Physics and Electronic Engineering, Sichuan Normal University, Chengdu 610068, China}

\author{Chu Guo}
\email{guochu604b@gmail.com}
\affiliation{Key Laboratory of Low-Dimensional Quantum Structures and Quantum Control of Ministry of Education, Department of Physics and Synergetic Innovation Center for Quantum Effects and Applications, Hunan Normal University, Changsha 410081, China}

\date{\today}

\begin{abstract}
The polaron problem is a very old problem in condensed matter physics that dates back to the thirties, but still remain largely unsolved today, especially when electron-electron interaction is taken into consideration. The presence of both electron-electron and electron-phonon interactions in the problem invalidates most existing numerical methods, either computationally too expensive or simply intractable. The continuous time quantum Monte Carlo (CTQMC) methods could tackle this problem, but are only effective on the imaginary-time axis. In this work we present a method based on tensor network and the path integral formalism to solve polaron impurity problems. 
As both the electron and phonon baths can be integrated out via the Feynman-Vernon influence functional in the path integral formalism, 
our method is free of bath discretization error. It can also flexibly work on the imaginary, Keldysh, and the L-shaped Kadanoff contour. In addition, our method can naturally resolve several long-existing challenges: (i) non-diagonal hybridization function; (ii) measuring multi-time correlations beyond the single particle Green's functions. We demonstrate the effectiveness and accuracy of our method with extensive numerical examples against analytic solutions, exact diagonalization and CTQMC. We also perform full-fledged real-time calculations that have never been done before to our knowledge, which could be a benchmarking baseline for future method developments.
\end{abstract}
\maketitle


\section{Introduction}
The polaron problem describes an itinerant electron with the phonon cloud created by it~\cite{landau1933-electron,landau1948-effective}, which is a ubiquitous phenomenon in physics, chemistry and materials science~\cite{AlexandrovDevreese2010,BredasStreet1985,FranchiniDiebold2021,thomas2025-theory,GrusdtArdila2024}. It is important to understand strongly correlated effects, such as the high-temperature superconductors~\cite{LanzaraShen2001} or the colossal magnetoresistance manganites~\cite{MillisShraiman1995,MillisMueller1996,Millis1998,YamasakiTokura2006}. An effective numerical approach to solve the polaron problem is the dynamical mean field theory (DMFT)~\cite{WernerMillis2007b,WernerMillis2010,Otsuki2013}, which, instead of solving the original lattices problem, solves a corresponding impurity problem that describes a localized impurity coupled to a free electron bath and a free phonon bath.
However, although being formally simple and elegant, the polaron impurity problem still remains a formidable challenge for numerical methods. 

The wave-function based methods are conceptually most straightforward to solve quantum impurity problems, in which the baths are discretized and parameterized together with the impurity as a global wave function.
The representative methods in this category include exact diagonalization~\cite{CaffarelKrauth1994,KochGunnarsson2008,GranathStrand2012,LuHaverkort2014,ZaeraLin2020,LuHaverkort2019,HeLu2014,HeLu2015}, numerical renormalization group~\cite{Wilson1975,Bulla1999,BullaPruschke2008,Frithjof2008,ZitkoPruschke2009,DengGeorges2013,StadlerWeichselbaum2015,LeeWeichselbaum2016,LeeWeichselbaum2017,CornagliaGrempel2004,PaaskeFlensberg2005,CornagliaNess2005,LaaksoMeden2014}, time-evolving matrix product states~\cite{WolfSchollwock2014b,GanahlEvertz2014,GanahlVerstraete2015,WolfSchollwock2015,GarciaRozenberg2004,NishimotoJeckelmann2006,WeichselbaumDelft2009,BauernfeindEvertz2017,WernerArrigoni2023,KohnSantoro2021,KohnSantoro2022}. 
However, for polaron impurity problems, this approach would quickly be overwhelmed by the large number of bosonic and fermionic modes from the discretized baths. 
Up to date, the continuous time quantum Monte Carlo methods are perhaps the only class of methods which can effectively deal with the polaron impurity problem~\cite{GullWerner2011,ProkofevSvistunov1998,HafermannGull2012,Otsuki2013}. 
For a long time, the Lang-Firsov transformation~\cite{lang1963-kinetic} is employed in CTQMC to eliminate the electron-phonon interaction term~\cite{WernerMillis2007b,Otsuki2013,Hafermann2014}, which is only possible for density-density couplings between impurity flavors. Very recently, a generalized CTQMC is proposed to handle non-diagonal couplings as well by ceasing this requirement~\cite{GuZhao2025}.
However, CTQMC could suffer from the sign problem~\cite{TroyerWiese2005}. Moreover, it is only effective on the imaginary-time axis, and numerical instability could occur during the analytic continuation from imaginary-time data to real time~\cite{WolfSchollwock2015,FeiGull2021}. There has been several improved quantum Monte Carlo methods which aim for real-time evolution~\cite{CohenGull2014,CohenMillis2014,CohenMillis2015,ChenReichman2017a,ChenReichman2017b,BertrandWaintal2019,ErpenbeckCohen2023}, but their ability to tackle the polaron impurity problem is yet to be demonstrated.

The origin of the powerfulness of CTQMC for quantum impurity problems, compared to its alternatives, can be understood as follows: it draws sample from the perturbative expansion of the impurity path integral (PI), in which the bath degrees of freedom are integrated out before hand via the Feynman-Vernon influence functional (IF)~\cite{FeynmanVernon1963,negele1998-quantum} and only the impurity degrees of freedom in the time direction persist. As such CTQMC is effectively dealing with a 1D problem on the time axis only. In comparison, the wave-function based methods are essentially dealing with a 1+1D problem (one space dimension plus the time dimension), which could be less efficient and will suffer from the bath discretization error.

An important progress made in recent years is to make use of the Feynman-Vernon IF in tensor network based methods. The basic idea is straightforward: if the problem is effectively 1D, then tensor networks, in particular the matrix product state (MPS), should be well suited for this case. A pioneering work in this direction is the time-evolving matrix product operator (TEMPO) method which represents the integrand of the impurity PI, referred to as the augmented density tensor (ADT), as an MPS for bosonic impurity problems~\cite{StrathearnLovett2018}. 
TEMPO is fundamentally different from the conventional MPS methods: in the latter case the MPSs are used to represent the impurity-bath wave function and the Feynman-Vernon IF is not used at all.
Similar ideas are explored later for fermionic impurity problems, by either using fermionic MPS in the Fock state representation~\cite{ThoennissAbanin2023b} or Grassmann MPS (GMPS) in the coherent state representation~\cite{ChenGuo2024a}.
In particular, as the analytic expression of the Feynman-Vernon IF for fermionic impurity problems is only available in the coherent state representation~\cite{negele1998-quantum}, the latter approach is a more natural and convenient choice, which also closely resembles TEMPO in that its core algorithms could be directly read off from the Feynman-Vernon IF without additional theoretical derivations. 
Due to this connection, the latter approach is referred to as the Grassmann time-evolving matrix product operator (GTEMPO) method.
Till now GTEMPO has been applied to solve fermionic impurity problems on the imaginary~\cite{ChenGuo2024b}, real (Keldysh)~\cite{ChenGuo2024c,GuoChen2024d}, and L-shaped Kadanoff contours~\cite{ChenGuo2024g}. 
The time-translational invariance of the Feynman-Vernon IF and the infinite MPS techniques have also been explored to speedup GTEMPO~\cite{GuoChen2024e,GuoChen2024f,SunGuo2025}.

Till now the idea to represent the impurity PI as an MPS has only been explored to solve pure bosonic~\cite{JorgensenPollock2019,popovic2021-quantum,fux2021-efficient,gribben2021-using,otterpohl2022-hidden,gribben2022-exact} or pure fermionic~\cite{NgReichman2023,ParkChan2024,ChenGuo2024b,ChenGuo2024c,GuoChen2024d,ChenGuo2024a} impurity problems. However, there is no fundamental limitation for this idea to be applied for polaron impurity problems, as the Feynman-Vernon IF is still applicable in this more general case. In fact, if one views the impurity PI as the partition function of an effective 1D Hamiltonian, then the electron bath contributes the quadratic terms (i.e., hybridization), while the phonon bath contributes the quartic terms (i.e., the retarded interaction). Therefore in principle, the core techniques of GTEMPO can be readily generalized to solve polaron impurity problems.



In this work, we extend GTEMPO to solve polaron impurity problems. In the PI formalism, we can separately represent the Feynman-Vernon IF for the phonon bath (phonon IF) as a bosonic MPS in the Fock state basis (the commutation relation between the density operators is bosonic) using TEMPO, and the Feynman-Vernon IF for the electron bath (electron IF) as a GMPS in the coherent state basis. The central challenge we manage to overcome is to transform phonon IF from a bosonic MPS into a GMPS, by realizing that the effect of the phonon IF is to reweight the corresponding components of the Grassmann tensor for the bare impurity dynamics. After this transformation, only GMPSs are left and we can use all the existing techniques of GTEMPO to calculate any multi-time impurity correlations. 
Our method inherits all the important properties of (G)TEMPO: it is completely general for any type of couplings between impurity flavors and can be used on any contour; it is free of the sign problem and the bath discretization error; it can be used to calculate any multi-time impurity correlations as the obtained ADT naturally encodes all these information; it contains very few hyperparameters, essentially only the bond dimension to control the MPS bond truncation error and the time step size to discretize the IF. We perform extensive numerical benchmarks against analytical solutions and exact diagonalization in very different application scenarios to demonstrate the validity and flexibility of our method. We also perform imaginary- and real-time calculations for a full-fledged polaron impurity problem with continuous phonon and electron baths, where the imaginary-time calculations are benchmarked against CTQMC. 

This paper is organized as follows. In Sec.~\ref{sec:method}, we provide necessary background materials and detailed description of our method. In Sec.~\ref{sec:independentbosons}, we validate our method against the analytical solvable independent bosons model in which the electron bath is absent. In Sec.~\ref{sec:toymodel}, we further validate our method against exact diagonalization with a toy model which contains a single electron bath and a single phonon bath. In Sec.~\ref{sec:fullmodel}, we study a full-fledged polaron impurity problem with continuous phonon and electron baths on both the imaginary- and real-time axis, and benchmark our imaginary-time results against CTQMC. We summarize in Sec.~\ref{sec:summary}.

\section{Background and method}\label{sec:method}

\subsection{Model}
The Hamiltonian for the polaron impurity problem can be generally written as
\begin{align}
\Hop = \Himp + \Hel + \Hph,
\end{align}
where $\Himp$ is the impurity Hamiltonian, $\Hel$ describes the coupling between the impurity and the free electron bath, and $\Hph$ describes the coupling between the impurity and the free phonon bath. $\Himp$ can be generally written as 
\begin{align}\label{eq:Himp}
\Himp = \sum_{p,q}t_{pq} \adop_p\aop_q + \sum_{p,q,r,s}\nu_{pqrs}\adop_p\adop_q\aop_r\aop_s,
\end{align}
where $p,q,r,s$ are the fermion flavor labels that could contain both the spin and orbital indices, $\adop_p$ and $\aop_p$ are the fermionic creation and annihilation operators of the
$p$th impurity flavor. The first term on the right hand side of Eq.(\ref{eq:Himp}) is the tunneling term and second term is the interaction term. $\Himp$ would be a diagonal matrix in the Fock state basis if only density-density coupling between impurity flavors is allowed. 
In this work we consider $\Hel$ of the form
\begin{align}\label{eq:Hel}
\Hel = \sum_{p, k} \epsilon_{p, k} \cdop_{p,k}\cop_{p,k} + \sum_{p, k}\lambda_{p,k}(\adop_p\cop_{p,k} + \cdop_{p,k}\aop_p),
\end{align}
where $\cdop_{p, k}$ and $\cop_{p, k}$ are the fermionic creation and annihilation operators of the electron bath that is coupled to the $p$th impurity flavor, $\epsilon_{p, k}$ is the band energy and $\lambda_{p,k}$ is the coupling strength. 
Here we have restricted to diagonal coupling between impurity and bath, as each impurity flavor is coupled to its own bath in Eq.(\ref{eq:Hel}).
We will further assume $\lambda_{p,k}=\lambda_k$ by default, unless particularly specified.
$\Hph$ describes the coupling between the free phonon bath and the impurity, which can be written as
\begin{align}\label{eq:Hph}
\Hph = \sum_k \omega_k \bdop_k\bop_k + \sum_p \nop_p \sum_k g_k (\bdop_k + \bop_k),
\end{align}
where $\omega_k$ is the phonon frequency, $g_k$ is the coupling strength between the impurity and the phonon bath, $\nop_p=\adop_p\aop_p$ is the electron density operator. Here the coupling between the phonon and the impurity is non-diagonal, i.e., all the impurity flavors are coupled to the same phonon bath. 
The spectral functions of the electron and phonon baths are defined as
\begin{align}\label{eq:spectrals}
  \Gamma (\epsilon) = \sum_k \lambda_k^2 \delta (\epsilon -
  \epsilon_k), \quad J (\omega) = \sum_k g_k^2 \delta (\omega - \omega_k) ,
\end{align}
which completely determine the effects of the two baths on the impurity dynamics respectively.

To this end, we note that in Eq.(\ref{eq:Hel}) and Eq.(\ref{eq:Hph}), the polaron impurity problem has been written in a ``star configuration'', where the impurity is coupled to all the bath modes, while the modes are decoupled from each other. However, one could also perform unitary transformations on the baths to turn the ``star configuration'' into an equivalent ``chain configuration'' with only short-ranged couplings~\cite{KohnSantoro2021}. Therefore impurity problems can also be view as (quasi)-1D problems.
We also note that we have chosen different types of couplings to different baths on purpose to demonstrate the general applicability of our method, apart from their physical relevance.

\subsection{Path integral formalism for the polaron impurity problem}

\begin{figure}
  \includegraphics[width=\columnwidth]{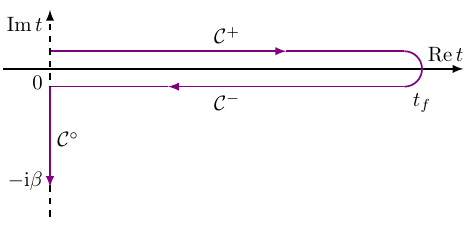} 
  \caption{Unified view of different contours that are commonly used in the path integral formalism. $\contour^\circ$ denotes the imaginary branch defined on the imaginary-time axis, $\contour^{\pm}$ denote the forward and backward branches defined on the real-time axis. The imaginary contour contains $\contour^\circ$ only, which can be used to calculate the imaginary-time impurity correlation functions such as the Matsubara Green's function. The Keldysh contour contains $\contour^{\pm}$, which can be used to calculate non-equilibrium impurity correlation functions starting from a separable state in Eq.(\ref{eq:separablestate}). The L-shaped Kadanoff contour contains all the three branches, which can be used to calculate equilibrium impurity correlation functions from the thermal state in Eq.(\ref{eq:thermalstate}). The arrow denotes the direction of integration along the contour. 
    }
    \label{fig:contour}
\end{figure}

The analytic expression of the impurity path integral for the polaron impurity problem is well-known in literatures~\cite{assaad2007-diagrammatic,Hafermann2014}, which can be written as
\begin{align}\label{eq:Z}
  Z = \Zph \Zel \int \mathcal{D} [\bar{\tmmathbf{a}}
  \tmmathbf{a}] \gK [\bar{\tmmathbf{a}} \tmmathbf{a}]
  \gIel [\bar{\tmmathbf{a}} \tmmathbf{a}]
  \gIph [\bar{\tmmathbf{a}} \tmmathbf{a}],
\end{align}
where $\Zph$ and $\Zel$ are the partition functions of the free phonon and electron baths respectively, $\bar{\tmmathbf{a}}_p = \{\abar_p(\tau)\}$, $\tmmathbf{a}_p = \{a_p(\tau)\}$ are the Grassmann trajectories for the $p$th flavor over the whole time interval, and $\bar{\tmmathbf{a}} = \{\boldabar_p, \boldabar_q, \cdots\}$, $\tmmathbf{a} = \{\bolda_p, \bolda_q, \cdots\}$. The measure $\mathcal{D} [\bar{\tmmathbf{a}}\tmmathbf{a}]$ is defined as
\begin{align}
\mathcal{D} [\bar{\tmmathbf{a}}\tmmathbf{a}] = \prod_{p, \tau} \dd\abar_p(\tau)\dd a_p(\tau) e^{-\abar_p(\tau)a_p(\tau)}.
\end{align}
The integrand in Eq.(\ref{eq:Z}) defines the \textit{augmented density tensor}, denoted as
\begin{align}\label{eq:adt}
\gA [\bar{\tmmathbf{a}} \tmmathbf{a}]= \gK [\bar{\tmmathbf{a}} \tmmathbf{a}] \gIel [\bar{\tmmathbf{a}} \tmmathbf{a}]\gIph [\bar{\tmmathbf{a}} \tmmathbf{a}],
\end{align}
where the three terms
$\gK [\bar{\tmmathbf{a}} \tmmathbf{a}]$, $\gIel [\bar{\tmmathbf{a}} \tmmathbf{a}]$ and $\gIph [\bar{\tmmathbf{a}} \tmmathbf{a}]$ 
represent the contributions from $\Himp$, $\Hel$ and $\Hph$ respectively. We can see that the effect of ADT is to put a weight on each Grassmann trajectory. In the PI formalism, three scenarios are usually considered: (1) imaginary time evolution on the imaginary contour, with which one can calculate, for example, the Matsubara Green's function
\begin{align}
G_{pq}(\tau) = -\langle \aop_p(\tau)\adop_q\rangle^{\rm eq}, \quad \forall \quad 0\leq \tau\leq \beta,
\end{align}
where $\langle \cdots\rangle^{\rm eq} = \mtrace[\rhoop^{\rm eq}\cdots]/\mtrace[\rhoop^{\rm eq}]$, and
\begin{align}\label{eq:thermalstate}
\rhoop^{\rm eq}=e^{-\beta\Hop}
\end{align}
is the equilibrium state with inverse temperature $\beta$; (2) real-time evolution on the Keldysh contour~\cite{Keldysh1965,LifshitzPitaevskii1981}, with which one can calculate, for example, the non-equilibrium greater and lesser Green's functions
\begin{align}
G_{pq}^{{\rm neq}, >}(t, t') &= -\im \langle \aop_p(t)\adop_q(t')\rangle^{\rm neq}; \\
G_{pq}^{{\rm neq}, <}(t, t') &= \im \langle \adop_q(t)\aop_p(t')\rangle^{\rm neq},
\end{align}
where $\langle \cdots\rangle^{\rm neq} = \mtrace[\rhoop^{\rm neq}\cdots]/\mtrace[\rhoop^{\rm neq}]$, and $\rhoop^{\rm neq}$ is a separable impurity-bath initial state defined as
\begin{align}\label{eq:separablestate}
\rhoop^{\rm neq} = \rhoimp \otimes \rhobath^{\rm eq},
\end{align}
with $\rhobath^{\rm eq}$ the bath equilibrium state (noticing the difference with $\rhoop^{\rm eq}$), and $\rhoimp$ some arbitrary impurity initial state; (3) L-shaped time evolution on the Kadanoff contour~\cite{kadanoff1962-quantum}, with which one can calculate not only the observables on the imaginary-time axis, but also the real-time equilibrium Green's functions such as  
\begin{align}
G_{pq}^{{\rm eq},>}(t) &= -\im \langle \aop_p(t)\adop_q\rangle^{\rm eq}; \\
G_{pq}^{{\rm eq}, <}(t) &= \im \langle \adop_q(t)\aop_p\rangle^{\rm eq}.
\end{align}
We note that the meaning of non-equilibrium and equilibrium Green's functions in the context of this work are only related to the different choices of the initial states, and has nothing to do with the subsequent dynamics, e.g., whether the Hamiltonian is time-dependent or not. The other observables we will calculate in the numerical examples of this work are the density-density correlation on different contours defined as:
\begin{align}
X_{pq}(\tau) &= \langle \nop_p(\tau)\nop_q\rangle^{\rm eq}; \label{eq:X1} \\
X^{{\rm neq}}_{pq}(t, t') &= \langle \nop_p(t)\nop_q(t')\rangle^{\rm neq}; \label{eq:X2} \\
X_{pq}^{\rm eq}(t) &= \langle \nop_p(t)\nop_q\rangle^{\rm eq}, \label{eq:X3}
\end{align}
which are useful in DMFT iterations in presence of electron-phonon interaction. 
In our numerical examples, we focus on observables with $p=q=1$, and their subscripts will be neglected. For our calculations on the Keldysh contour, we will further fix $\rhoimp = e^{-\beta \Himp}$ (local thermal state of the impurity) and $t'=0$, therefore we will also neglect the second argument in the non-equilibrium Green's functions and density-density correlation. 
The three different contours are schematically illustrated in Fig.~\ref{fig:contour}. 

In the following we will explicitly show the expressions of $\gK [\bar{\tmmathbf{a}} \tmmathbf{a}]$, $\gIel [\bar{\tmmathbf{a}} \tmmathbf{a}]$ and $\gIph [\bar{\tmmathbf{a}} \tmmathbf{a}]$ on the imaginary contour first, and then we briefly address the generalization to other contours.
On the imaginary-time contour, $\gK[\bar{\tmmathbf{a}} \tmmathbf{a}]$ can be formally written as
\begin{align}\label{eq:K}
\gK[\bar{\tmmathbf{a}} \tmmathbf{a}] = e^{-\int_{0}^{\beta} \dd \tau\gHimp(\tau)},
\end{align}
where $\gHimp(\tau)$ is obtained from $\Himp$ by making the substitutions $\aop_p\rightarrow a_p(\tau)$, $\adop_p\rightarrow \abar_p(\tau)$. The electron IF for the $p$th flavor is
\begin{align}\label{eq:elif}
\gIel [\boldabar_p \bolda_p] &= e^{-\int_0^{\beta}\dd\tau'\int_0^{\beta}\dd\tau''\abar_p(\tau')\Delta(\tau', \tau'')a_p(\tau'')},
\end{align}
and $\gIel [\boldabar \bolda] = \prod_p \gIel [\boldabar_p \bolda_p]$.
Here $\Delta(\tau', \tau'')$ is usually referred to as the hybridization function which is determined by the temperature and the electron spectral function $\Gamma(\epsilon)$:
\begin{align}\label{eq:Delta}
 \Delta (\tau', \tau'') = \int\dd\epsilon \Gamma(\epsilon) \Del_{\epsilon}(\tau', \tau''),
\end{align}
where $\Del_{\epsilon}(\tau', \tau'')$ is the free electron bath Matsubara Green’s function:
\begin{align}\label{eq:Del}
\Del_{\epsilon}(\tau', \tau'') = -[\Theta(\tau'-\tau'') -\bar{f}(\epsilon)] e^{-\epsilon(\tau'-\tau'')},
\end{align}
with $\bar{f}(\epsilon)=(e^{\beta\epsilon}+1)^{-1}$ the Fermi–Dirac distribution and $\Theta$ the Heaviside step function. 
The phonon IF is
\begin{align}\label{eq:phifg}
&\gIph [\bar{\tmmathbf{a}} \tmmathbf{a}] \nonumber \\ 
= &e^{- \int_0^{\beta} \dd \tau'
  \int_0^{\beta} \dd \tau'' [\sum_p \abar_p(\tau')a_p(\tau')] \Lambda (\tau', \tau'') [\sum_q \abar_q(\tau'')a_q(\tau'')]}.
\end{align}
Here $\Lambda (\tau', \tau'')$ is usually referred to as the correlation function, which is determined by the temperature and the phonon spectral function $J(\omega)$:
\begin{align}\label{eq:Lambda}
 \Lambda (\tau', \tau'') = \int\dd\omega J(\omega) \Dph_{\omega}(\tau', \tau''),
\end{align}
where $\Dph_{\omega}(\tau', \tau'')$ is the free phonon bath Matsubara Green’s function:
\begin{align}\label{eq:Dph}
\Dph_{\omega}(\tau', \tau'') = -[\Theta(\tau'-\tau'')+\bar{n}(\omega)]e^{-\omega(\tau'-\tau'')}.
\end{align}
with $\bar{n}(\omega)=(e^{\beta\omega}-1)^{-1}$ the Bose-Einstein distribution.

For other contours, one only needs to make the substitution $\int_0^{\beta} \rightarrow \int_{\contour}$ in Eq.(\ref{eq:K}), Eq.(\ref{eq:elif}) and Eq.(\ref{eq:phifg}), where $\contour$ is the specific contour, and change free bath Matsubara Green's functions in Eq.(\ref{eq:Del}) and Eq.(\ref{eq:Dph}) by the free bath contour-ordered Green's functions accordingly \cite{Chen2025}.


\subsection{Brief introduction to TEMPO and GTEMPO}
In this section, we will first review the major steps of TEMPO, and then we briefly review GTEMPO by drawing connections to TEMPO. 
For notational simplicity we will restrict our discussions to the imaginary contour.

The central goal of (G)TEMPO is to construct the augmented density tensor as a (G)MPS. 
For pure bosonic impurity problems, 
the phonon IF can be most straightforwardly written in the Fock state basis as
\begin{align}\label{eq:phif}
\gIph [\boldn] = e^{- \int_0^{\beta} \dd \tau'
  \int_0^{\beta} \dd \tau'' [\sum_p n_p (\tau')] \Lambda (\tau', \tau'') [\sum_q n_q (\tau'')]} ,
\end{align}
where $\boldn_p = \{n_p(\tau)\}$ and $\boldn = \{\boldn_p, \boldn_q, \cdots\}$ denotes the Fock space trajectory.
The ADT in this case can be calculated as $\gA[\boldn] = \gK[\boldn]\gIph[\boldn]$. For pure fermionic impurity problems, we have $\gA[\bar{\tmmathbf{a}} \tmmathbf{a}] = \gK[\bar{\tmmathbf{a}} \tmmathbf{a}]\gIel[\bar{\tmmathbf{a}} \tmmathbf{a}]$ instead.
In fact, if one neglects $\Hel$ in the polaron impurity problem, then it can be viewed as a pure bosonic impurity problem and solved with TEMPO, as the density operators obey bosonic communication relation and the impurity Hamiltonian can be mapped into a bosonic one via Jordan-Wigner transformation~\cite{JordanWigner1928}. 
For brevity, we will consider only a single flavor and neglect the flavor indices in the following derivations.

The first step of TEMPO is to discretize the continuous trajectory $\boldn$ into $M$ discrete variables with equal-distant step size $\delta\tau = \beta/M$, denoted as $\boldn = \{n_0, n_1, \cdots, n_{M-1}\}$ with $n_k = n(k\delta\tau)$. 
After that, the double integral in the exponent of phonon IF in Eq.(\ref{eq:phif}) becomes 
\begin{align}
\int_0^{\beta} \dd \tau'
  \int_0^{\beta} \dd \tau'' n (\tau') \Lambda (\tau', \tau'') n (\tau'') \approx \sum_{j,k=0}^{M-1} n_j \Lambda_{jk} n_k,
\end{align}
where $\Lambda_{jk} = \int_{j\delta\tau}^{(j+1)\delta\tau}\dd\tau' \int_{k\delta\tau}^{(k+1)\delta\tau}\dd\tau''\Lambda (\tau', \tau'')$. This procedure is referred to as the quasi-adiabatic propagator path integral (QuAPI) in literatures~\cite{makarov1994-path,makri1995-numerical}.
Denoting $\gF[\boldn] = \sum_{j,k=0}^{M-1} n_j \Lambda_{jk} n_k$, we can see that the discretized phonon IF $\gIph[\boldn] = e^{-\gF[\boldn]}$ has exactly the same form as the partition function of a classical Hamiltonian $\gF[\boldn]$: $\gF[\boldn]$ is a normal rank-$M$ tensor with $2^M$ elements, and $\gI[\boldn]$ simply takes the element-wise exponential of it. As the terms in $\gF[\tmmathbf{n}]$ all commute with each other, the discretized $\gIph[\boldn]$ can be exactly decomposed as
\begin{align}
\gIph[\boldn] = e^{-\sum_{j,k=0}^{M-1} n_j \Lambda_{jk} n_k} = \prod_{j,k=0}^{M-1}e^{-n_j \Lambda_{jk} n_k}.
\end{align}
As a result, $\gIph[\boldn]$ can be systematically built as an MPS by applying a series of one-body (noticing that $n_j \Lambda_{jk} n_k = \Lambda_{jj}n_j$ for $j=k$) and two-body gate operations onto a ``vacuum state'' with all elements $1$ (i.e., all the paths gives the same contribution). However, this scheme requires $O(M^2)$ (long-range) two-body gate operations, which could be expensive in practice. A better scheme is to regroup the terms in $\gIph[\boldn]$ as:
\begin{align}
\gIph[\boldn] = \prod_{j=0}^{M-1} \left(\prod_{k=1}^{M-1} e^{-n_j \Lambda_{jk} n_k}\right), 
\end{align}
where the term inside the bracket on the right hand side is referred to as a \textit{partial IF}. It is advantageous to built each partial IF as an MPS before hand and then multiply the $M$ partial IFs together, since the gates in one partial IF share a common variable $n_j$ and could usually be built together as an MPS with very small bond dimension. The partial IF algorithm to build $\gIph[\boldn]$ as an MPS is shown in Algorithm.~\ref{alg:partialIF}. 

\begin{algorithm}[H]
  \caption{Partial IF algorithm to build $\gIph[\boldn]$ as an MPS}\label{alg:partialIF}
  \begin{algorithmic}[1]
  \Function{vacuumstate}{$M$} 
  \State \Return a product MPS $\vert\Phi_0\rangle = \sum_{\boldn} \vert\boldn\rangle$ with $M$ sites 
  \EndFunction
  \State $\vert \Phi\rangle=$ VACUUMSTATE($M$)
  \For{$j=0:M-1$}
    \State $\vert \Phi_0\rangle=$ VACUUMSTATE($M$) \Comment{The following for-loop builds $\vert \Phi_0\rangle$ as the $j$th partial IF}
    \For{$k=0:M-1$}
      \If{$j==k$}
      \State Apply a one-body gate $\left[\begin{array}{cc}e^{\Lambda_{kk}} & 0 \\ 0 & 1 \end{array}\right]$ on the $k$th site of $\vert \Phi_0\rangle$ \Comment{the bond dimension of $\vert\Phi_0\rangle$ remains the same}
      \Else
      \State Apply a two-body gate $\left[\begin{array}{cccc}e^{\Lambda_{jk}} & 0 & 0 & 0 \\ 0 & 1 & 0 & 0 \\ 0 & 0 & 1 & 0 \\ 0 & 0 & 0 & 1 \end{array}\right]$ on the $j$th and $k$th sites of $\vert \Phi_0\rangle$ \Comment{the bond dimension of $\vert\Phi_0\rangle$ doubles}
      \EndIf
      \State $\vert \Phi_0\rangle = {\rm COMPRESS}(\vert \Phi_0\rangle, \chi)$ \Comment{Bond truncation to reduce the bond dimension of $\vert\Phi_0\rangle$ down to $\chi$}
    \EndFor
    \State $\vert \Phi\rangle= {\rm MULTIPLY}(\vert \Phi\rangle, \vert \Phi_0\rangle)$ \Comment{element-wise product}
    \State $\vert \Phi\rangle = {\rm COMPRESS}(\vert \Phi\rangle, \chi)$ 
  \EndFor
  \State \Return $\vert \Phi\rangle$
  \end{algorithmic}
  \end{algorithm}

We note that this algorithm is slightly different from the original one proposed in Ref.~\cite{StrathearnLovett2018}, in that the causality relation in the time axis is not explicitly preserved here (there is no causality relation on the imaginary-time axis either). 
It is also possible to directly write the partial IF as an MPS with a small bond dimension ($2$ in this case)~\cite{StrathearnLovett2018}, instead of using the numerical approach shown in the inner for-loop of Algorithm.~\ref{alg:partialIF}. 
The element-wise product between two MPSs can be implemented, for example, by converting one MPS into a matrix product operator (MPO) by copying its physical indices (for example, one could copy the index of a rank-$1$ tensor $X_j$ to obtain a rank-$2$ tensor $Y_{jk} = X_j\delta_{jk}$) and then performing standard MPO-MPS multiplication as done in Ref.~\cite{StrathearnLovett2018} (one can refer to the excellent reviews such as Refs.~\cite{Schollwock2011,orus2014-practical} for details of the standard MPO and MPS algorithms). For two MPSs 
with bond dimensions $\chi_1$ and $\chi_2$, the resulting MPS will have a bond dimension $\chi_1\chi_2$ if bond truncation is not performed, similar to the tensor product between two MPSs. One can also explore the time-translational invariance to speedup the construction of $\gIph[\boldn]$, for which one could refer to Ref.~\cite{GuoChen2024d} for details. In case there are multiple flavors, if all the flavors are coupled to the same bath, then one should modify the two for-loops in Algorithm.~\ref{alg:partialIF} to also iterate over the flavor indices; if each flavor is coupled to its own bath, then one can build each $\gIph[\boldn_p]$ as an MPS and then multiply all those MPSs together. In the latter case, the bond dimension of the resulting MPS will generally grow exponentially with the number of flavors as $\gIph[\boldn_p]$ are completely independent. 
Nevertheless, this issue could be resolved by using a multi-flavor extension of the GTEMPO method~\cite{SunGuo2025b}, which is particularly effective for the latter case.

After discretization, the contribution of the bare impurity dynamics can be written as (denoting $\Uimp=e^{-\delta\tau\Himp}$)
\begin{align}\label{eq:gK}
\gK[\boldn] = \langle n_{0}\vert \Uimp\vert n_{M-1}\rangle \cdots \langle n_{1}\vert \Uimp\vert n_0\rangle,
\end{align}
where we have imposed the periodic boundary condition $\langle n_{M}\vert =\langle n_{0}\vert $ on the imaginary contour. Each propagator $\langle n_{k+1}\vert \Uimp\vert n_{k}\rangle$ is a rank-$2$ tensor with two variables $n_{k+1}$ and $n_{k}$ on two successive time steps, which can be transformed into a nearest-neighbour two-body gate operation (if the variables $n_k$ are ordered in the time axis) by copying its physical indices and then applied onto a given MPS. Alternatively, one can also built $\gK[\boldn]$ as a separate MPS by applying those nearest-neighbour two-body gates onto a vacuum state, and then multiply this MPS with $\gIph[\boldn]$ to obtain $\gA[\boldn]$.

The GTEMPO algorithm for fermionic impurity problems faithfully follows TEMPO. Roughly speaking, the only differences are: (1) Grassmann MPSs are used instead of normal MPSs, as $\gK[\boldabar\bolda]$ and $\gIel[\boldabar\bolda]$ are Grassmann tensors, and the element-wise product between normal tensors needs to be reformulated for Grassmann tensors~\cite{GuoChen2024d}; (2) the contents of gate operations need to be changed accordingly, for example, there will be no one-body gates, and four-body gates could appear as the interaction term contains $4$ creation and annihilation operators in total.
One could also refer to the review~\cite{XuChen2024} for more details of the GTEMPO method.

\subsection{Extended GTEMPO for the polaron impurity problem}
As the analytical expression for the impurity PI of the polaron impurity problem is only available in terms of Grassmann variables, ultimately one needs to represent the ADT in Eq.(\ref{eq:adt}) as a GMPS instead of a normal MPS. 
If one looks at the expression in Eq.(\ref{eq:Z}), the only difference to pure fermionic impurity problems is the additional term $\gIph[\boldabar\bolda]$. Similar to $\gIel[\boldabar\bolda]$, we could still interpret $\gIph[\boldabar\bolda]$ as the partition function of an effective Hamiltonian, which contains only quartic terms instead of quadratic terms. In principle, this difference does not affect the application of the partial IF algorithm at all. However, the real difficulty comes with the discretization with a finite time step size: if we discretize $\gIph[\boldabar\bolda]$ in Eq.(\ref{eq:phifg}) in the same way as $\gIel[\boldabar\bolda]$, significant numerical errors will occur.

To understand the origin of this issue, we take a step back by explicitly deriving the path integral formula for a simplified polaron impurity problem: the imaginary-time evolution of a single-flavor impurity that is coupled to a single electron bath and a single phonon bath, i.e., we consider $\Himp = \epsilon_a \adop\aop$, $\Hel = \epsilon_0 \cdop\cop + \lambda(\adop\cop+\cdop\aop)$, $\Hph = \omega_0\bdop\bop + g\nop(\bdop+\bop)$. In this case, the spectral functions in Eq.(\ref{eq:spectrals}) are simplified into
\begin{align}\label{eq:deltaspectrums}
\Gamma (\epsilon) = \lambda^2 \delta (\epsilon - \epsilon_0),
   \quad J (\omega) = g^2 \delta (\omega - \omega_0).
\end{align}
The partition function for the imaginary time evolution can be written as
\begin{align}\label{eq:Zdef}
Z = \mtrace e^{-\beta \Hop} = \mtrace[\Uimp\cdots \Uimp].
\end{align}
Now we choose the Fock state basis $\vert n\rangle$ and $\vert m\rangle$ for the impurity and electron bath mode, and the coherent state basis $\vert \varphi\rangle$ for the phonon bath mode. In this basis, the identity operator can be written as
\begin{align}\label{eq:Iop}
\Iop=\sum_{m, n=0}^1\int\mathcal{D}[\bar{\varphi}\varphi]\vert mn\varphi\rangle\langle mn\varphi\vert,
\end{align}
where the measure for the bosonic coherent state is $\mathcal{D}[\bar{\varphi}\varphi] = \frac{\dd \bar{\varphi}\dd\varphi}{2\pi \im}e^{-\bar{\varphi}\varphi}$.
Inserting Eq.(\ref{eq:Iop}) in between each pair of $\Uimp$ on the right hand side of Eq.(\ref{eq:Zdef}), we get
\begin{align}\label{eq:Z1}
  Z = & \sum_{\tmmathbf{m} \tmmathbf{n}} \int \mathcal{D}
  [\bar{\tmmathbf{\varphi}} \tmmathbf{\varphi}] e^{-
  \bar{\tmmathbf{\varphi}} \tmmathbf{\varphi}} \langle m n \varphi
  |_M e^{- \delta \tau \Hop}  {}_{M-1}|m n \varphi\rangle \nonumber\\
  &  \times \cdots \times \langle m n \varphi |_1 e^{- \delta \tau \Hop}{}_{0}|m n \varphi \rangle, 
\end{align}
where we have used $\tmmathbf{m} = \{m_0, \cdots, m_{M-1}\}$ (similar for $\boldn$ and $\tmmathbf{\varphi}$), $\langle mn\varphi\vert_j = \langle m_jn_j\varphi_j\vert $ (and similar for${}_{j}\vert mn\varphi\rangle$) for notational convenience, and the boundary condition is
\begin{align}
\langle m_M\vert = \langle m_0\vert, \langle n_M\vert = \langle n_0\vert, \langle \varphi_M\vert = \langle \varphi_0\vert.
\end{align}
Inserting the first order approximation
\begin{align}\label{eq:trotter}
e^{-\delta\tau \Hop} \approx e^{-\delta\tau \Himp} e^{-\delta\tau \Hel} e^{-\delta\tau \Hph}
\end{align}
into Eq.(\ref{eq:Z1}), and using $\bop\vert \varphi\rangle = \varphi\vert\varphi\rangle$, $\nop\vert n\rangle = n\vert n\rangle$, we obtain
\begin{widetext}
\begin{align}\label{eq:Z2}
  Z = & \sum_{\tmmathbf{m n}} \langle m n |_M \Uimp e^{- \delta \tau \Hel} {}_{M-1}|m n \rangle \times\cdots\times \langle m n |_1 \Uimp
  e^{- \delta \tau \Hel} {}_{0}|m n \rangle \nonumber\\
   & \times \int \mathcal{D} [\bar{\tmmathbf{\varphi}}
  \tmmathbf{\varphi}] e^{- \bar{\tmmathbf{\varphi}} \tmmathbf{\varphi}}
  e^{- \delta \tau [g n_{M - 1} (\bar{\varphi}_M + \varphi_{M - 1}) +
  \omega_0 \bar{\varphi}_M \varphi_{M - 1}]} \cdots e^{- \delta \tau [g
  n_0 (\bar{\varphi}_1 + \varphi_0) + \omega_0 \bar{\varphi}_1
  \varphi_0]} \nonumber \\
  =& \Zph \sum_{\tmmathbf{m n}} \langle m n |_M \Uimp e^{- \delta \tau \Hel} {}_{M-1}|m n \rangle \cdots \langle m n |_1 \Uimp
  e^{- \delta \tau \Hel} {}_{0}|m n \rangle
  \gIph [\boldn] ,
\end{align}
\end{widetext}
where we have integrated out the second line of Eq.(\ref{eq:Z2}) involving $\bar{\tmmathbf{\varphi}}$, $\tmmathbf{\varphi}$ via Gaussian integrals~\cite{chen2023-heat,Chen2025}.
The last term $\gIph[\boldn]$ in the third line of Eq.(\ref{eq:Z2}) is the phonon IF, which is often referred to as the \textit{retarded interaction}, and the explicit expression of it is exactly Eq.(\ref{eq:phif}). For subsequent derivations, we will replace $\gIph[\boldn]$ in Eq.(\ref{eq:Z2}) by its operator form:
\begin{align}\label{eq:phifoperator}
\gIph [\hat{\tmmathbf{n}}] = e^{- \int_0^{\beta} \dd \tau'
  \int_0^{\beta} \dd \tau'' \nop (\tau') \Lambda (\tau', \tau'') \nop (\tau'')}, 
\end{align}
where $\nop(\tau)$ acts on $\vert n_k\rangle$ with $\tau=k\delta\tau$, so that we can freely change the choice of basis for the impurity.

The first line in Eq.(\ref{eq:Z2}) only contains fermionic degrees of freedom, and it is more convenient to be written in the Grassmann coherent state basis defined as $\aop\vert a\rangle=a\vert a\rangle$ and $\cop\vert c\rangle=c\vert c\rangle$,  with $a, c$ Grassmann variables. Replacing the identity operator in the first line of Eq.(\ref{eq:Z2}) as:
\begin{align}
 \sum_{m,n=0}^1\vert mn\rangle\langle mn\vert \rightarrow \int \mathcal{D} [\bar{a} a] \int \mathcal{D} [\bar{c} c] |a c
  \rangle \langle a c|,
\end{align}
we obtain
\begin{widetext}
\begin{align}\label{eq:Z3}
  Z =& Z_{\tmop{ph}} \int \mathcal{D} [\bar{\tmmathbf{a}} \tmmathbf{a}] \int
  \mathcal{D} [\bar{\tmmathbf{c}} \tmmathbf{c}] \langle (- a) (- c) |_M
  e^{- \delta \tau \Himp} e^{- \delta \tau \Hel} {}_{M-1}|a
  c \rangle\times \cdots\times \langle a c |_1 e^{- \delta \tau \Himp} e^{- \delta \tau \Hel} {}_{0}|a c \rangle
  \gIph [\hat{\tmmathbf{n}}] \nonumber \\
=& \Zph \Zel \int \mathcal{D}[\boldabar\bolda] \langle - a_M |
  \Uimp |a_{M-1}
  \rangle\times 
  \cdots\times \langle a_1 | \Uimp |a_0 \rangle 
  \gIph [\hat{\tmmathbf{n}}] \gIel[\boldabar\bolda],
\end{align}
\end{widetext}
where we have integrated out the free electron bath via fermionic Gaussian integrals in the second line.
We stress that in Eq.(\ref{eq:Z3}) $\gIph [\hat{\tmmathbf{n}}]$ is an operator that should be applied on the bra $\langle \bolda\vert$ and ket $\vert \bolda\rangle$, it is placed at the end of the integrand only for notational brevity. The minus sign in $\langle (- a) (- c) |_M$ is due to interchanges of boundary Grassmann variables. 

In the integrand on the second line of Eq.(\ref{eq:Z3}), we recognize the contribution of the bare impurity dynamics in terms of Grassmann variables:
\begin{align}\label{eq:gK}
\gK[\boldabar\bolda] = \langle - a_M |
  \Uimp |a_{M-1}
  \rangle\times 
  \cdots\times \langle a_1 | \Uimp |a_0 \rangle .
\end{align}
Therefore Eq.(\ref{eq:Z3}) differs from Eq.(\ref{eq:Z}) only in the term $\gIph[\hat{\boldn}]$. These two equations can only become the same if one can make the replacement $\nop\rightarrow \abar a$ in $\gIph[\hat{\boldn}]$.
However, such a replacement is not correct in general, which can be easily seen by evaluating a simple exponential of the density operator on the coherent states:
\begin{align}
  \langle a_{k + 1} | e^{\alpha \hat{n}} |a_k \rangle &= \langle a_{k + 1} | 1 + (e^{\alpha}-1)\nop |a_k \rangle \nonumber \\  
  &= \langle a_{k + 1} | 1 + (e^{\alpha}-1)\abar_{k+1}a_k |a_k \rangle \nonumber \\
  &= \langle a_{k + 1} |
  e^{(e^{\alpha} - 1) \bar{a}_{k + 1} a_k} |a_k \rangle \nonumber \\ 
  &\neq \langle a_{k
  + 1} | e^{\alpha \bar{a}_{k + 1} a_k} |a_k \rangle.
\end{align}
$\langle a_{k
  + 1} | e^{\alpha \bar{a}_{k + 1} a_k} |a_k \rangle$ is only a first-order approximation of $ \langle a_{k + 1} | e^{\alpha \hat{n}} |a_k \rangle$, and they are only equal in the limit $\alpha\rightarrow 0$.
In fact, the correct way to evaluate $\gIph[\hat{\boldn}]$ on the Grassmann coherent states is: (i) explicitly expanding the exponential form of $\gIph[\hat{\boldn}]$ in Eq.(\ref{eq:phifoperator}) into the summation of a series of operator products; (ii) transforming each operator product in the series into normal order (creation operators placed to the left); (iii) applying fermionic creation operators on the bra and annihilation operators on the ket. With these three steps, one could in principle convert $\gIph[\hat{\boldn}]$ into an expression of Grassmann variables only. However, the resulting Grassmann expression will generally not be the same as Eq.(\ref{eq:phifg}) with finite $\delta\tau$.

This is the origin of the issue that we can not directly use Eq.(\ref{eq:phifg}) as the starting point for GTEMPO which takes a finite value of $\delta\tau$: Eq.(\ref{eq:phifg}) is not as accurate Eq.(\ref{eq:phif}) under a finite $\delta\tau$. From the above discussions, we can see that to derive Eq.(\ref{eq:phif}) one only needs to make the first-order Trotter decomposition in Eq.(\ref{eq:trotter}), while to obtain Eq.(\ref{eq:phifg}) one needs to make an additional first-order approximation of $\gIph[\hat{\boldn}]$, which is similar to further approximate $e^{-\delta\tau\Hph} \approx 1-\delta\tau\Hph$ in Eq.(\ref{eq:trotter}). The latter approximation could easily result in significant numerical errors.
In comparison, a major reason for the success of QuAPI is that it can usually result in very accurate solutions, even with a relatively large time step size. 

Therefore we face a dilemma when dealing with the polaron impurity problem: the electron IF favors the coherent state basis while the phonon IF favors the Fock state basis. To resolve this dilemma, we recall the treatment of the bare impurity dynamics in GTEMPO. In fact, it has been shown that if one simply uses a first-order approximation for the propagator as
$\langle a_{k+1} | \Uimp |a_k \rangle \approx \langle a_{k+1} | e^{-\delta\tau\gHimp} |a_k \rangle$, then one could result in qualitative errors in the observables~\cite{ChenGuo2024b}. To resolve this issue, Ref.~\cite{ChenGuo2024b} proposed a simple trick: one could use a finer time discretization $\delta\tau' = \delta\tau/N$ (and we denote $\Uimp' = e^{-\delta\tau'\Himp}$) to further expand each propagator and then integrate out the intermediate Grassmann variables:
\begin{align}
\langle a_{k+1} | \Uimp |a_k \rangle \approx &\int\mathcal{D}[\bar{\tmmathbf{f}}\tmmathbf{f}] \langle a_{k+1} | \Uimp' f_{N-1}\rangle \times \nonumber \\ 
&\cdots\times \langle f_1 \vert \Uimp'\vert f_0\rangle \langle f_0\vert \Uimp' |a_k \rangle,
\end{align}
which will result in an accurate propagator for very small $\delta\tau'$. 
For $\gK[\boldabar\bolda]$, the computational cost of this trick is negligible since the Hilbert space size of the impurity is usually very small. However, it could be very expensive if we apply the same trick to deal with $\gIph[\boldabar\bolda]$.

To this end, we go back to Eq.(\ref{eq:Z3}) and notice that
\begin{align}
  |a \rangle \langle a| =& (|0 \rangle - a| 1 \rangle) (\langle 0| - \langle 1|
  \bar{a}) \nonumber \\ 
  =& |0 \rangle \langle 0| - a |1 \rangle \langle 0| - |0 \rangle
  \langle 1| \bar{a} + a\bar{a}|1 \rangle \langle 1| \nonumber \\
  =& \sum_{\nbar n} (-a)^n\vert n\rangle\langle \nbar\vert (-\abar)^{\nbar}.
\end{align}
Substituting the above equation into Eq.(\ref{eq:Z3}), we obtain
\begin{widetext}
\begin{align}\label{eq:Z4}
Z =& \Zph \Zel \int \mathcal{D}[\boldabar\bolda] 
\left(\sum_{\bar{\boldn}\boldn}\eta[\bar{\tmmathbf{n}}\tmmathbf{n}] \langle \nbar_0\vert \Uimp\vert n_{M-1}\rangle \times 
\cdots\times \langle \nbar_{1}\vert \Uimp\vert n_{0}\rangle  \gIph[\tmmathbf{n}]
 a_{M-1}^{n_{M-1}} \abar_{M-1}^{\bar{n}_{M-1}}\cdots a_0^{n_0}  \abar_0^{\bar{n}_0} \right) \gIel[\boldabar\bolda],
\end{align}
\end{widetext}
where $\eta[\bar{\boldn}\boldn]$ is a sign factor due to the interchange of Grassmann variables, and we have replaced the operator $\gIph[\hat{\boldn}]$ with $\gIph[\boldn]$ by applying it onto the Fock state $\vert \boldn\rangle$. The term inside the big bracket of Eq.(\ref{eq:Z4}) contains all the contributions from $\Himp$ and $\Hph$, which we denote as $\gK\gIph[\boldabar\bolda]$. Importantly, we can see that $\gK\gIph[\boldabar\bolda]$ is a Grassmann tensor, in which the component for the basis $a_{M-1}^{n_{M-1}} \abar_{M-1}^{\bar{n}_{M-1}}\cdots a_0^{n_0}  \abar_0^{\bar{n}_0}$, denoted as $\gK\gIph[\boldabar\bolda;\bar{\boldn} \boldn]$, is
\begin{align}\label{eq:KIweight}
\gK\gIph[\boldabar\bolda; \bar{\boldn}\boldn] \nonumber =& \eta[\bar{\tmmathbf{n}}\tmmathbf{n}] \langle \nbar_0\vert \Uimp\vert n_{M-1}\rangle \times \nonumber \\
&\cdots\times \langle \nbar_{1}\vert \Uimp\vert n_{0}\rangle  \gIph[\boldn].
\end{align}
Therefore the effect of $\gIph[\hat{\boldn}]$ is to multiply each component of the Grassmann tensor $\gK[\boldabar\bolda]$ by a scalar $\gIph[\boldn]$.

As a result, $\gK\gIph[\boldabar\bolda]$ can be built as a GMPS as follows: one first builds $\gK[\boldabar\bolda]$ as a GMPS, and $\gIph[\boldn]$ as a normal MPS using Algorithm.~\ref{alg:partialIF}, then $\gK\gIph[\boldabar\bolda]$ is the element-wise product between $\gK[\boldabar\bolda]$ and $\gIph[\boldn]$ according to Eq.(\ref{eq:KIweight}), during which $\gK[\boldabar\bolda]$ is treated as a normal MPS (i.e., neglecting the Grassmann basis information). 
In the meantime, $\gIel[\boldabar\bolda]$ can be built as a GMPS independently. 
After that, one could either directly multiply $\gK\gIph[\boldabar\bolda]$ and $\gIel[\boldabar\bolda]$ to obtain $\gA[\boldabar\bolda]$ as a GMPS, or use the zipup algorithm to avoid explicit construction of $\gA[\boldabar\bolda]$. In this work we choose the latter approach.
In our numerical simulations, we will set the maximum bond dimension of both $\gK\gIph[\boldabar\bolda]$ and $\gIel[\boldabar\bolda]$ to be $\chi$, and we find that the actual bond dimension of $\gIel[\boldabar\bolda]$ is generally smaller than that of $\gK\gIph[\boldabar\bolda]$.

In the following, we will demonstrate our method on extensive numerical examples. We note that the main purpose of our numerical examples is not to explore the physics, but to illustrate the accuracy and flexibility of our method. Therefore we will consider very different impurity Hamiltonians, different types of impurity-bath couplings, different parameter regimes as well as different contours.

\section{Independent bosons model}\label{sec:independentbosons}

\begin{figure}
  \includegraphics[width=\columnwidth]{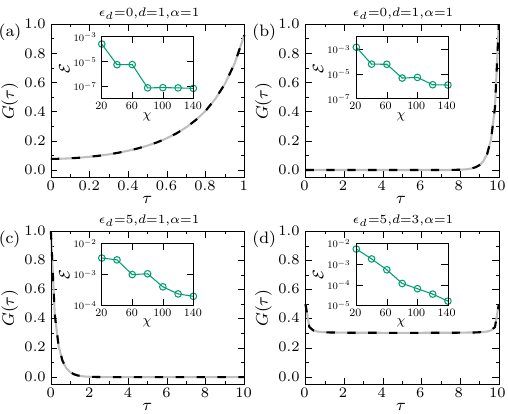} 
  \caption{The Matsubara Green's function $G(\tau)$ as a function of the imaginary time $\tau$ for the independent bosons model with a single-flavor impurity Hamiltonian in Eq.(\ref{eq:singleflavorHimp}) under different parameter settings, as shown in the title of each panel. The gray solid lines in all the panels are the exact analytical solutions, while the black dashed lines are extended GTEMPO results calculated with $\chi=140$. We have used $\beta=1$, $\delta\tau=0.05$ in (a) and $\beta=10$, $\delta\tau=0.2$ in (b,c,d). The insets show the errors between extended GTEMPO results and the analytical solutions as functions of $\chi$.
    }
    \label{fig:ib_imag_noint}
\end{figure}

\begin{figure}
  \includegraphics[width=\columnwidth]{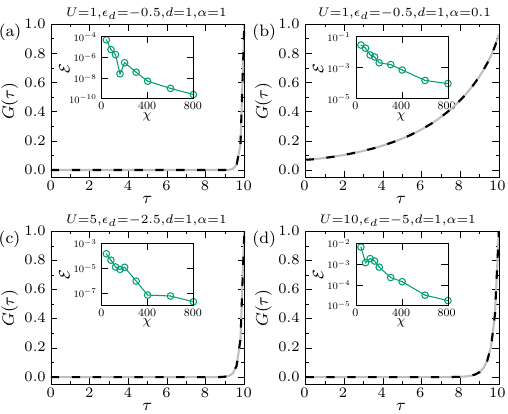} 
  \caption{The Matsubara Green's function as a function of the imaginary time $\tau$ for the independent bosons model with a two-flavor impurity Hamiltonian in Eq.(\ref{eq:twoflavorHimp}) under different parameter settings, as shown in the title of each panel. The gray solid lines in all the panels are the exact analytical solutions, while the black dashed lines are extended GTEMPO results with $\chi=140$.
  We have used $\beta=10$, $\delta\tau=0.2$ in all panels. The insets show the errors between extended GTEMPO results and the analytical solutions as functions of $\chi$.
    }
    \label{fig:ib_imag_int}
\end{figure}

We first consider the independent bosons model, in which $\Hel$ is absent~\cite{mahan2000-many}.  
This model can be analytically solved by employing the Lang-Firsov canonical transformation~\cite{lang1963-kinetic,mahan2000-many}.
Without the electron baths, the model can be directly solved with TEMPO only, but here we will still solve it with our extended GTEMPO in the coherent state basis, to test our construction of $\gK\gIph[\boldabar\bolda]$ as a GMPS in particular.
Another advantage of this model which is useful for our test is that in this special case 
GTEMPO is free of the Trotter decomposition error (essentially because that QuAPI is free of the Trotter decomposition error, which can be seen in Appendix.~\ref{app:ib} from the alternative derivation of the analytic solutions for the independent bosons model via the Feynman-Vernon IF). 
Thus the only source of error left in extended GTEMPO is the MPS bond truncation error, which is fully characterized by the bond dimension $\chi$ of $\gK\gIph[\boldabar\bolda]$. 

In the following, 
we will apply extended GTEMPO to solve the independent bosons model, for both spinless (a single flavor) and spin (two flavors, spin up and spin down) impurity Hamiltonians. We will also consider both the imaginary-time contour and the L-shaped Kadanoff contour, to calculate the Matsubara Green's function and the real-time equilibrium Green's functions respectively. The convergence of the extended GTEMPO results with respect to $\chi$ will be analyzed throughout this section. As the Lang-Firsov transformation is not applicable to a separable initial state, the Keldysh contour will not be considered.

For the phonon bath, we choose a continuous spectral function as (which is also used, for example, in Refs.~\cite{StrathearnLovett2018,XuStockburger2022})
\begin{align}\label{eq:continuousbosonspectrum}
J(\omega) = \frac{\alpha}{2} \frac{\omega^d}{\omega_c^{d-1}}e^{-\frac{\omega}{\omega_c}},
\end{align}
where $\omega_c$ is a high-frequency cut off, $\alpha$ is the interaction strength between the impurity and the phonon bath, $d$ is the dimensionality of the bath. Throughout this work we fix $\omega_c = 5$.
We will use the mean square error, defined as $\mathcal{E}(\tmmathbf{x}, \tmmathbf{y}) = ||\tmmathbf{x}-\tmmathbf{y}||^2/L$, between two vectors $\tmmathbf{x}$ and $\tmmathbf{y}$ of size $L$, to characterize the error between our extended GTEMPO results and other results.

We first consider the imaginary-time evolution of the single-flavor case, with
\begin{align}\label{eq:singleflavorHimp}
\Himp = \epsilon_a \nop,
\end{align}
where $\epsilon_a$ is the on-site energy.
In Fig.~\ref{fig:ib_imag_noint}, we plot $G(\tau)$ calculated by extended GTEMPO (the dashed line) as a function of $\tau$
for several very different parameters combinations: panels (a, b) differ only by the inverse temperature $\beta$, panels (b, c) differ only by $\epsilon_a$, panels (c, d) differ only by the phonon environment dimensionality $d$. 
The solid line are the corresponding analytical solutions.
In the insets we show the errors between the extended GTEMPO results and the analytical solutions as functions of $\chi$. In panel (a) we have used $\delta\tau=0.05$ and in the rest panels we have used $\delta\tau=0.2$ (for this model the choice of $\delta\tau$ is irrelevant for extended GTEMPO, and we have chosen those values of $\delta\tau$ only for visual purpose).
We can see that indeed our method can reach extremely high precision in this case (with $\chi=140$, the errors are less than $10^{-4}$ in panels (a,b,d), and less than $10^{-3}$ in panel (c)), and becomes more accurate as $\chi$ increases. 

\begin{figure}
  \includegraphics[width=\columnwidth]{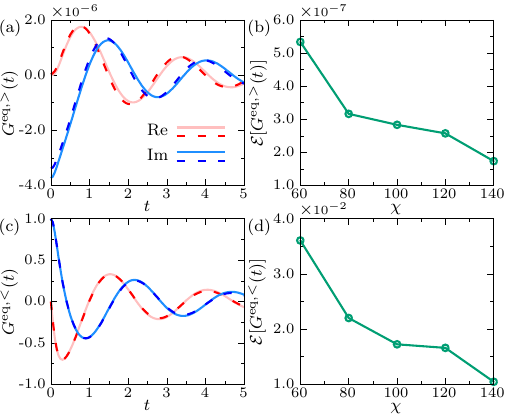} 
  \caption{(a) The greater Green's function $G^>(t)$ and (c) the lesser Green's function $G^<(t)$ as a function of the real time $t$, for the independent bosons model with a single-flavor impurity Hamiltonian in Eq.(\ref{eq:singleflavorHimp}). The red and blue dashed lines are the extended GTEMPO results for the real and imaginary parts of the two Green's functions respectively, and the solid lines with similar but lighter colors are the corresponding analytical solutions. (b,d) The error of the Green's function against the analytical solution as a function of $\chi$, corresponding to (a,c) respectively. We have used $\epsilon_a=0$, $\beta=5$, $\delta\tau=0.1$, $t=5$, $\delta t=0.05$, $d=1$ (ohmic spectral function), $\alpha=1$ in these simulations. For (a,c) we have used $\chi=140$.
    }
    \label{fig:ib_mixed_noint}
\end{figure}

\begin{figure}
  \includegraphics[width=\columnwidth]{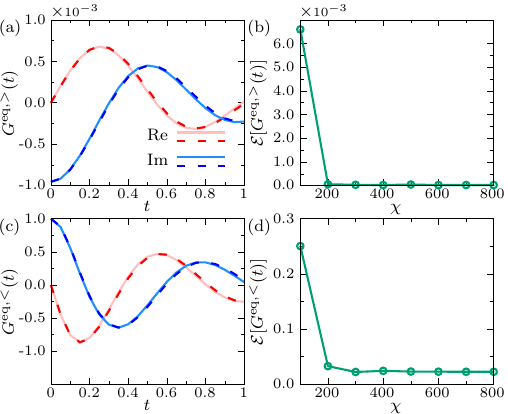} 
  \caption{Similar to Fig.~\ref{fig:ib_mixed_noint}, but for the independent bosons model with a two-flavor interacting impurity Hamiltonian in Eq.(\ref{eq:twoflavorHimp}). We have used $U=1$, $\epsilon_a=-0.5$, $\beta=1$, $\delta\tau=0.1$, $t=1$, $\delta t=0.05$, $d=1$, $\alpha=1$ in these simulations. For (a, c) we have used $\chi=800$.
    }
    \label{fig:ib_mixed_int}
\end{figure}

In Fig.~\ref{fig:ib_imag_int}, we calculate the Matsubara Green's function for the two-flavor impurity Hamiltonian
\begin{align}\label{eq:twoflavorHimp}
\Himp = \epsilon_a (\nop_{\uparrow} + \nop_{\downarrow}) + U\nop_{\uparrow}\nop_{\downarrow},
\end{align}
where $U$ is the Coulomb interaction strength between the two flavors.
Similar to the single-flavor case, we consider very different parameter settings: panels (a, b) differ only by $\alpha$, panels (a,c,d) differ only by the values of $U$ and $\epsilon_a$. The insets show the errors as functions of the bond dimension $\chi$.
We can see that in the two-flavor case, with the presence of interaction between the two flavors, the extended GTEMPO results still match perfectly with the analytical solutions and improves with larger $\chi$, and the errors in all parameter settings are of the order $10^{-4}$ or less with $\chi=140$.

In Fig.~\ref{fig:ib_mixed_noint} and Fig.~\ref{fig:ib_mixed_int}, we perform extended GTEMPO calculations on the L-shaped Kadanoff contour to obtain the equilibrium greater and lesser Green's functions, for the single-flavor and two-flavor cases respectively. 
The extended GTEMPO results are shown in dashed lines in panels (a,c) of both figures, and the corresponding analytical solutions are shown in solid lines with the same colors for comparison. 
In Fig.~\ref{fig:ib_mixed_noint} we have used $\epsilon_a=0$, $\beta=5$, $t=5$, while in Fig.~\ref{fig:ib_mixed_int} we have used $U=1$, $\epsilon_a=-0.5$, $\beta=1$, $t=1$. We can see that in both cases we can reach an accuracy of the order $10^{-2}$ (roughly the same order as the discrete time step sizes we have used), and that the error goes down with larger $\chi$. We note that for calculations on the L-shaped contour, there are two time step sizes, $\delta\tau$ and $\delta t$, and it is generally harder to achieve perfect convergence due to the interplay between these two time scales. For example, if we make one time scale much larger than the other, then the MPS bond truncation error would be highly unbalanced on the two time axis, which would likely result in large errors during MPS bond truncation, even though QuAPI is exact for this model. Therefore, for calculations on the Kadanoff contour one would like to change $\delta t$ and $\delta \tau$ simultaneously, but keep their ratio roughly the same.
As we have fixed $\delta\tau$ in our calculations we are not able to achieve extremely high accuracy as in the case of the imaginary contour. Nevertheless, the main purpose of this calculation is to illustrate the flexibility of the extended GTEMPO method, and we are not supposed to reach arbitrarily high precision in general as the first-order Trotter decomposition error will eventually come into play. Here we also note that for the two-flavor case we have used smaller $\beta$ and $t$, but larger $\chi$ (we can see from Fig.~\ref{fig:ib_mixed_int} that the results in the two-flavor case have converged at about $\chi=300$), which is because that the number of terms in $\gIph[\boldn]$ in this case is $4$ times larger than that in the single-flavor case, therefore it is natural that a larger bond dimension is required (the number of partial IFs in Algorithm.~\ref{alg:partialIF} will be two times more than the single-flavor case).

\section{Toy model with a single electron bath and a single phonon bath}\label{sec:toymodel}

\begin{figure}
  \includegraphics[width=\columnwidth]{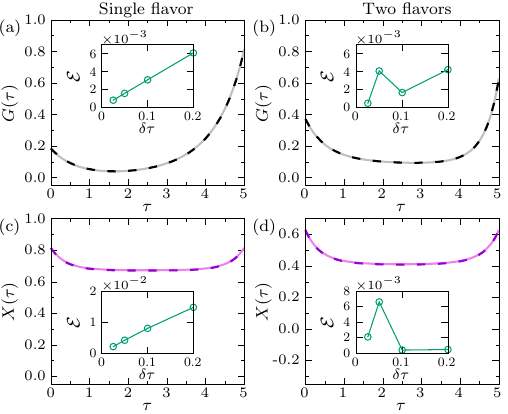} 
  \caption{(a, c) The Matsubara Green's function $G(\tau)$ (a) and the density-density correlation $X(\tau)$ (c) as functions of the imaginary time $\tau$, for the single-flavor case with impurity Hamiltonian in Eq.(\ref{eq:singleflavorHimp}). (b, d) The Matsubara Green's function (b) and the density-density correlation (d) for the two-flavor case with impurity Hamiltonian in Eq.(\ref{eq:twoflavorHimp2}). The dashed lines in all the panels are extended GTEMPO results while the solid lines are the corresponding ED results. The other parameters used which are common for all these simulations are $\beta=5$, $\delta\tau=0.025$, $\chi=100$. The insets show the errors between the extended GTEMPO results and the ED results as functions of $\delta\tau$.
    }
    \label{fig:toy_imag}
\end{figure}

\begin{figure}
  \includegraphics[]{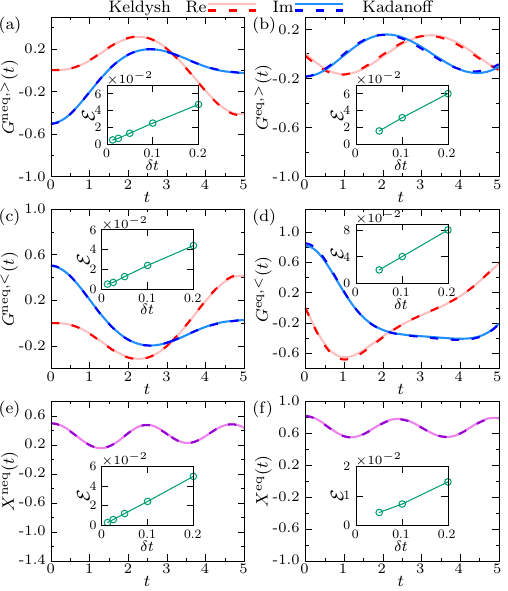} 
  \caption{(a,c,e) The non-equilibrium greater Green's function $G^{{\rm neq},>}(t)$ (a), lesser Green's function $G^{{\rm neq}, <}(t)$ (c), density-density correlation $X^{\rm neq}(t)$ (e) as functions of $t$, for real-time evolution starting from the separable state in Eq.(\ref{eq:separablestate}) (the Keldysh contour) in the single-flavor case. The red and blue dashed lines in (a) represent the real and imaginary parts of $G^{{\rm neq},>}(t)$ and similar for (c). The solid lines with similar but lighter colors are the corresponding ED results. (b,d,f) The equilibrium greater Green's function $G^{{\rm eq},>}(t)$ (b), lesser Green's function $G^{{\rm eq},<}(t)$ (d), density-density correlation $X^{\rm eq}(t)$ (f) for real-time evolution starting from the thermal state in Eq.(\ref{eq:thermalstate}) (the Kadanoff contour) in the single-flavor case. The insets show the errors of the extended GTEMPO results against ED as functions of the real-time step size $\delta t$. In (a,c,e) we have used $\delta t=0.0125$. In (b,d,f) we have used $\delta t=0.05$ and $\delta \tau =0.05$.
  We have used $\beta=5$, $\chi=300$ in all these simulations.
    }
    \label{fig:toy_real_mixed_noint}
\end{figure}

\begin{figure*}
  \includegraphics[width=2\columnwidth]{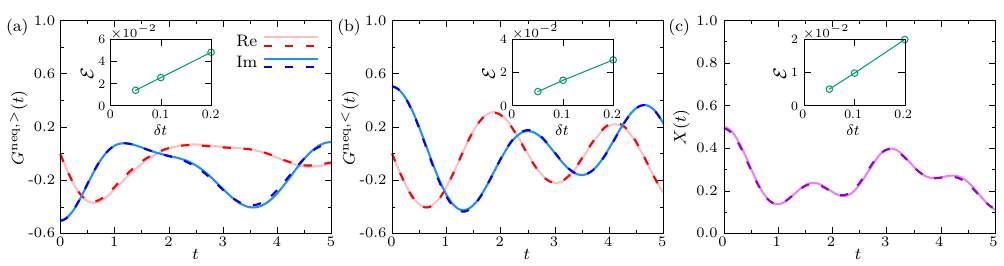} 
  \caption{The non-equilibrium greater Green's function $G^{{\rm neq},>}(t)$ (a), lesser Green's function $G^{{\rm neq}, <}(t)$ (b), density-density correlation $X^{\rm neq}(t)$ (c) as functions of $t$, for the real-time evolution starting from the separable state in Eq.(\ref{eq:separablestate}) for the two-flavor impurity Hamiltonian in Eq.(\ref{eq:twoflavorHimp2}). In these simulations we have used $\chi=800$, $\beta=5$, $t=1$, $\delta t=0.05$. The red and blue dashed lines in (a) represent the real and imaginary parts of $G^{{\rm neq},>}(t)$ and similar for (b). The solid lines with similar but lighter colors are the corresponding ED results. The insets show the errors of these observables against ED as functions of the real-time step size $\delta t$.
    }
    \label{fig:toy_real_int}
\end{figure*}

In the next we will consider the presence of both the electron and the phonon baths. As analytical solutions are typically not available in this general case, we will first consider toy models with a single phonon and a single electron bath in this section, in which case we can easily compare our extended GTEMPO results against exact diagonalization (ED).
Concretely, in this section we will restrict ourselves to the delta spectral functions in Eq.(\ref{eq:deltaspectrums})
with $\lambda=1$, $\epsilon_0=1$ and $g=1/\sqrt{2}$, $\omega_0=1$ respectively.
The only approximation that needs to be made in ED is to use a finite Hilbert space size for the phonon mode. In our simulations we choose it as $50$ and we have checked that our ED results have well converged under this value.
Again we will consider two cases for the impurity Hamiltonian: (i) a single-flavor case with impurity Hamiltonian in Eq.(\ref{eq:singleflavorHimp}) and we set $\epsilon_a=0$; (ii) a two-flavor case with 
\begin{align}\label{eq:twoflavorHimp2}
\Himp = \epsilon_a (\nop_1+\nop_2) + U\nop_1\nop_2 + J(\adop_1\aop_2 + \adop_2\aop_1),
\end{align}
which describes two spinless fermions with on-site energy $\epsilon_a$, interaction strength $U$ and tunneling strength $J$, 
and we set $U=2$, $J=1$, $\epsilon_a=-1$.
In the two-flavor case we have added a tunneling term on top of the one used in Eq.(\ref{eq:twoflavorHimp}), to illustrate that we can freely treat generic impurity Hamiltonians with non-diagonal couplings between flavors as well.
In addition, we assume that the electron bath is only coupled to the first flavor, i.e. $\lambda_{2, k}=0$ in Eq.(\ref{eq:Hel}), to further illustrate the flexibility of our method (in extended GTEMPO, we can simply set $\gIel[\boldabar_2\bolda_2]=1$ in this case).
In this section we will focus on analyzing the errors against the time step sizes, and use a large enough bond dimension for extended GTEMPO such that the results have mostly converged with respect to it.

In Fig.~\ref{fig:toy_imag}, we show the extended GTEMPO results on the imaginary-time contour. In panels (a,c), we show $G(\tau)$ and $X(\tau)$ (defined in Eq.(\ref{eq:X1})) calculated for the single-flavor case and in panels (b,d) we show these two observables calculated for the two-flavor case. In the insets we show the errors between extended GTEMPO results and the ED results as functions of $\delta\tau$. We can see that the extended GTEMPO results generally agree very well with ED, with errors of the order $10^{-2}$ or less, even for $\delta\tau=0.2$. There are spurious jumps of error in the two-flavor case when decreasing $\delta\tau$ beyond $0.1$, but the errors are already of the order $10^{-3}$.
These results also illustrate the power of QuAPI: even with a relatively large time step size, one could still obtain accurate results.

In Fig.~\ref{fig:toy_real_mixed_noint}, we calculate the real-time Green's functions and density-density correlation for the single-flavor case with extended GTEMPO on both the Keldysh (the non-equilibrium scenario) and the Kadanoff (the equilibrium scenario) contours. 
In panels (a,c,e), we show $G^{{\rm neq}, >}(t)$, $G^{{\rm neq},<}(t)$ and $X^{\rm neq}(t)$ as functions of the real time $t$ calculated on the Keldysh contour respectively, while in panels (b,d,f), we show $G^{{\rm eq},>}(t)$, $G^{{\rm eq},<}(t)$ and $X^{\rm eq}(t)$ calculated on the Kadanoff contour. In the insets, we plot the errors of these results against ED as functions of the real-time step size $\delta t$. We can see that our extended GTEMPO results in both the non-equilibrium and equilibrium scenarios match very well with ED, with the errors of the order $10^{-2}$ which are comparable to the time step sizes we have used. Interestingly, the errors decrease almost linearly against $\delta t$ in all these simulations, which also reflects that the bond dimension we have used is large enough in these simulations and the dominant error is the time discretization error. For extended GTEMPO calculations on the Kadanoff contour, we have only considered $\delta t$ till $0.05$, which is because that we have fixed $\delta\tau$ (one should generally change $\delta t$ and $\delta\tau$ simultaneously on the Kadanoff contour as have been discussed).

In Fig.~\ref{fig:toy_real_int}, we calculate the non-equilibrium Green's functions and density-density correlation for the two-flavor case with extended GTEMPO. In panels (a,b,c) we show $G^{{\rm neq},>}(t)$, $G^{{\rm neq},>}(t)$ and $X^{\rm neq}(t)$ as functions of the real-time $t$ respectively, and in the insets we show the their errors against ED as functions of $\delta t$. In this case we have only considered $t=1$ as a larger bond dimension is required in extended GTEMPO ($\chi=800$ is used in this case). We can see that the behaviors of these non-equilibrium real-time observables are similar to the results in Fig.~\ref{fig:toy_real_mixed_noint}: the errors are still of the order $10^{-2}$ and decrease almost linearly with smaller $\delta t$. 
These results illustrate that our extended GTEMPO can well handle generic couplings between impurity flavors.

\section{Full-fledged model with continuous baths}\label{sec:fullmodel}
In this section, we study a full-fledged model where the impurity is coupled to continuous phonon and electron baths. We will still consider the two cases of impurity Hamiltonians that are studied in Sec.~\ref{sec:independentbosons}. For the single-flavor case, we choose $\epsilon_a=0$, while for the two-flavor case we choose $U=1$ and $\epsilon_a=-0.5$. For the free phonon bath we use the spectral function in Eq.(\ref{eq:continuousbosonspectrum}) and fix $d=1$, for the free electron bath we will use a semi-circular spectrum (which is also used, for example, in Refs.~\cite{WolfSchollwock2014,WolfSchollwock2015})
\begin{align}
\Gamma(\epsilon) = \frac{\pi}{2} \sqrt{1-\epsilon^2}.
\end{align}
We note that if the phonon bath contains only a single mode, then the single-flavor case we consider here reduces to the Holstein polaron impurity model~\cite{holstein1959-studies,holstein1959-studies-II}, while the two-flavor case reduces to the Anderson-Holstein model~\cite{Hafermann2014,chen2016-anderson}.

We will perform extended GTEMPO calculation on both the imaginary and the Keldysh contours. For imaginary-time calculations, we benchmark our extended GTEMPO results against CTQMC. For real-time calculations on the Keldysh contour, there does not exist any reliable methods for benchmarking to our knowledge, therefore we will only perform error analysis of our extended GTEMPO results against the bond dimension $\chi$ and the real-time step size $\delta t$.

\subsection{Imaginary-time evolution}

\begin{figure}
  \includegraphics[width=\columnwidth]{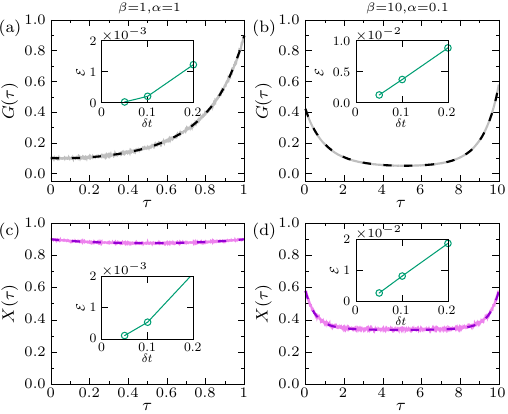} 
  \caption{(a,c) Matsubara Green's function $G(\tau)$ (a) and the density-density correlation $X(\tau)$ as functions of the imaginary time $\tau$, for the single-flavor case with $\epsilon_a=0$, $\beta=1$, $\alpha=1$. (b,d) Similar to (a,c) but for $\beta=10$, $\alpha=0.1$. The dashed lines are the extended GTEMPO results, and the solid lines with similar but lighter colors are the corresponding CTQMC results. For (a,c) we have used $\chi=100$ while for (b,d) we have used $\chi=200$. For all these simulations we have used $\delta\tau=0.025$. In the insets, we use the extended GTEMPO results calculated with $\delta\tau=0.025$ as the baseline, and plot the errors between the baseline and the other extended GTEMPO results calculated with $\delta\tau=0.2, 0.1, 0.05$ respectively, as a function of $\delta\tau$. Here the CTQMC results are calculated using $8$ Markov chains with $10^7$ samples generated in each chain.
    }
    \label{fig:full_imag_noint}
\end{figure}

\begin{figure}
  \includegraphics[width=\columnwidth]{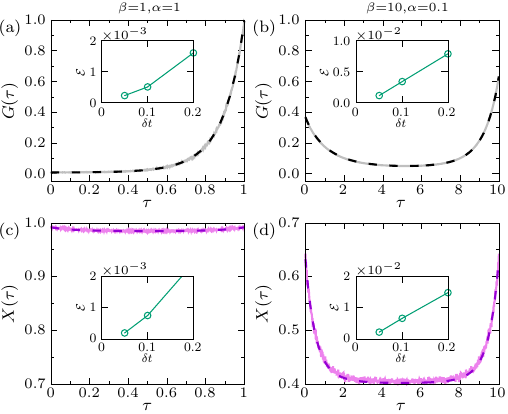} 
  \caption{Similar to Fig.~\ref{fig:full_imag_noint}, but for the two-flavor case with $U=1$, $\epsilon_a=-0.5$. For all these simulations we have used $\delta\tau=0.025$. For (a,c) we have used $\chi=100$ while for (b,d) we have used $\chi=150$. Here the CTQMC results are calculated using $8$ Markov chains with $10^7$ samples generated in each chain.
    }
    \label{fig:full_imag_int}
\end{figure}

We first consider imaginary-time evolution, and the results for the two cases are shown in Fig.~\ref{fig:full_imag_noint} and Fig.~\ref{fig:full_imag_int} respectively. In Fig.~\ref{fig:full_imag_noint}(a, c), we fix $\beta=1$ and $\alpha=1$ and plot $G(\tau)$ and $X(\tau)$ as functions of the imaginary time $\tau$ respectively. The dash lines are extended GTEMPO results calculated with $\delta\tau=0.025$, while the solid lines with lighter colors are the corresponding CTQMC results. In Fig.~\ref{fig:full_imag_noint}(b, d), we plot the same observables as in Fig.~\ref{fig:full_imag_noint}(a, c), but for $\beta=10$, $\alpha=0.1$ (lower temperature and weaker electron-phonon interaction). We can see that the Matsubara Green's functions calculated by extended GTEMPO and CTQMC match very well with each other. While for the density-density correlation, the fluctuations of the CTQMC results are much stronger than the case of Matsubara Green's function, even though we have already used a total of $8\times 10^7$ samples, which indicates that for this model the density-density correlation is a much harder observable to compute than the Matsubara Green's function by CTQMC. In fact, we have also observed that the fluctuation of the CTQMC results becomes even stronger at lower temperature, which is the reason that we have used $\alpha=0.1$ for the case $\beta=10$. In the insets, we take the extended GTEMPO results calculated with $\delta\tau=0.025$ as the baseline, then we compute the errors between the baseline and the other extended GTEMPO results calculated with $\delta\tau=0.2, 0.1, 0.05$ respectively, and plot these errors as a function of $\delta\tau$. We can see that errors become monotonically smaller with smaller $\delta\tau$.

In Fig.~\ref{fig:full_imag_int}, we plot the same observables as in Fig.~\ref{fig:full_imag_noint}, but for the two-flavor case instead. In the two-flavor case, we observe a similar error behavior of the extended GTEMPO results. Again we can see that the fluctuation of $X(\tau)$ is significantly stronger than that of $G(\tau)$ in CTQMC.

From the above benchmarks on the imaginary contour, we can see that on the one hand, the extended GTEMPO results well match the CTQMC results (although the CTQMC results for $X(\tau)$ have strong fluctuations), on the other hand, our extended GTEMPO results could be advantageous even for imaginary-time calculations, as it is free of the sampling noise. In addition, for all the extended GTEMPO calculations carried out in this work, we observe a similar level of accuracy for both the Green's functions and the density-density correlation, which is expected as they are calculated using the same ADT we have obtained (the computational costs for these two observables are also similar in our method). This is also very different from CTQMC, where the latter seems to be quite sensitive to the types of observables been calculated, to the parameter settings as well as the temperature.

\subsection{Non-equilibrium real-time evolution}

Finally, we study the non-equilibrium real-time evolution of the full-fledged model starting from a separable state in Eq.(\ref{eq:separablestate}) and calculated the non-equilibrium Green's functions with our extended GTEMPO. We will consider both the single-flavor and the two-flavor cases. In each case, we will consider two very different parameter settings with $\alpha=1$ (strong electron-phonon interaction) and $\alpha=0.1$ (weak electron-phonon interaction) respectively.

\begin{figure*}
  \includegraphics[width=2\columnwidth]{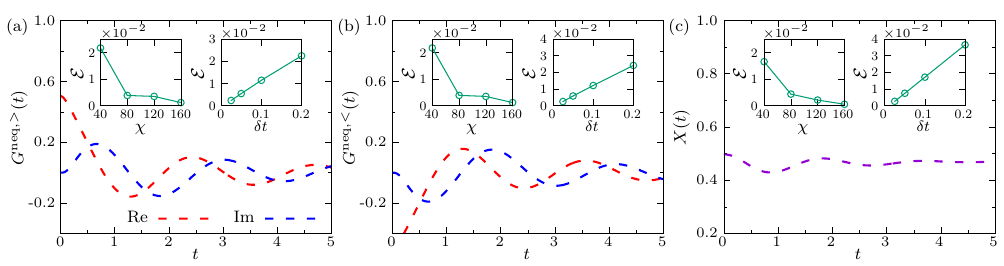} 
  \caption{The non-equilibrium greater Green's function $G^{{\rm neq},>}(t)$ (a), lesser Green's function $G^{{\rm neq}, <}(t)$ (b), density-density correlation $X^{\rm neq}(t)$ (c) as functions of $t$, for real-time evolution starting from the separable state in Eq.(\ref{eq:separablestate}) (the Keldysh contour) in the single-flavor case with $\alpha=1$. The red and blue dashed lines in (a) represent the real and imaginary parts of $G^{{\rm neq},>}(t)$ respectively, and similar for (b). The observables in the main panels are calculated using extended GTEMPO with $\chi=200$, $\beta=10$, $t=5$ and $\delta t=0.0125$, which are used as the baselines for the error analysis in the insets. 
  In each panel, the left inset shows the error analysis against the bond dimension $\chi$, where the errors are between the baseline and the extended GTEMPO results calculated with different values of $\chi$, the right inset shows the error analysis against the real-time step size $\delta t$, where the errors are between the baseline and the extended GTEMPO results calculated with different values of $\delta t$. In the left insets, we have fixed $\delta t=0.0125$ and in the right insets we have fixed $\chi=200$.
    }
    \label{fig:full_real_noint_a}
\end{figure*}

\begin{figure*}
  \includegraphics[width=2\columnwidth]{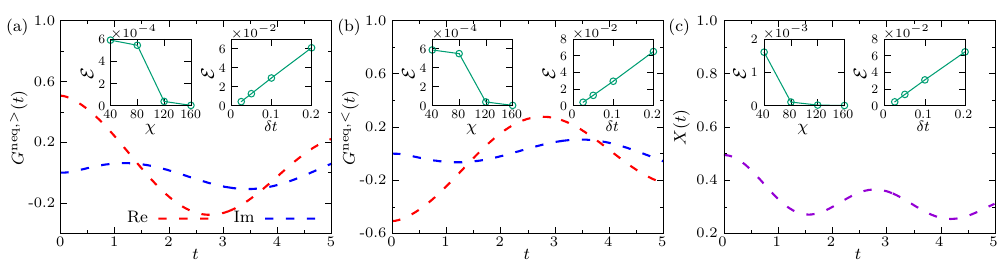} 
  \caption{Similar to Fig.~\ref{fig:full_real_noint_a}, but for the single-flavor case with $\alpha=0.1$. The hyperparameters used in these simulation are also the same as those used in Fig.~\ref{fig:full_real_noint_a}. 
    }
    \label{fig:full_real_noint_b}
\end{figure*}

In Fig.~\ref{fig:full_real_noint_a} and Fig.~\ref{fig:full_real_noint_b}, we plot the extended GTEMPO results for the single-flavor case with $\alpha=1$ and $\alpha=0.1$ respectively. In panels (a, b) we plot the non-equilibrium greater and lesser Green's functions, and in panel (c) we plot the non-equilibrium density-density correlation, as functions of the real time $t$, where we have used $\chi=200$, $\beta=10$, $t=5$ and $\delta t=0.0125$ in both figures. The extended GTEMPO results shown in the main panels of both figures are also used as the baselines for the error analysis in the insets. 
In each panel, we show two insets, where in the left inset we analyze the error with respect to baseline against different bond dimensions used in the extended GTEMPO calculations, while in the right inset we analyze the error with respect to the baseline against different real-time step sizes.
We can see that for both the strong and weak electron-phonon interaction cases, the errors decreases (i.e., become closer to the baseline case) monotonically with larger $\chi$ or smaller $\delta t$. Comparing the error analysis against $\chi$ in these two figures, we can see that the results for the weaker interaction case have better converged with $\chi$. From the error analysis against $\delta t$ in both figures, we can see that the errors decrease almost linearly against $\delta t$, which indicates that the errors have mostly saturated against $\chi$ under $\chi=200$ and is dominated by the time discretization.
In both figures, we observe that the errors between the extended GTEMPO results with $\delta t=0.025$ and $\delta t=0.0125$ are of the order $10^{-3}$ at $\chi=200$, indicating that the extended GTEMPO results could already be very accurate.

\begin{figure*}
  \includegraphics[width=2\columnwidth]{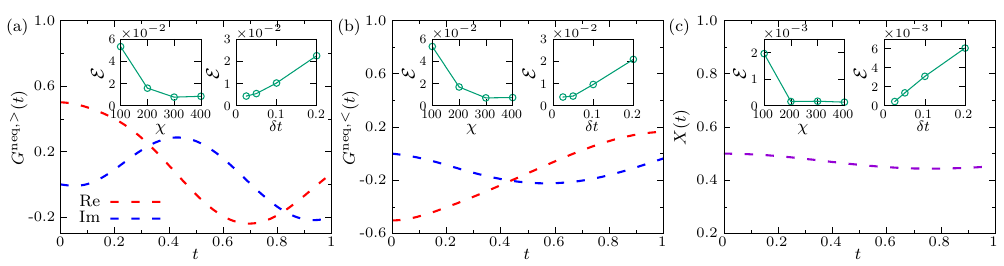} 
  \caption{The non-equilibrium greater Green's function $G^{{\rm neq},>}(t)$ (a), lesser Green's function $G^{{\rm neq}, <}(t)$ (b), density-density correlation $X^{\rm neq}(t)$ (c) as functions of $t$, for real-time evolution starting from the separable state in Eq.(\ref{eq:separablestate}) (the Keldysh contour) in the two-flavor case with $\alpha=1$. The red and blue dashed lines in (a) represent the real and imaginary parts of $G^{{\rm neq},>}(t)$, and similar for (b). The observables in the main panels are calculated using extended GTEMPO with $\chi=500$, $\beta=10$, $t=1$ and $\delta t=0.0125$, which are used as the baselines for the error analysis in the insets. 
  In each panel, the left inset shows the error analysis against the bond dimension $\chi$, where the errors are between the baseline and the extended GTEMPO results calculated with different values of $\chi$, the right inset shows the error analysis against the real-time step size $\delta t$, where the errors are between the baseline and the extended GTEMPO results calculated with different values of $\delta t$.  In the left insets, we have fixed $\delta t=0.0125$ and in the right insets we have fixed $\chi=500$.
    }
    \label{fig:full_real_int_a}
\end{figure*}

\begin{figure*}
  \includegraphics[width=2\columnwidth]{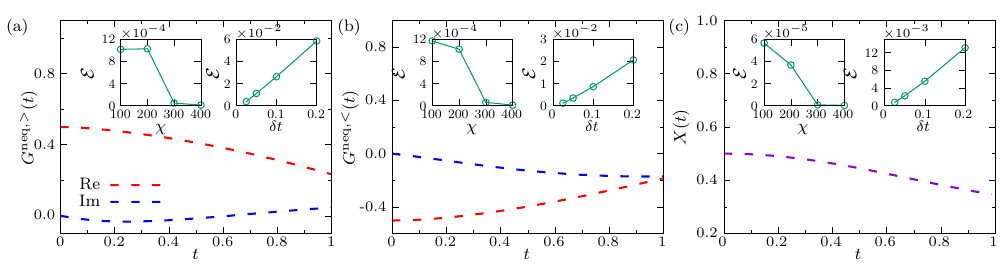} 
  \caption{Similar to Fig.~\ref{fig:full_real_int_a}, but for the two-flavor case with $\alpha=0.1$. The hyperparameters used in these simulation are also the same as those used in Fig.~\ref{fig:full_real_int_a}.
    }
    \label{fig:full_real_int_b}
\end{figure*}

In Fig.~\ref{fig:full_real_int_a} and Fig.~\ref{fig:full_real_int_b}, we plot the extended GTEMPO results for the two-flavor case with $\alpha=1$ and $\alpha=0.1$ respectively, similar to Fig.~\ref{fig:full_real_noint_a} and Fig.~\ref{fig:full_real_noint_b}. In the main panels, we have used a larger $\chi=500$ and a smaller $t=1$, as the simulation for the two-flavor case is significantly more demanding than the single-flavor case (also because we want to perform error analysis against the hyperparameters, thus very large $\chi$ and very small $\delta t$ are considered). Again the data from the main panels are used as the baseline for the error analysis in the insets. From the left insets in all the panels of these two figures, we can see that the errors decrease monotonically with respect to the baseline against $\chi$, and have more or less saturated beyond $\chi=300$.
From the right insets in all the panels of both figures, we also observe monotonically decrease of errors with respect to the baseline against $\delta t$. In both figures, we observe that the error between the extended GTEMPO results with $\delta t=0.025$ and $\delta t=0.0125$ is of the order $10^{-3}$ or less at $\chi=500$, indicating that the extended GTEMPO results could already be very accurate.

To this end, we note that
with the error analysis against the only two hyperparameters $\chi$ and $\delta t$, it is fair to believe that our extended GTEMPO results on the Keldysh contour for full-fledged model are already quite accurate. The confidence in these results could also be built from our benchmarks with ED in Sec.~\ref{sec:toymodel}, since there is no obvious simplification for extended GTEMPO when the spectral functions are chosen as delta functions. In fact, for delta spectral function the hybridization function in Eq.(\ref{eq:Delta}) or the correlation function in Eq.(\ref{eq:Lambda}) become purely oscillating on the Keldysh contour, in which case the IF can be viewed as the partition function of an effective Hamiltonian with \textit{non-decaying long-range} couplings, which is more likely to be a harder case for extended GTEMPO, although this case can be trivially solved by ED (in comparison, the hybridization function or the correlation function will generally decay polynomially with $t$ for continuous spectral functions).

\section{Conclusion and discussions}\label{sec:summary}

In summary, we have proposed an extended Grassmann time-evolving matrix product operator method to solve the polaron impurity problem, in which the impurity is coupled to both a free electron bath and a free phonon bath. The central challenge we overcome in this work is to find an accurate way to take into account the effects of the Feynman-Vernon influence functional of the phonon bath in the Grassmann coherent state basis, such that the only two sources of errors in our method are the first-order discretization error of the impurity path integral (i.e., the error from QuAPI) and the MPS bond truncation error, the same as vanilla TEMPO and GTEMPO. 
The extended GTEMPO method inherits all the advantages of the (G)TEMPO methods, and could be as accurate as the latter. The implication is enormous: we can now directly solve polaron impurity problems on the Keldysh contour or even the L-shaped Kadanoff contour to obtain the real-time Green's functions or higher-order impurity correlation functions, while the only two sources of errors can both be well-controlled. 
As our method essentially handles the impurity path integrals with retarded interaction, it can also be readily used in the extended DMFT which takes into account long-range electron-electron couplings by solving an effective impurity problem with retarded interaction~\cite{SenguptaGeorges1995,SiSmith1996,SmithSi2000,ChitraKotliar2000,SunKotliar2002,HuangWerner2014}.
Extensive numerical examples are provided with up to two impurity flavors, in which we benchmark the accuracy of our extended GTEMPO calculations comprehensively on the imaginary, Keldysh and Kadanoff contours, and for different types of impurity-bath couplings and impurity Hamiltonians. 
Scaling up to more flavors is possible, by employing the multi-flavor extension of the GTEMPO method proposed in Ref.~\cite{SunGuo2025b}, which further reduces the computational cost of GTEMPO by integrating out all the flavors before hand except the one used to compute observables.

\begin{acknowledgments}
This work is supported by National Natural Science Foundation of China under Grant No. 12104328. 
\end{acknowledgments}

\appendix

\section{Analytic solution for the independent bosons model}\label{app:ib}

In this section, we will give the analytic solutions of the independent bosons model for the Matsubara Green's function on the imaginary-time contour, and the equilibrium greater and lesser Green's functions on L-shaped Kadanoff contour. We will also rederive those analytical solutions based on the path integral formalism, from which we can see that the QuAPI discretization for this particular problem is exact irrespective of the time step size.

\subsection{The Matsubara Green's function}

We first consider the single-flavor case with impurity Hamiltonian in Eq.(\ref{eq:singleflavorHimp}). 
In this case, the analytical solution of the independent bosons model is often derived in literatures by employing the Lang-Firsov canonical transformation~\cite{lang1963-kinetic,mahan2000-many},
and the Matsubara Green's function is given as
\begin{align} \label{eq:im-G-appendix}
  G (\tau) = - \langle T \hat{a} (\tau) \hat{a}^{\dag} \rangle^{\rm eq} = \tilde{G}_0
  (\tau) F (\tau), 
\end{align}
where $\tilde{G}_0 (\tau)$ is the renormalized free electron Green's function:
\begin{align}\label{eq:G0}
\tilde{G}_0 (\tau) = - [1 - \bar{f} (\tilde{\epsilon}_a)] e^{- \tau\tilde{\epsilon}_a}, 
\end{align}
with $\tilde{\epsilon}_a = \epsilon_a - \Sigma$ and 
\begin{align}
  \Sigma = \int \mathd \omega \frac{J (\omega)}{\omega}.
\end{align}
Here $\bar{f} (\epsilon) = (e^{\beta \epsilon} + 1)^{- 1}$ is the Fermi-Dirac distribution.
The function $F (\tau)$ is defined as
\begin{align}
  F (\tau) = e^{-\int \mathd \omega \frac{J (\omega)}{\omega^2} \{ [1 +
  \bar{n} (\omega)] (1 - e^{- \tau \omega}) + \bar{n} (\omega) (1 -
  e^{\tau \omega}) \}} ,
\end{align}
with $\bar{n} (\omega)$ the Bose-Einstein distribution.

In the two-flavor case with impurity Hamiltonian in Eq.(\ref{eq:twoflavorHimp}) and that both flavors are coupled to the same phonon bath as in Eq.(\ref{eq:Hph}), 
one could still solve the problem analytically in a similar way via the Lang-Firsov canonical transformation. The final result is still of the form in Eq.(\ref{eq:im-G-appendix}), but 
the renormalized free electron Green's function in Eq.(\ref{eq:G0}) should be replaced by the following form:
\begin{align}
  \tilde{G}_0 (\tau) = - \frac{e^{- \tau \tilde{\epsilon}_a} + e^{- \beta
  \tilde{\epsilon}_a} e^{- \tau (\tilde{\epsilon}_a + \tilde{U})}}{1 + 2
  e^{- \beta \tilde{\epsilon}_a} + e^{- \beta (2 \tilde{\epsilon}_a +
  \tilde{U})}},
\end{align}
where $\tilde{U} = U - 2 \Sigma$ is the
renormalized Coulomb interaction.

Interestingly, the analytical solution can also be derived based only on the path integral formalism, which is shown in the following. Due to the absence of $\Hel$, the impurity path integral of the independent bosons model can be completely expressed in the Fock space, in which the 
partition function can be written as
\begin{align}
Z = \sum_{\tmmathbf{n}} \mathcal{A}[\tmmathbf{n}]
\end{align}
with $\mathcal{A} [\tmmathbf{n}] =\mathcal{K} [\tmmathbf{n}].
\mathcal{I}_{\tmop{ph}} [\tmmathbf{n}]$, and
\begin{align}
  \mathcal{I}_{\tmop{ph}} [\tmmathbf{n}] = e^{- \int_0^{\beta} \mathd \tau'
  \int_0^{\beta} \mathd \tau'' n (\tau') \Lambda (\tau', \tau'') n (\tau'')} .
\end{align}
The explicit form of $\Lambda (\tau', \tau'')$ is shown in Eq.(\ref{eq:Lambda}).
Since there is no off-diagonal term in the impurity Hamiltonian, only
two paths: $n_1 (\tau') = 1$ and $n_0 (\tau') = 0$ in the range $0 \leqslant \tau'
\leqslant \beta$ can give nonzero contributions in $Z$. Along the path $n_0 (\tau') = 0$, we have
\begin{align}
  \mathcal{K} [\tmmathbf{n}_0] = 1, \quad \mathcal{I}_{\tmop{ph}} [\tmmathbf{n}_0]
  = 1,
\end{align}
and along the path $n_1 (\tau') = 1$, we have
\begin{align}
  \mathcal{K} [\tmmathbf{n}_1] = e^{- \beta \epsilon_a}, \quad
  \mathcal{I}_{\tmop{ph}} [\tmmathbf{n}_1] = e^{\beta \Sigma} .
\end{align}
Therefore the partition function is  
\begin{align}
Z \equallim \sum_{d=0}^1  \mathcal{K} [\tmmathbf{n}_d] \mathcal{I}_{\tmop{ph}} [\tmmathbf{n}_d] =  1 + e^{- \beta
\tilde{\epsilon}_a}.
\end{align}

To evaluate the Matsubara Green's function, we notice that
the creation
operator $\hat{a}^{\dag}$ at imaginary time $0$ eliminates the state $|1 \rangle$ and
changes the state $|0 \rangle$ to $|1 \rangle$, while the annihilation operator
$\hat{a}$ at imaginary time $\tau$ brings $|1 \rangle$ back to $|0 \rangle$, therefore after inserting $\aop(\tau)$ and $\adop$ into the path integral, we
only need to consider a single path (denoted as $\tilde{n}(\tau)$) in the numerator of $G(\tau)$: $\tilde{n} (\tau') = 1$ for $0 \leqslant \tau'
\leqslant \tau$ and $\tilde{n} (\tau') = 0$ for $\tau < \tau' \leqslant \beta$. For
this path, we have $\gK[\tilde{\boldn}] = e^{-\tau\epsilon_a}$ and
\begin{align}
  &\int_0^{\beta} \mathd \tau' \int_0^{\beta} \mathd \tau'' \tilde{n} (\tau')
  \Lambda (\tau', \tau'') \tilde{n} (\tau'') \nonumber\\
  = & \int_0^{\tau} \mathd \tau' \int_0^{\tau} \mathd \tau'' 1 \times
  \Lambda (\tau', \tau'') \times 1 \nonumber \\ 
  &+ \int_0^{\tau} \mathd \tau'
  \int_{\tau}^{\beta} 1 \times \Lambda (\tau', \tau'') \times 0 \nonumber\\
  & + \int_{\tau}^{\beta} \mathd \tau' \int_0^{\tau} \mathd \tau'' 0
  \times \Lambda (\tau', \tau'') \times 1 \nonumber \\ 
  &+ \int_{\tau}^{\beta} \mathd \tau'
  \int_{\tau}^{\beta} 0 \times \Lambda (\tau', \tau'') \times 0 \nonumber\\
  = & \int_0^{\tau} \mathd \tau' \int_0^{\tau} \mathd \tau'' \Lambda
  (\tau', \tau''),
\end{align}
which gives
\begin{align}
\gIph[\tilde{\boldn}] =  e^{\tau \Sigma - \int \mathd \omega
  \frac{J (\omega)}{\omega^2} \{ [1 + \bar{n} (\omega)] (1 - e^{- \tau
  \omega}) + \bar{n} (\omega) (1 - e^{\tau \omega}) \}}.
\end{align}
Then the Matsubara Green's function is evaluated as $G (\tau) = - 
\mathcal{K} [\tilde{\boldn}] \mathcal{I}_{\tmop{ph}} [\tilde{\boldn}] / Z$, which
gives the desired result in Eq.(\ref{eq:im-G-appendix}). For the two-flavor case, there is still no
off-diagonal term in the impurity Hamiltonian, thus repeating the same procedure would
lead to the desired result.

In the QuAPI discretization scheme, the inverse temperature is split into $M$ equal-distant segments 
with $\delta \tau =\beta / M $. The approximation made in QuAPI is that the path remains unchanged within each segment, namely 
$n (\tau') = n_k$ if $k \delta \tau \leqslant \tau' < (k + 1) \delta \tau$. 
Therefore the two paths $\boldn_0$ and $\boldn_1$ is exactly included when evaluating $Z$. In addition, as we only evaluate the Green's functions at the time points $n_k = k\delta\tau$, the path $\tilde{\nop}$ is also included when evaluating the numerator of $G(\tau)$. Therefore for this particular case, QuAPI exactly includes all the nontrivial paths without any approximations.


\subsection{The equilibrium Green's functions}

The analytic solutions for the real-time Green's functions can also be obtained through the Lang\mbox{-}Firsov transformation. The concretely form for the equilibrium greater and lesser Green's functions in the single-flavor case are
\begin{align}
  G^{{\rm eq},>}(t) &= - \im e^{- \im \tilde{\epsilon}_a t} [1 - \bar{f}
  (\tilde{\epsilon}_a)] F (t); \label{eq:ib_greater} \\
  G^{{\rm eq},<}(t) &= \im e^{-\im\tilde{\epsilon}_a t}\bar{f}(\tilde{\epsilon}_a)F(t),
\end{align}
where
\begin{align}
  F (t) = e^{- \int \mathd \omega \frac{J (\omega)}{\omega^2} \{ [1 +
  \bar{n} (\omega)] (1 - e^{- \im \omega t}) + \bar{n} (\omega) (1 -
  e^{\im \omega t}) \}} .
\end{align}
In the two-flavor case, the equilibrium greater and lesser Green's functions are
\begin{align}
  G^{{\rm eq},>}(t) &= -\im \frac{e^{- \im \tilde{\epsilon}_a t} + e^{- \beta
  \tilde{\epsilon}_a} e^{- \im (\tilde{\epsilon}_a + \tilde{U})t}}{1 + 2
  e^{- \beta \tilde{\epsilon}_a} + e^{- \beta (2 \tilde{\epsilon}_a +
  \tilde{U})}}; \\
  G^{{\rm eq},<}(t) &= \im \frac{e^{- \beta \tilde{\epsilon}_a}e^{- \im \tilde{\epsilon}_a t} +
           e^{- \beta (2 \tilde{\epsilon}_a +\tilde{U})} e^{- \im (\tilde{\epsilon}_a + \tilde{U})t}}{1 + 2
  e^{- \beta \tilde{\epsilon}_a} + e^{- \beta (2 \tilde{\epsilon}_a +\tilde{U})}}.
\end{align}
As an example, we rederive Eq.(\ref{eq:ib_greater}) in the single-flavor case based on the impurity path integral only. And the rest equations can be derived similarly.

The partition function $Z = \tmop{Tr} [e^{- \im \hat{H} t} e^{- \beta \hat{H}}
e^{\im \hat{H} t}]$ can be expressed as a path integral along the Kadanoff contour,
where the Feynman-Vernon IF can be written as (here $\mathcal{C}$ refers to the
Kadanoff contour)
\begin{align}
  \mathcal{I}_{\tmop{ph}} [\tmmathbf{n}] = e^{- \int_{\mathcal{C}} \mathd t'
  \int_{\mathcal{C}} \mathd t'' n (t') \Lambda (t', t'') n (t'')} . 
\end{align}
Again, there are only two paths $n_0 (t') = 0$ and $n_1 (t') = 1$ which can give nonzero contributions. Along $n_0 (t')
= 0$, we have
\begin{align}
  \mathcal{K} [\tmmathbf{n}_0] = 1, \quad \mathcal{I}_{\tmop{ph}} [\tmmathbf{n}_0]
  = 1,
\end{align}
and along $n_1 (t') = 1$ we have
\begin{align}
  \mathcal{K} [\tmmathbf{n}_1] = e^{- \beta \epsilon_a}, \quad
  \mathcal{I}_{\tmop{ph}} [\tmmathbf{n}_1] = e^{\beta \Sigma} .
\end{align}
This gives
\begin{align}
  Z = 1 + e^{- \beta \tilde{\epsilon}_a} .
\end{align}

To evaluate $G^{{\rm eq},>}(t)$, we notice that 
the creation operator
$\hat{a}^{\dag}$ rises $|0 \rangle$ to $|1 \rangle$ at real time $0$ and the
annihilation operator $\hat{a} (t)$ brings $|1 \rangle$ back to $|0 \rangle$.
Therefore we only need to consider the path $\tilde{n} (t') = 1$ for $0 \preccurlyeq
t' \preccurlyeq t^+$, $\tilde{n} (t') = 0$ for $t^- \preccurlyeq t' \preccurlyeq - \im 
\beta$. Here $t' \preccurlyeq t''$ means $t''$ succeeds $t'$ on the Kadanoff contour (see Fig.~\ref{fig:contour}).
On the interval $0 \preccurlyeq t' \preccurlyeq t^+$, we have
\begin{align}
  \Lambda (t', t'') = \left\{\begin{array}{ll}
    {}[1 + \bar{n} (\omega)] e^{- \im \omega (t' - t'')} & t' \geqslant
    t''\\
    \bar{n} (\omega) e^{- \im \omega (t' - t'')} & t' < t''.
  \end{array}\right.
\end{align}
For this path, we only need to consider a single path $\tilde{\boldn}$ to evaluate the numerator of $G^{{\rm eq},>}(t)$, for which we have
\begin{align}
\gK[\tilde{\boldn}] =& e^{-\im \epsilon_a t}  ; \\
  \gIph[\tilde{\boldn}] =&e^{-\int_0^t \mathd t' \int_0^t \mathd t'' \tilde{n}(t') \Lambda (t', t'') \nonumber \tilde{n}(t'') } \\ 
  =&  e^{\im \Sigma t -\int \mathd \omega \frac{J (\omega)}{\omega^2} \{ [1 + \bar{n}
  (\omega)] (1 - e^{- \im \omega t}) + \bar{n} (\omega) (1 - e^{\im \omega
  t}) \}} .
\end{align}
Then greater Green's function is evaluated as $G^{{\rm eq},>}(t)=-\im \mathcal{K} [\tilde{\boldn}]
\mathcal{I}_{\tmop{ph}} [\tilde{\boldn}] / Z$, which gives the desire result in Eq.(\ref{eq:ib_greater}). 

\bibliography{refs}

\begin{thebibliography}{107}%
\makeatletter
\providecommand \@ifxundefined [1]{%
 \@ifx{#1\undefined}
}%
\providecommand \@ifnum [1]{%
 \ifnum #1\expandafter \@firstoftwo
 \else \expandafter \@secondoftwo
 \fi
}%
\providecommand \@ifx [1]{%
 \ifx #1\expandafter \@firstoftwo
 \else \expandafter \@secondoftwo
 \fi
}%
\providecommand \natexlab [1]{#1}%
\providecommand \enquote  [1]{``#1''}%
\providecommand \bibnamefont  [1]{#1}%
\providecommand \bibfnamefont [1]{#1}%
\providecommand \citenamefont [1]{#1}%
\providecommand \href@noop [0]{\@secondoftwo}%
\providecommand \href [0]{\begingroup \@sanitize@url \@href}%
\providecommand \@href[1]{\@@startlink{#1}\@@href}%
\providecommand \@@href[1]{\endgroup#1\@@endlink}%
\providecommand \@sanitize@url [0]{\catcode `\\12\catcode `\$12\catcode
  `\&12\catcode `\#12\catcode `\^12\catcode `\_12\catcode `\%12\relax}%
\providecommand \@@startlink[1]{}%
\providecommand \@@endlink[0]{}%
\providecommand \url  [0]{\begingroup\@sanitize@url \@url }%
\providecommand \@url [1]{\endgroup\@href {#1}{\urlprefix }}%
\providecommand \urlprefix  [0]{URL }%
\providecommand \Eprint [0]{\href }%
\providecommand \doibase [0]{https://doi.org/}%
\providecommand \selectlanguage [0]{\@gobble}%
\providecommand \bibinfo  [0]{\@secondoftwo}%
\providecommand \bibfield  [0]{\@secondoftwo}%
\providecommand \translation [1]{[#1]}%
\providecommand \BibitemOpen [0]{}%
\providecommand \bibitemStop [0]{}%
\providecommand \bibitemNoStop [0]{.\EOS\space}%
\providecommand \EOS [0]{\spacefactor3000\relax}%
\providecommand \BibitemShut  [1]{\csname bibitem#1\endcsname}%
\let\auto@bib@innerbib\@empty
\bibitem [{\citenamefont {Landau}(1933)}]{landau1933-electron}%
  \BibitemOpen
  \bibfield  {author} {\bibinfo {author} {\bibfnamefont {L.~D.}\ \bibnamefont
  {Landau}},\ }\bibfield  {title} {\bibinfo {title} {Electron motion in crystal
  lattices},\ }\href@noop {} {\bibfield  {journal} {\bibinfo  {journal} {Phys.
  Z. Sowjet.}\ }\textbf {\bibinfo {volume} {3}},\ \bibinfo {pages} {664}
  (\bibinfo {year} {1933})}\BibitemShut {NoStop}%
\bibitem [{\citenamefont {Landau}\ and\ \citenamefont
  {Pekar}(1948)}]{landau1948-effective}%
  \BibitemOpen
  \bibfield  {author} {\bibinfo {author} {\bibfnamefont {L.~D.}\ \bibnamefont
  {Landau}}\ and\ \bibinfo {author} {\bibfnamefont {S.~I.}\ \bibnamefont
  {Pekar}},\ }\bibfield  {title} {\bibinfo {title} {Effective mass of a
  polaron},\ }\href@noop {} {\bibfield  {journal} {\bibinfo  {journal} {Zh.
  Eksp. Teor. Fiz.}\ }\textbf {\bibinfo {volume} {18}},\ \bibinfo {pages} {419}
  (\bibinfo {year} {1948})}\BibitemShut {NoStop}%
\bibitem [{\citenamefont {Alexandrov}\ and\ \citenamefont
  {Devreese}(2010)}]{AlexandrovDevreese2010}%
  \BibitemOpen
  \bibfield  {author} {\bibinfo {author} {\bibfnamefont {A.~S.}\ \bibnamefont
  {Alexandrov}}\ and\ \bibinfo {author} {\bibfnamefont {J.~T.}\ \bibnamefont
  {Devreese}},\ }\href@noop {} {\emph {\bibinfo {title} {Advances in polaron
  physics}}},\ Vol.\ \bibinfo {volume} {159}\ (\bibinfo  {publisher}
  {Springer},\ \bibinfo {year} {2010})\BibitemShut {NoStop}%
\bibitem [{\citenamefont {Bredas}\ and\ \citenamefont
  {Street}(1985)}]{BredasStreet1985}%
  \BibitemOpen
  \bibfield  {author} {\bibinfo {author} {\bibfnamefont {J.~L.}\ \bibnamefont
  {Bredas}}\ and\ \bibinfo {author} {\bibfnamefont {G.~B.}\ \bibnamefont
  {Street}},\ }\bibfield  {title} {\bibinfo {title} {Polarons, bipolarons, and
  solitons in conducting polymers},\ }\href
  {https://doi.org/10.1021/ar00118a005} {\bibfield  {journal} {\bibinfo
  {journal} {Accounts of Chemical Research}\ }\textbf {\bibinfo {volume}
  {18}},\ \bibinfo {pages} {309} (\bibinfo {year} {1985})},\ \Eprint
  {https://arxiv.org/abs/https://doi.org/10.1021/ar00118a005}
  {https://doi.org/10.1021/ar00118a005} \BibitemShut {NoStop}%
\bibitem [{\citenamefont {Franchini}\ \emph {et~al.}(2021)\citenamefont
  {Franchini}, \citenamefont {Reticcioli}, \citenamefont {Setvin},\ and\
  \citenamefont {Diebold}}]{FranchiniDiebold2021}%
  \BibitemOpen
  \bibfield  {author} {\bibinfo {author} {\bibfnamefont {C.}~\bibnamefont
  {Franchini}}, \bibinfo {author} {\bibfnamefont {M.}~\bibnamefont
  {Reticcioli}}, \bibinfo {author} {\bibfnamefont {M.}~\bibnamefont {Setvin}},\
  and\ \bibinfo {author} {\bibfnamefont {U.}~\bibnamefont {Diebold}},\
  }\bibfield  {title} {\bibinfo {title} {Polarons in materials},\ }\href
  {https://doi.org/10.1038/s41578-021-00289-w} {\bibfield  {journal} {\bibinfo
  {journal} {Nature Reviews Materials}\ }\textbf {\bibinfo {volume} {6}},\
  \bibinfo {pages} {560} (\bibinfo {year} {2021})}\BibitemShut {NoStop}%
\bibitem [{\citenamefont {Thomas}\ \emph {et~al.}(2025)\citenamefont {Thomas},
  \citenamefont {BanerJee}, \citenamefont {Nocera},\ and\ \citenamefont
  {Johnston}}]{thomas2025-theory}%
  \BibitemOpen
  \bibfield  {author} {\bibinfo {author} {\bibfnamefont {J.}~\bibnamefont
  {Thomas}}, \bibinfo {author} {\bibfnamefont {D.}~\bibnamefont {BanerJee}},
  \bibinfo {author} {\bibfnamefont {A.}~\bibnamefont {Nocera}},\ and\ \bibinfo
  {author} {\bibfnamefont {S.}~\bibnamefont {Johnston}},\ }\bibfield  {title}
  {\bibinfo {title} {Theory of electron-phonon interactions in extended
  correlated systems probed by resonant inelastic x-ray scattering},\ }\href
  {https://doi.org/10.1103/PhysRevX.15.021030} {\bibfield  {journal} {\bibinfo
  {journal} {Phys. Rev. X}\ }\textbf {\bibinfo {volume} {15}},\ \bibinfo
  {pages} {021030} (\bibinfo {year} {2025})}\BibitemShut {NoStop}%
\bibitem [{\citenamefont {Grusdt}\ \emph {et~al.}(2024)\citenamefont {Grusdt},
  \citenamefont {Mostaan}, \citenamefont {Demler},\ and\ \citenamefont
  {Ardila}}]{GrusdtArdila2024}%
  \BibitemOpen
  \bibfield  {author} {\bibinfo {author} {\bibfnamefont {F.}~\bibnamefont
  {Grusdt}}, \bibinfo {author} {\bibfnamefont {N.}~\bibnamefont {Mostaan}},
  \bibinfo {author} {\bibfnamefont {E.}~\bibnamefont {Demler}},\ and\ \bibinfo
  {author} {\bibfnamefont {L.~A.~P.}\ \bibnamefont {Ardila}},\ }\href
  {https://arxiv.org/abs/2410.09413} {\bibinfo {title} {Impurities and polarons
  in bosonic quantum gases: a review on recent progress}} (\bibinfo {year}
  {2024}),\ \Eprint {https://arxiv.org/abs/2410.09413} {arXiv:2410.09413
  [cond-mat.quant-gas]} \BibitemShut {NoStop}%
\bibitem [{\citenamefont {Lanzara}\ \emph {et~al.}(2001)\citenamefont
  {Lanzara}, \citenamefont {Bogdanov}, \citenamefont {Zhou}, \citenamefont
  {Kellar}, \citenamefont {Feng}, \citenamefont {Lu}, \citenamefont {Yoshida},
  \citenamefont {Eisaki}, \citenamefont {Fujimori}, \citenamefont {Kishio}
  \emph {et~al.}}]{LanzaraShen2001}%
  \BibitemOpen
  \bibfield  {author} {\bibinfo {author} {\bibfnamefont {A.}~\bibnamefont
  {Lanzara}}, \bibinfo {author} {\bibfnamefont {P.}~\bibnamefont {Bogdanov}},
  \bibinfo {author} {\bibfnamefont {X.}~\bibnamefont {Zhou}}, \bibinfo {author}
  {\bibfnamefont {S.}~\bibnamefont {Kellar}}, \bibinfo {author} {\bibfnamefont
  {D.}~\bibnamefont {Feng}}, \bibinfo {author} {\bibfnamefont {E.}~\bibnamefont
  {Lu}}, \bibinfo {author} {\bibfnamefont {T.}~\bibnamefont {Yoshida}},
  \bibinfo {author} {\bibfnamefont {H.}~\bibnamefont {Eisaki}}, \bibinfo
  {author} {\bibfnamefont {A.}~\bibnamefont {Fujimori}}, \bibinfo {author}
  {\bibfnamefont {K.}~\bibnamefont {Kishio}}, \emph {et~al.},\ }\bibfield
  {title} {\bibinfo {title} {Evidence for ubiquitous strong electron--phonon
  coupling in high-temperature superconductors},\ }\href
  {https://doi.org/10.1038/35087518} {\bibfield  {journal} {\bibinfo  {journal}
  {Nature}\ }\textbf {\bibinfo {volume} {412}},\ \bibinfo {pages} {510}
  (\bibinfo {year} {2001})}\BibitemShut {NoStop}%
\bibitem [{\citenamefont {Millis}\ \emph {et~al.}(1995)\citenamefont {Millis},
  \citenamefont {Littlewood},\ and\ \citenamefont
  {Shraiman}}]{MillisShraiman1995}%
  \BibitemOpen
  \bibfield  {author} {\bibinfo {author} {\bibfnamefont {A.~J.}\ \bibnamefont
  {Millis}}, \bibinfo {author} {\bibfnamefont {P.~B.}\ \bibnamefont
  {Littlewood}},\ and\ \bibinfo {author} {\bibfnamefont {B.~I.}\ \bibnamefont
  {Shraiman}},\ }\bibfield  {title} {\bibinfo {title} {Double exchange alone
  does not explain the resistivity of
  ${{\mathrm{La}}_{1}}_{\ensuremath{-}\mathit{x}}{\mathrm{sr}}_{\mathit{x}}{\mathrm{mno}}_{3}$},\
  }\href {https://doi.org/10.1103/PhysRevLett.74.5144} {\bibfield  {journal}
  {\bibinfo  {journal} {Phys. Rev. Lett.}\ }\textbf {\bibinfo {volume} {74}},\
  \bibinfo {pages} {5144} (\bibinfo {year} {1995})}\BibitemShut {NoStop}%
\bibitem [{\citenamefont {Millis}\ \emph {et~al.}(1996)\citenamefont {Millis},
  \citenamefont {Shraiman},\ and\ \citenamefont {Mueller}}]{MillisMueller1996}%
  \BibitemOpen
  \bibfield  {author} {\bibinfo {author} {\bibfnamefont {A.~J.}\ \bibnamefont
  {Millis}}, \bibinfo {author} {\bibfnamefont {B.~I.}\ \bibnamefont
  {Shraiman}},\ and\ \bibinfo {author} {\bibfnamefont {R.}~\bibnamefont
  {Mueller}},\ }\bibfield  {title} {\bibinfo {title} {Dynamic jahn-teller
  effect and colossal magnetoresistance inla1-xsrxmno3},\ }\href
  {https://doi.org/10.1103/physrevlett.77.175} {\bibfield  {journal} {\bibinfo
  {journal} {Phys. Rev. Lett.}\ }\textbf {\bibinfo {volume} {77}},\ \bibinfo
  {pages} {175} (\bibinfo {year} {1996})}\BibitemShut {NoStop}%
\bibitem [{\citenamefont {Millis}(1998)}]{Millis1998}%
  \BibitemOpen
  \bibfield  {author} {\bibinfo {author} {\bibfnamefont {A.~J.}\ \bibnamefont
  {Millis}},\ }\bibfield  {title} {\bibinfo {title} {Lattice effects in
  magnetoresistive manganese perovskites},\ }\href
  {https://doi.org/10.1038/32348} {\bibfield  {journal} {\bibinfo  {journal}
  {Nature}\ }\textbf {\bibinfo {volume} {392}},\ \bibinfo {pages} {147}
  (\bibinfo {year} {1998})}\BibitemShut {NoStop}%
\bibitem [{\citenamefont {Yamasaki}\ \emph {et~al.}(2006)\citenamefont
  {Yamasaki}, \citenamefont {Miyasaka}, \citenamefont {Kaneko}, \citenamefont
  {He}, \citenamefont {Arima},\ and\ \citenamefont
  {Tokura}}]{YamasakiTokura2006}%
  \BibitemOpen
  \bibfield  {author} {\bibinfo {author} {\bibfnamefont {Y.}~\bibnamefont
  {Yamasaki}}, \bibinfo {author} {\bibfnamefont {S.}~\bibnamefont {Miyasaka}},
  \bibinfo {author} {\bibfnamefont {Y.}~\bibnamefont {Kaneko}}, \bibinfo
  {author} {\bibfnamefont {J.-P.}\ \bibnamefont {He}}, \bibinfo {author}
  {\bibfnamefont {T.}~\bibnamefont {Arima}},\ and\ \bibinfo {author}
  {\bibfnamefont {Y.}~\bibnamefont {Tokura}},\ }\bibfield  {title} {\bibinfo
  {title} {Magnetic reversal of the ferroelectric polarization in a
  multiferroic spinel oxide},\ }\href
  {https://doi.org/10.1103/PhysRevLett.96.207204} {\bibfield  {journal}
  {\bibinfo  {journal} {Phys. Rev. Lett.}\ }\textbf {\bibinfo {volume} {96}},\
  \bibinfo {pages} {207204} (\bibinfo {year} {2006})}\BibitemShut {NoStop}%
\bibitem [{\citenamefont {Werner}\ and\ \citenamefont
  {Millis}(2007)}]{WernerMillis2007b}%
  \BibitemOpen
  \bibfield  {author} {\bibinfo {author} {\bibfnamefont {P.}~\bibnamefont
  {Werner}}\ and\ \bibinfo {author} {\bibfnamefont {A.~J.}\ \bibnamefont
  {Millis}},\ }\bibfield  {title} {\bibinfo {title} {Efficient dynamical mean
  field simulation of the holstein-hubbard model},\ }\href
  {https://doi.org/10.1103/PhysRevLett.99.146404} {\bibfield  {journal}
  {\bibinfo  {journal} {Phys. Rev. Lett.}\ }\textbf {\bibinfo {volume} {99}},\
  \bibinfo {pages} {146404} (\bibinfo {year} {2007})}\BibitemShut {NoStop}%
\bibitem [{\citenamefont {Werner}\ and\ \citenamefont
  {Millis}(2010)}]{WernerMillis2010}%
  \BibitemOpen
  \bibfield  {author} {\bibinfo {author} {\bibfnamefont {P.}~\bibnamefont
  {Werner}}\ and\ \bibinfo {author} {\bibfnamefont {A.~J.}\ \bibnamefont
  {Millis}},\ }\bibfield  {title} {\bibinfo {title} {Dynamical screening in
  correlated electron materials},\ }\href
  {https://doi.org/10.1103/PhysRevLett.104.146401} {\bibfield  {journal}
  {\bibinfo  {journal} {Phys. Rev. Lett.}\ }\textbf {\bibinfo {volume} {104}},\
  \bibinfo {pages} {146401} (\bibinfo {year} {2010})}\BibitemShut {NoStop}%
\bibitem [{\citenamefont {Otsuki}(2013)}]{Otsuki2013}%
  \BibitemOpen
  \bibfield  {author} {\bibinfo {author} {\bibfnamefont {J.}~\bibnamefont
  {Otsuki}},\ }\bibfield  {title} {\bibinfo {title} {Spin-boson coupling in
  continuous-time quantum monte carlo},\ }\href
  {https://doi.org/10.1103/PhysRevB.87.125102} {\bibfield  {journal} {\bibinfo
  {journal} {Phys. Rev. B}\ }\textbf {\bibinfo {volume} {87}},\ \bibinfo
  {pages} {125102} (\bibinfo {year} {2013})}\BibitemShut {NoStop}%
\bibitem [{\citenamefont {Caffarel}\ and\ \citenamefont
  {Krauth}(1994)}]{CaffarelKrauth1994}%
  \BibitemOpen
  \bibfield  {author} {\bibinfo {author} {\bibfnamefont {M.}~\bibnamefont
  {Caffarel}}\ and\ \bibinfo {author} {\bibfnamefont {W.}~\bibnamefont
  {Krauth}},\ }\bibfield  {title} {\bibinfo {title} {Exact diagonalization
  approach to correlated fermions in infinite dimensions: Mott transition and
  superconductivity},\ }\href {https://doi.org/10.1103/PhysRevLett.72.1545}
  {\bibfield  {journal} {\bibinfo  {journal} {Phys. Rev. Lett.}\ }\textbf
  {\bibinfo {volume} {72}},\ \bibinfo {pages} {1545} (\bibinfo {year}
  {1994})}\BibitemShut {NoStop}%
\bibitem [{\citenamefont {Koch}\ \emph {et~al.}(2008)\citenamefont {Koch},
  \citenamefont {Sangiovanni},\ and\ \citenamefont
  {Gunnarsson}}]{KochGunnarsson2008}%
  \BibitemOpen
  \bibfield  {author} {\bibinfo {author} {\bibfnamefont {E.}~\bibnamefont
  {Koch}}, \bibinfo {author} {\bibfnamefont {G.}~\bibnamefont {Sangiovanni}},\
  and\ \bibinfo {author} {\bibfnamefont {O.}~\bibnamefont {Gunnarsson}},\
  }\bibfield  {title} {\bibinfo {title} {Sum rules and bath parametrization for
  quantum cluster theories},\ }\href
  {https://doi.org/10.1103/PhysRevB.78.115102} {\bibfield  {journal} {\bibinfo
  {journal} {Phys. Rev. B}\ }\textbf {\bibinfo {volume} {78}},\ \bibinfo
  {pages} {115102} (\bibinfo {year} {2008})}\BibitemShut {NoStop}%
\bibitem [{\citenamefont {Granath}\ and\ \citenamefont
  {Strand}(2012)}]{GranathStrand2012}%
  \BibitemOpen
  \bibfield  {author} {\bibinfo {author} {\bibfnamefont {M.}~\bibnamefont
  {Granath}}\ and\ \bibinfo {author} {\bibfnamefont {H.~U.~R.}\ \bibnamefont
  {Strand}},\ }\bibfield  {title} {\bibinfo {title} {Distributional exact
  diagonalization formalism for quantum impurity models},\ }\href
  {https://doi.org/10.1103/PhysRevB.86.115111} {\bibfield  {journal} {\bibinfo
  {journal} {Phys. Rev. B}\ }\textbf {\bibinfo {volume} {86}},\ \bibinfo
  {pages} {115111} (\bibinfo {year} {2012})}\BibitemShut {NoStop}%
\bibitem [{\citenamefont {Lu}\ \emph {et~al.}(2014)\citenamefont {Lu},
  \citenamefont {H\"oppner}, \citenamefont {Gunnarsson},\ and\ \citenamefont
  {Haverkort}}]{LuHaverkort2014}%
  \BibitemOpen
  \bibfield  {author} {\bibinfo {author} {\bibfnamefont {Y.}~\bibnamefont
  {Lu}}, \bibinfo {author} {\bibfnamefont {M.}~\bibnamefont {H\"oppner}},
  \bibinfo {author} {\bibfnamefont {O.}~\bibnamefont {Gunnarsson}},\ and\
  \bibinfo {author} {\bibfnamefont {M.~W.}\ \bibnamefont {Haverkort}},\
  }\bibfield  {title} {\bibinfo {title} {Efficient real-frequency solver for
  dynamical mean-field theory},\ }\href
  {https://doi.org/10.1103/PhysRevB.90.085102} {\bibfield  {journal} {\bibinfo
  {journal} {Phys. Rev. B}\ }\textbf {\bibinfo {volume} {90}},\ \bibinfo
  {pages} {085102} (\bibinfo {year} {2014})}\BibitemShut {NoStop}%
\bibitem [{\citenamefont {Mejuto-Zaera}\ \emph {et~al.}(2020)\citenamefont
  {Mejuto-Zaera}, \citenamefont {Zepeda-N\'u\~nez}, \citenamefont {Lindsey},
  \citenamefont {Tubman}, \citenamefont {Whaley},\ and\ \citenamefont
  {Lin}}]{ZaeraLin2020}%
  \BibitemOpen
  \bibfield  {author} {\bibinfo {author} {\bibfnamefont {C.}~\bibnamefont
  {Mejuto-Zaera}}, \bibinfo {author} {\bibfnamefont {L.}~\bibnamefont
  {Zepeda-N\'u\~nez}}, \bibinfo {author} {\bibfnamefont {M.}~\bibnamefont
  {Lindsey}}, \bibinfo {author} {\bibfnamefont {N.}~\bibnamefont {Tubman}},
  \bibinfo {author} {\bibfnamefont {B.}~\bibnamefont {Whaley}},\ and\ \bibinfo
  {author} {\bibfnamefont {L.}~\bibnamefont {Lin}},\ }\bibfield  {title}
  {\bibinfo {title} {Efficient hybridization fitting for dynamical mean-field
  theory via semi-definite relaxation},\ }\href
  {https://doi.org/10.1103/PhysRevB.101.035143} {\bibfield  {journal} {\bibinfo
   {journal} {Phys. Rev. B}\ }\textbf {\bibinfo {volume} {101}},\ \bibinfo
  {pages} {035143} (\bibinfo {year} {2020})}\BibitemShut {NoStop}%
\bibitem [{\citenamefont {Lu}\ \emph {et~al.}(2019)\citenamefont {Lu},
  \citenamefont {Cao}, \citenamefont {Hansmann},\ and\ \citenamefont
  {Haverkort}}]{LuHaverkort2019}%
  \BibitemOpen
  \bibfield  {author} {\bibinfo {author} {\bibfnamefont {Y.}~\bibnamefont
  {Lu}}, \bibinfo {author} {\bibfnamefont {X.}~\bibnamefont {Cao}}, \bibinfo
  {author} {\bibfnamefont {P.}~\bibnamefont {Hansmann}},\ and\ \bibinfo
  {author} {\bibfnamefont {M.~W.}\ \bibnamefont {Haverkort}},\ }\bibfield
  {title} {\bibinfo {title} {Natural-orbital impurity solver and projection
  approach for green's functions},\ }\href
  {https://doi.org/10.1103/PhysRevB.100.115134} {\bibfield  {journal} {\bibinfo
   {journal} {Phys. Rev. B}\ }\textbf {\bibinfo {volume} {100}},\ \bibinfo
  {pages} {115134} (\bibinfo {year} {2019})}\BibitemShut {NoStop}%
\bibitem [{\citenamefont {He}\ and\ \citenamefont {Lu}(2014)}]{HeLu2014}%
  \BibitemOpen
  \bibfield  {author} {\bibinfo {author} {\bibfnamefont {R.-Q.}\ \bibnamefont
  {He}}\ and\ \bibinfo {author} {\bibfnamefont {Z.-Y.}\ \bibnamefont {Lu}},\
  }\bibfield  {title} {\bibinfo {title} {Quantum renormalization groups based
  on natural orbitals},\ }\href {https://doi.org/10.1103/PhysRevB.89.085108}
  {\bibfield  {journal} {\bibinfo  {journal} {Phys. Rev. B}\ }\textbf {\bibinfo
  {volume} {89}},\ \bibinfo {pages} {085108} (\bibinfo {year}
  {2014})}\BibitemShut {NoStop}%
\bibitem [{\citenamefont {He}\ \emph {et~al.}(2015)\citenamefont {He},
  \citenamefont {Dai},\ and\ \citenamefont {Lu}}]{HeLu2015}%
  \BibitemOpen
  \bibfield  {author} {\bibinfo {author} {\bibfnamefont {R.-Q.}\ \bibnamefont
  {He}}, \bibinfo {author} {\bibfnamefont {J.}~\bibnamefont {Dai}},\ and\
  \bibinfo {author} {\bibfnamefont {Z.-Y.}\ \bibnamefont {Lu}},\ }\bibfield
  {title} {\bibinfo {title} {Natural orbitals renormalization group approach to
  the two-impurity kondo critical point},\ }\href
  {https://doi.org/10.1103/PhysRevB.91.155140} {\bibfield  {journal} {\bibinfo
  {journal} {Phys. Rev. B}\ }\textbf {\bibinfo {volume} {91}},\ \bibinfo
  {pages} {155140} (\bibinfo {year} {2015})}\BibitemShut {NoStop}%
\bibitem [{\citenamefont {Wilson}(1975)}]{Wilson1975}%
  \BibitemOpen
  \bibfield  {author} {\bibinfo {author} {\bibfnamefont {K.~G.}\ \bibnamefont
  {Wilson}},\ }\bibfield  {title} {\bibinfo {title} {The renormalization group:
  Critical phenomena and the kondo problem},\ }\href
  {https://doi.org/10.1103/RevModPhys.47.773} {\bibfield  {journal} {\bibinfo
  {journal} {Rev. Mod. Phys.}\ }\textbf {\bibinfo {volume} {47}},\ \bibinfo
  {pages} {773} (\bibinfo {year} {1975})}\BibitemShut {NoStop}%
\bibitem [{\citenamefont {Bulla}(1999)}]{Bulla1999}%
  \BibitemOpen
  \bibfield  {author} {\bibinfo {author} {\bibfnamefont {R.}~\bibnamefont
  {Bulla}},\ }\bibfield  {title} {\bibinfo {title} {Zero temperature
  metal-insulator transition in the infinite-dimensional hubbard model},\
  }\href {https://doi.org/10.1103/PhysRevLett.83.136} {\bibfield  {journal}
  {\bibinfo  {journal} {Phys. Rev. Lett.}\ }\textbf {\bibinfo {volume} {83}},\
  \bibinfo {pages} {136} (\bibinfo {year} {1999})}\BibitemShut {NoStop}%
\bibitem [{\citenamefont {Bulla}\ \emph {et~al.}(2008)\citenamefont {Bulla},
  \citenamefont {Costi},\ and\ \citenamefont {Pruschke}}]{BullaPruschke2008}%
  \BibitemOpen
  \bibfield  {author} {\bibinfo {author} {\bibfnamefont {R.}~\bibnamefont
  {Bulla}}, \bibinfo {author} {\bibfnamefont {T.~A.}\ \bibnamefont {Costi}},\
  and\ \bibinfo {author} {\bibfnamefont {T.}~\bibnamefont {Pruschke}},\
  }\bibfield  {title} {\bibinfo {title} {Numerical renormalization group method
  for quantum impurity systems},\ }\href
  {https://doi.org/10.1103/RevModPhys.80.395} {\bibfield  {journal} {\bibinfo
  {journal} {Rev. Mod. Phys.}\ }\textbf {\bibinfo {volume} {80}},\ \bibinfo
  {pages} {395} (\bibinfo {year} {2008})}\BibitemShut {NoStop}%
\bibitem [{\citenamefont {Anders}(2008)}]{Frithjof2008}%
  \BibitemOpen
  \bibfield  {author} {\bibinfo {author} {\bibfnamefont {F.~B.}\ \bibnamefont
  {Anders}},\ }\bibfield  {title} {\bibinfo {title} {A numerical
  renormalization group approach to non-equilibrium green functions for quantum
  impurity models},\ }\href {https://doi.org/10.1088/0953-8984/20/19/195216}
  {\bibfield  {journal} {\bibinfo  {journal} {J. Phys. Condens. Matter}\
  }\textbf {\bibinfo {volume} {20}},\ \bibinfo {pages} {195216} (\bibinfo
  {year} {2008})}\BibitemShut {NoStop}%
\bibitem [{\citenamefont {\ifmmode~\check{Z}\else \v{Z}\fi{}itko}\ and\
  \citenamefont {Pruschke}(2009)}]{ZitkoPruschke2009}%
  \BibitemOpen
  \bibfield  {author} {\bibinfo {author} {\bibfnamefont {R.}~\bibnamefont
  {\ifmmode~\check{Z}\else \v{Z}\fi{}itko}}\ and\ \bibinfo {author}
  {\bibfnamefont {T.}~\bibnamefont {Pruschke}},\ }\bibfield  {title} {\bibinfo
  {title} {Energy resolution and discretization artifacts in the numerical
  renormalization group},\ }\href {https://doi.org/10.1103/PhysRevB.79.085106}
  {\bibfield  {journal} {\bibinfo  {journal} {Phys. Rev. B}\ }\textbf {\bibinfo
  {volume} {79}},\ \bibinfo {pages} {085106} (\bibinfo {year}
  {2009})}\BibitemShut {NoStop}%
\bibitem [{\citenamefont {Deng}\ \emph {et~al.}(2013)\citenamefont {Deng},
  \citenamefont {Mravlje}, \citenamefont {\ifmmode~\check{Z}\else
  \v{Z}\fi{}itko}, \citenamefont {Ferrero}, \citenamefont {Kotliar},\ and\
  \citenamefont {Georges}}]{DengGeorges2013}%
  \BibitemOpen
  \bibfield  {author} {\bibinfo {author} {\bibfnamefont {X.}~\bibnamefont
  {Deng}}, \bibinfo {author} {\bibfnamefont {J.}~\bibnamefont {Mravlje}},
  \bibinfo {author} {\bibfnamefont {R.}~\bibnamefont {\ifmmode~\check{Z}\else
  \v{Z}\fi{}itko}}, \bibinfo {author} {\bibfnamefont {M.}~\bibnamefont
  {Ferrero}}, \bibinfo {author} {\bibfnamefont {G.}~\bibnamefont {Kotliar}},\
  and\ \bibinfo {author} {\bibfnamefont {A.}~\bibnamefont {Georges}},\
  }\bibfield  {title} {\bibinfo {title} {How bad metals turn good:
  Spectroscopic signatures of resilient quasiparticles},\ }\href
  {https://doi.org/10.1103/PhysRevLett.110.086401} {\bibfield  {journal}
  {\bibinfo  {journal} {Phys. Rev. Lett.}\ }\textbf {\bibinfo {volume} {110}},\
  \bibinfo {pages} {086401} (\bibinfo {year} {2013})}\BibitemShut {NoStop}%
\bibitem [{\citenamefont {Stadler}\ \emph {et~al.}(2015)\citenamefont
  {Stadler}, \citenamefont {Yin}, \citenamefont {von Delft}, \citenamefont
  {Kotliar},\ and\ \citenamefont {Weichselbaum}}]{StadlerWeichselbaum2015}%
  \BibitemOpen
  \bibfield  {author} {\bibinfo {author} {\bibfnamefont {K.~M.}\ \bibnamefont
  {Stadler}}, \bibinfo {author} {\bibfnamefont {Z.~P.}\ \bibnamefont {Yin}},
  \bibinfo {author} {\bibfnamefont {J.}~\bibnamefont {von Delft}}, \bibinfo
  {author} {\bibfnamefont {G.}~\bibnamefont {Kotliar}},\ and\ \bibinfo {author}
  {\bibfnamefont {A.}~\bibnamefont {Weichselbaum}},\ }\bibfield  {title}
  {\bibinfo {title} {Dynamical mean-field theory plus numerical
  renormalization-group study of spin-orbital separation in a three-band hund
  metal},\ }\href {https://doi.org/10.1103/PhysRevLett.115.136401} {\bibfield
  {journal} {\bibinfo  {journal} {Phys. Rev. Lett.}\ }\textbf {\bibinfo
  {volume} {115}},\ \bibinfo {pages} {136401} (\bibinfo {year}
  {2015})}\BibitemShut {NoStop}%
\bibitem [{\citenamefont {Lee}\ and\ \citenamefont
  {Weichselbaum}(2016)}]{LeeWeichselbaum2016}%
  \BibitemOpen
  \bibfield  {author} {\bibinfo {author} {\bibfnamefont {S.-S.~B.}\
  \bibnamefont {Lee}}\ and\ \bibinfo {author} {\bibfnamefont {A.}~\bibnamefont
  {Weichselbaum}},\ }\bibfield  {title} {\bibinfo {title} {Adaptive broadening
  to improve spectral resolution in the numerical renormalization group},\
  }\href {https://doi.org/10.1103/PhysRevB.94.235127} {\bibfield  {journal}
  {\bibinfo  {journal} {Phys. Rev. B}\ }\textbf {\bibinfo {volume} {94}},\
  \bibinfo {pages} {235127} (\bibinfo {year} {2016})}\BibitemShut {NoStop}%
\bibitem [{\citenamefont {Lee}\ \emph {et~al.}(2017)\citenamefont {Lee},
  \citenamefont {von Delft},\ and\ \citenamefont
  {Weichselbaum}}]{LeeWeichselbaum2017}%
  \BibitemOpen
  \bibfield  {author} {\bibinfo {author} {\bibfnamefont {S.-S.~B.}\
  \bibnamefont {Lee}}, \bibinfo {author} {\bibfnamefont {J.}~\bibnamefont {von
  Delft}},\ and\ \bibinfo {author} {\bibfnamefont {A.}~\bibnamefont
  {Weichselbaum}},\ }\bibfield  {title} {\bibinfo {title} {Doublon-holon origin
  of the subpeaks at the hubbard band edges},\ }\href
  {https://doi.org/10.1103/PhysRevLett.119.236402} {\bibfield  {journal}
  {\bibinfo  {journal} {Phys. Rev. Lett.}\ }\textbf {\bibinfo {volume} {119}},\
  \bibinfo {pages} {236402} (\bibinfo {year} {2017})}\BibitemShut {NoStop}%
\bibitem [{\citenamefont {Cornaglia}\ \emph {et~al.}(2004)\citenamefont
  {Cornaglia}, \citenamefont {Ness},\ and\ \citenamefont
  {Grempel}}]{CornagliaGrempel2004}%
  \BibitemOpen
  \bibfield  {author} {\bibinfo {author} {\bibfnamefont {P.~S.}\ \bibnamefont
  {Cornaglia}}, \bibinfo {author} {\bibfnamefont {H.}~\bibnamefont {Ness}},\
  and\ \bibinfo {author} {\bibfnamefont {D.~R.}\ \bibnamefont {Grempel}},\
  }\bibfield  {title} {\bibinfo {title} {Many-body effects on the transport
  properties of single-molecule devices},\ }\href
  {https://doi.org/10.1103/PhysRevLett.93.147201} {\bibfield  {journal}
  {\bibinfo  {journal} {Phys. Rev. Lett.}\ }\textbf {\bibinfo {volume} {93}},\
  \bibinfo {pages} {147201} (\bibinfo {year} {2004})}\BibitemShut {NoStop}%
\bibitem [{\citenamefont {Paaske}\ and\ \citenamefont
  {Flensberg}(2005)}]{PaaskeFlensberg2005}%
  \BibitemOpen
  \bibfield  {author} {\bibinfo {author} {\bibfnamefont {J.}~\bibnamefont
  {Paaske}}\ and\ \bibinfo {author} {\bibfnamefont {K.}~\bibnamefont
  {Flensberg}},\ }\bibfield  {title} {\bibinfo {title} {Vibrational sidebands
  and the kondo effect in molecular transistors},\ }\href
  {https://doi.org/10.1103/PhysRevLett.94.176801} {\bibfield  {journal}
  {\bibinfo  {journal} {Phys. Rev. Lett.}\ }\textbf {\bibinfo {volume} {94}},\
  \bibinfo {pages} {176801} (\bibinfo {year} {2005})}\BibitemShut {NoStop}%
\bibitem [{\citenamefont {Cornaglia}\ \emph {et~al.}(2005)\citenamefont
  {Cornaglia}, \citenamefont {Grempel},\ and\ \citenamefont
  {Ness}}]{CornagliaNess2005}%
  \BibitemOpen
  \bibfield  {author} {\bibinfo {author} {\bibfnamefont {P.~S.}\ \bibnamefont
  {Cornaglia}}, \bibinfo {author} {\bibfnamefont {D.~R.}\ \bibnamefont
  {Grempel}},\ and\ \bibinfo {author} {\bibfnamefont {H.}~\bibnamefont
  {Ness}},\ }\bibfield  {title} {\bibinfo {title} {Quantum transport through a
  deformable molecular transistor},\ }\href
  {https://doi.org/10.1103/PhysRevB.71.075320} {\bibfield  {journal} {\bibinfo
  {journal} {Phys. Rev. B}\ }\textbf {\bibinfo {volume} {71}},\ \bibinfo
  {pages} {075320} (\bibinfo {year} {2005})}\BibitemShut {NoStop}%
\bibitem [{\citenamefont {Laakso}\ \emph {et~al.}(2014)\citenamefont {Laakso},
  \citenamefont {Kennes}, \citenamefont {Jakobs},\ and\ \citenamefont
  {Meden}}]{LaaksoMeden2014}%
  \BibitemOpen
  \bibfield  {author} {\bibinfo {author} {\bibfnamefont {M.~A.}\ \bibnamefont
  {Laakso}}, \bibinfo {author} {\bibfnamefont {D.~M.}\ \bibnamefont {Kennes}},
  \bibinfo {author} {\bibfnamefont {S.~G.}\ \bibnamefont {Jakobs}},\ and\
  \bibinfo {author} {\bibfnamefont {V.}~\bibnamefont {Meden}},\ }\bibfield
  {title} {\bibinfo {title} {Functional renormalization group study of the
  anderson–holstein model},\ }\href
  {https://doi.org/10.1088/1367-2630/16/2/023007} {\bibfield  {journal}
  {\bibinfo  {journal} {New Journal of Physics}\ }\textbf {\bibinfo {volume}
  {16}},\ \bibinfo {pages} {023007} (\bibinfo {year} {2014})}\BibitemShut
  {NoStop}%
\bibitem [{\citenamefont {Wolf}\ \emph
  {et~al.}(2014{\natexlab{a}})\citenamefont {Wolf}, \citenamefont {McCulloch},
  \citenamefont {Parcollet},\ and\ \citenamefont
  {Schollw\"ock}}]{WolfSchollwock2014b}%
  \BibitemOpen
  \bibfield  {author} {\bibinfo {author} {\bibfnamefont {F.~A.}\ \bibnamefont
  {Wolf}}, \bibinfo {author} {\bibfnamefont {I.~P.}\ \bibnamefont {McCulloch}},
  \bibinfo {author} {\bibfnamefont {O.}~\bibnamefont {Parcollet}},\ and\
  \bibinfo {author} {\bibfnamefont {U.}~\bibnamefont {Schollw\"ock}},\
  }\bibfield  {title} {\bibinfo {title} {Chebyshev matrix product state
  impurity solver for dynamical mean-field theory},\ }\href
  {https://doi.org/10.1103/PhysRevB.90.115124} {\bibfield  {journal} {\bibinfo
  {journal} {Phys. Rev. B}\ }\textbf {\bibinfo {volume} {90}},\ \bibinfo
  {pages} {115124} (\bibinfo {year} {2014}{\natexlab{a}})}\BibitemShut
  {NoStop}%
\bibitem [{\citenamefont {Ganahl}\ \emph {et~al.}(2014)\citenamefont {Ganahl},
  \citenamefont {Thunstr\"om}, \citenamefont {Verstraete}, \citenamefont
  {Held},\ and\ \citenamefont {Evertz}}]{GanahlEvertz2014}%
  \BibitemOpen
  \bibfield  {author} {\bibinfo {author} {\bibfnamefont {M.}~\bibnamefont
  {Ganahl}}, \bibinfo {author} {\bibfnamefont {P.}~\bibnamefont {Thunstr\"om}},
  \bibinfo {author} {\bibfnamefont {F.}~\bibnamefont {Verstraete}}, \bibinfo
  {author} {\bibfnamefont {K.}~\bibnamefont {Held}},\ and\ \bibinfo {author}
  {\bibfnamefont {H.~G.}\ \bibnamefont {Evertz}},\ }\bibfield  {title}
  {\bibinfo {title} {Chebyshev expansion for impurity models using matrix
  product states},\ }\href {https://doi.org/10.1103/PhysRevB.90.045144}
  {\bibfield  {journal} {\bibinfo  {journal} {Phys. Rev. B}\ }\textbf {\bibinfo
  {volume} {90}},\ \bibinfo {pages} {045144} (\bibinfo {year}
  {2014})}\BibitemShut {NoStop}%
\bibitem [{\citenamefont {Ganahl}\ \emph {et~al.}(2015)\citenamefont {Ganahl},
  \citenamefont {Aichhorn}, \citenamefont {Evertz}, \citenamefont
  {Thunstr\"om}, \citenamefont {Held},\ and\ \citenamefont
  {Verstraete}}]{GanahlVerstraete2015}%
  \BibitemOpen
  \bibfield  {author} {\bibinfo {author} {\bibfnamefont {M.}~\bibnamefont
  {Ganahl}}, \bibinfo {author} {\bibfnamefont {M.}~\bibnamefont {Aichhorn}},
  \bibinfo {author} {\bibfnamefont {H.~G.}\ \bibnamefont {Evertz}}, \bibinfo
  {author} {\bibfnamefont {P.}~\bibnamefont {Thunstr\"om}}, \bibinfo {author}
  {\bibfnamefont {K.}~\bibnamefont {Held}},\ and\ \bibinfo {author}
  {\bibfnamefont {F.}~\bibnamefont {Verstraete}},\ }\bibfield  {title}
  {\bibinfo {title} {Efficient dmft impurity solver using real-time dynamics
  with matrix product states},\ }\href
  {https://doi.org/10.1103/PhysRevB.92.155132} {\bibfield  {journal} {\bibinfo
  {journal} {Phys. Rev. B}\ }\textbf {\bibinfo {volume} {92}},\ \bibinfo
  {pages} {155132} (\bibinfo {year} {2015})}\BibitemShut {NoStop}%
\bibitem [{\citenamefont {Wolf}\ \emph {et~al.}(2015)\citenamefont {Wolf},
  \citenamefont {Go}, \citenamefont {McCulloch}, \citenamefont {Millis},\ and\
  \citenamefont {Schollw\"ock}}]{WolfSchollwock2015}%
  \BibitemOpen
  \bibfield  {author} {\bibinfo {author} {\bibfnamefont {F.~A.}\ \bibnamefont
  {Wolf}}, \bibinfo {author} {\bibfnamefont {A.}~\bibnamefont {Go}}, \bibinfo
  {author} {\bibfnamefont {I.~P.}\ \bibnamefont {McCulloch}}, \bibinfo {author}
  {\bibfnamefont {A.~J.}\ \bibnamefont {Millis}},\ and\ \bibinfo {author}
  {\bibfnamefont {U.}~\bibnamefont {Schollw\"ock}},\ }\bibfield  {title}
  {\bibinfo {title} {Imaginary-time matrix product state impurity solver for
  dynamical mean-field theory},\ }\href
  {https://doi.org/10.1103/PhysRevX.5.041032} {\bibfield  {journal} {\bibinfo
  {journal} {Phys. Rev. X}\ }\textbf {\bibinfo {volume} {5}},\ \bibinfo {pages}
  {041032} (\bibinfo {year} {2015})}\BibitemShut {NoStop}%
\bibitem [{\citenamefont {Garc\'{\i}a}\ \emph {et~al.}(2004)\citenamefont
  {Garc\'{\i}a}, \citenamefont {Hallberg},\ and\ \citenamefont
  {Rozenberg}}]{GarciaRozenberg2004}%
  \BibitemOpen
  \bibfield  {author} {\bibinfo {author} {\bibfnamefont {D.~J.}\ \bibnamefont
  {Garc\'{\i}a}}, \bibinfo {author} {\bibfnamefont {K.}~\bibnamefont
  {Hallberg}},\ and\ \bibinfo {author} {\bibfnamefont {M.~J.}\ \bibnamefont
  {Rozenberg}},\ }\bibfield  {title} {\bibinfo {title} {Dynamical mean field
  theory with the density matrix renormalization group},\ }\href
  {https://doi.org/10.1103/PhysRevLett.93.246403} {\bibfield  {journal}
  {\bibinfo  {journal} {Phys. Rev. Lett.}\ }\textbf {\bibinfo {volume} {93}},\
  \bibinfo {pages} {246403} (\bibinfo {year} {2004})}\BibitemShut {NoStop}%
\bibitem [{\citenamefont {Nishimoto}\ \emph {et~al.}(2006)\citenamefont
  {Nishimoto}, \citenamefont {Gebhard},\ and\ \citenamefont
  {Jeckelmann}}]{NishimotoJeckelmann2006}%
  \BibitemOpen
  \bibfield  {author} {\bibinfo {author} {\bibfnamefont {S.}~\bibnamefont
  {Nishimoto}}, \bibinfo {author} {\bibfnamefont {F.}~\bibnamefont {Gebhard}},\
  and\ \bibinfo {author} {\bibfnamefont {E.}~\bibnamefont {Jeckelmann}},\
  }\bibfield  {title} {\bibinfo {title} {Dynamical mean-field theory
  calculation with the dynamical density-matrix renormalization group},\ }\href
  {https://doi.org/https://doi.org/10.1016/j.physb.2006.01.104} {\bibfield
  {journal} {\bibinfo  {journal} {Physica B Condens. Matter}\ }\textbf
  {\bibinfo {volume} {378-380}},\ \bibinfo {pages} {283} (\bibinfo {year}
  {2006})}\BibitemShut {NoStop}%
\bibitem [{\citenamefont {Weichselbaum}\ \emph {et~al.}(2009)\citenamefont
  {Weichselbaum}, \citenamefont {Verstraete}, \citenamefont {Schollw\"ock},
  \citenamefont {Cirac},\ and\ \citenamefont {von
  Delft}}]{WeichselbaumDelft2009}%
  \BibitemOpen
  \bibfield  {author} {\bibinfo {author} {\bibfnamefont {A.}~\bibnamefont
  {Weichselbaum}}, \bibinfo {author} {\bibfnamefont {F.}~\bibnamefont
  {Verstraete}}, \bibinfo {author} {\bibfnamefont {U.}~\bibnamefont
  {Schollw\"ock}}, \bibinfo {author} {\bibfnamefont {J.~I.}\ \bibnamefont
  {Cirac}},\ and\ \bibinfo {author} {\bibfnamefont {J.}~\bibnamefont {von
  Delft}},\ }\bibfield  {title} {\bibinfo {title} {Variational
  matrix-product-state approach to quantum impurity models},\ }\href
  {https://doi.org/10.1103/PhysRevB.80.165117} {\bibfield  {journal} {\bibinfo
  {journal} {Phys. Rev. B}\ }\textbf {\bibinfo {volume} {80}},\ \bibinfo
  {pages} {165117} (\bibinfo {year} {2009})}\BibitemShut {NoStop}%
\bibitem [{\citenamefont {Bauernfeind}\ \emph {et~al.}(2017)\citenamefont
  {Bauernfeind}, \citenamefont {Zingl}, \citenamefont {Triebl}, \citenamefont
  {Aichhorn},\ and\ \citenamefont {Evertz}}]{BauernfeindEvertz2017}%
  \BibitemOpen
  \bibfield  {author} {\bibinfo {author} {\bibfnamefont {D.}~\bibnamefont
  {Bauernfeind}}, \bibinfo {author} {\bibfnamefont {M.}~\bibnamefont {Zingl}},
  \bibinfo {author} {\bibfnamefont {R.}~\bibnamefont {Triebl}}, \bibinfo
  {author} {\bibfnamefont {M.}~\bibnamefont {Aichhorn}},\ and\ \bibinfo
  {author} {\bibfnamefont {H.~G.}\ \bibnamefont {Evertz}},\ }\bibfield  {title}
  {\bibinfo {title} {Fork tensor-product states: Efficient multiorbital
  real-time dmft solver},\ }\href {https://doi.org/10.1103/PhysRevX.7.031013}
  {\bibfield  {journal} {\bibinfo  {journal} {Phys. Rev. X}\ }\textbf {\bibinfo
  {volume} {7}},\ \bibinfo {pages} {031013} (\bibinfo {year}
  {2017})}\BibitemShut {NoStop}%
\bibitem [{\citenamefont {Werner}\ \emph {et~al.}(2023)\citenamefont {Werner},
  \citenamefont {Lotze},\ and\ \citenamefont {Arrigoni}}]{WernerArrigoni2023}%
  \BibitemOpen
  \bibfield  {author} {\bibinfo {author} {\bibfnamefont {D.}~\bibnamefont
  {Werner}}, \bibinfo {author} {\bibfnamefont {J.}~\bibnamefont {Lotze}},\ and\
  \bibinfo {author} {\bibfnamefont {E.}~\bibnamefont {Arrigoni}},\ }\bibfield
  {title} {\bibinfo {title} {Configuration interaction based nonequilibrium
  steady state impurity solver},\ }\href
  {https://doi.org/10.1103/PhysRevB.107.075119} {\bibfield  {journal} {\bibinfo
   {journal} {Phys. Rev. B}\ }\textbf {\bibinfo {volume} {107}},\ \bibinfo
  {pages} {075119} (\bibinfo {year} {2023})}\BibitemShut {NoStop}%
\bibitem [{\citenamefont {Kohn}\ and\ \citenamefont
  {Santoro}(2021)}]{KohnSantoro2021}%
  \BibitemOpen
  \bibfield  {author} {\bibinfo {author} {\bibfnamefont {L.}~\bibnamefont
  {Kohn}}\ and\ \bibinfo {author} {\bibfnamefont {G.~E.}\ \bibnamefont
  {Santoro}},\ }\bibfield  {title} {\bibinfo {title} {Efficient mapping for
  anderson impurity problems with matrix product states},\ }\href
  {https://doi.org/10.1103/PhysRevB.104.014303} {\bibfield  {journal} {\bibinfo
   {journal} {Phys. Rev. B}\ }\textbf {\bibinfo {volume} {104}},\ \bibinfo
  {pages} {014303} (\bibinfo {year} {2021})}\BibitemShut {NoStop}%
\bibitem [{\citenamefont {Kohn}\ and\ \citenamefont
  {Santoro}(2022)}]{KohnSantoro2022}%
  \BibitemOpen
  \bibfield  {author} {\bibinfo {author} {\bibfnamefont {L.}~\bibnamefont
  {Kohn}}\ and\ \bibinfo {author} {\bibfnamefont {G.~E.}\ \bibnamefont
  {Santoro}},\ }\bibfield  {title} {\bibinfo {title} {Quench dynamics of the
  anderson impurity model at finite temperature using matrix product states:
  entanglement and bath dynamics},\ }\href
  {https://doi.org/10.1088/1742-5468/ac729b} {\bibfield  {journal} {\bibinfo
  {journal} {J. Stat. Mech. Theory Exp.}\ }\textbf {\bibinfo {volume} {2022}},\
  \bibinfo {pages} {063102} (\bibinfo {year} {2022})}\BibitemShut {NoStop}%
\bibitem [{\citenamefont {Gull}\ \emph {et~al.}(2011)\citenamefont {Gull},
  \citenamefont {Millis}, \citenamefont {Lichtenstein}, \citenamefont
  {Rubtsov}, \citenamefont {Troyer},\ and\ \citenamefont
  {Werner}}]{GullWerner2011}%
  \BibitemOpen
  \bibfield  {author} {\bibinfo {author} {\bibfnamefont {E.}~\bibnamefont
  {Gull}}, \bibinfo {author} {\bibfnamefont {A.~J.}\ \bibnamefont {Millis}},
  \bibinfo {author} {\bibfnamefont {A.~I.}\ \bibnamefont {Lichtenstein}},
  \bibinfo {author} {\bibfnamefont {A.~N.}\ \bibnamefont {Rubtsov}}, \bibinfo
  {author} {\bibfnamefont {M.}~\bibnamefont {Troyer}},\ and\ \bibinfo {author}
  {\bibfnamefont {P.}~\bibnamefont {Werner}},\ }\bibfield  {title} {\bibinfo
  {title} {Continuous-time monte carlo methods for quantum impurity models},\
  }\href {https://doi.org/10.1103/RevModPhys.83.349} {\bibfield  {journal}
  {\bibinfo  {journal} {Rev. Mod. Phys.}\ }\textbf {\bibinfo {volume} {83}},\
  \bibinfo {pages} {349} (\bibinfo {year} {2011})}\BibitemShut {NoStop}%
\bibitem [{\citenamefont {Prokof'ev}\ and\ \citenamefont
  {Svistunov}(1998)}]{ProkofevSvistunov1998}%
  \BibitemOpen
  \bibfield  {author} {\bibinfo {author} {\bibfnamefont {N.~V.}\ \bibnamefont
  {Prokof'ev}}\ and\ \bibinfo {author} {\bibfnamefont {B.~V.}\ \bibnamefont
  {Svistunov}},\ }\bibfield  {title} {\bibinfo {title} {Polaron problem by
  diagrammatic quantum monte carlo},\ }\href
  {https://doi.org/10.1103/PhysRevLett.81.2514} {\bibfield  {journal} {\bibinfo
   {journal} {Phys. Rev. Lett.}\ }\textbf {\bibinfo {volume} {81}},\ \bibinfo
  {pages} {2514} (\bibinfo {year} {1998})}\BibitemShut {NoStop}%
\bibitem [{\citenamefont {Hafermann}\ \emph {et~al.}(2013)\citenamefont
  {Hafermann}, \citenamefont {Werner},\ and\ \citenamefont
  {Gull}}]{HafermannGull2012}%
  \BibitemOpen
  \bibfield  {author} {\bibinfo {author} {\bibfnamefont {H.}~\bibnamefont
  {Hafermann}}, \bibinfo {author} {\bibfnamefont {P.}~\bibnamefont {Werner}},\
  and\ \bibinfo {author} {\bibfnamefont {E.}~\bibnamefont {Gull}},\ }\bibfield
  {title} {\bibinfo {title} {Efficient implementation of the continuous-time
  hybridization expansion quantum impurity solver},\ }\href
  {https://doi.org/https://doi.org/10.1016/j.cpc.2012.12.013} {\bibfield
  {journal} {\bibinfo  {journal} {Computer Physics Communications}\ }\textbf
  {\bibinfo {volume} {184}},\ \bibinfo {pages} {1280} (\bibinfo {year}
  {2013})}\BibitemShut {NoStop}%
\bibitem [{\citenamefont {Lang}\ and\ \citenamefont
  {Firsov}(1963)}]{lang1963-kinetic}%
  \BibitemOpen
  \bibfield  {author} {\bibinfo {author} {\bibfnamefont {I.~G.}\ \bibnamefont
  {Lang}}\ and\ \bibinfo {author} {\bibfnamefont {Y.~A.}\ \bibnamefont
  {Firsov}},\ }\bibfield  {title} {\bibinfo {title} {Kinetic theory of
  semiconductors with low mobility},\ }\href@noop {} {\bibfield  {journal}
  {\bibinfo  {journal} {Soviet Physics JETP}\ }\textbf {\bibinfo {volume}
  {16}},\ \bibinfo {pages} {1301} (\bibinfo {year} {1963})}\BibitemShut
  {NoStop}%
\bibitem [{\citenamefont {Hafermann}(2014)}]{Hafermann2014}%
  \BibitemOpen
  \bibfield  {author} {\bibinfo {author} {\bibfnamefont {H.}~\bibnamefont
  {Hafermann}},\ }\bibfield  {title} {\bibinfo {title} {Self-energy and vertex
  functions from hybridization-expansion continuous-time quantum monte carlo
  for impurity models with retarded interaction},\ }\href
  {https://doi.org/10.1103/PhysRevB.89.235128} {\bibfield  {journal} {\bibinfo
  {journal} {Phys. Rev. B}\ }\textbf {\bibinfo {volume} {89}},\ \bibinfo
  {pages} {235128} (\bibinfo {year} {2014})}\BibitemShut {NoStop}%
\bibitem [{\citenamefont {Gu}\ \emph {et~al.}(2025)\citenamefont {Gu},
  \citenamefont {Luo}, \citenamefont {Wu},\ and\ \citenamefont
  {Zhao}}]{GuZhao2025}%
  \BibitemOpen
  \bibfield  {author} {\bibinfo {author} {\bibfnamefont {L.}~\bibnamefont
  {Gu}}, \bibinfo {author} {\bibfnamefont {J.}~\bibnamefont {Luo}}, \bibinfo
  {author} {\bibfnamefont {R.}~\bibnamefont {Wu}},\ and\ \bibinfo {author}
  {\bibfnamefont {G.}~\bibnamefont {Zhao}},\ }\bibfield  {title} {\bibinfo
  {title} {Extended hybridization expansion solver for impurity models with
  retarded interactions},\ }\href {https://doi.org/10.48550/arXiv.2504.02274}
  {\bibfield  {journal} {\bibinfo  {journal} {arXiv:2504.02274}\ } (\bibinfo
  {year} {2025})}\BibitemShut {NoStop}%
\bibitem [{\citenamefont {Troyer}\ and\ \citenamefont
  {Wiese}(2005)}]{TroyerWiese2005}%
  \BibitemOpen
  \bibfield  {author} {\bibinfo {author} {\bibfnamefont {M.}~\bibnamefont
  {Troyer}}\ and\ \bibinfo {author} {\bibfnamefont {U.-J.}\ \bibnamefont
  {Wiese}},\ }\bibfield  {title} {\bibinfo {title} {Computational complexity
  and fundamental limitations to fermionic quantum monte carlo simulations},\
  }\href {https://doi.org/10.1103/PhysRevLett.94.170201} {\bibfield  {journal}
  {\bibinfo  {journal} {Phys. Rev. Lett.}\ }\textbf {\bibinfo {volume} {94}},\
  \bibinfo {pages} {170201} (\bibinfo {year} {2005})}\BibitemShut {NoStop}%
\bibitem [{\citenamefont {Fei}\ \emph {et~al.}(2021)\citenamefont {Fei},
  \citenamefont {Yeh},\ and\ \citenamefont {Gull}}]{FeiGull2021}%
  \BibitemOpen
  \bibfield  {author} {\bibinfo {author} {\bibfnamefont {J.}~\bibnamefont
  {Fei}}, \bibinfo {author} {\bibfnamefont {C.-N.}\ \bibnamefont {Yeh}},\ and\
  \bibinfo {author} {\bibfnamefont {E.}~\bibnamefont {Gull}},\ }\bibfield
  {title} {\bibinfo {title} {Nevanlinna analytical continuation},\ }\href
  {https://doi.org/10.1103/PhysRevLett.126.056402} {\bibfield  {journal}
  {\bibinfo  {journal} {Phys. Rev. Lett.}\ }\textbf {\bibinfo {volume} {126}},\
  \bibinfo {pages} {056402} (\bibinfo {year} {2021})}\BibitemShut {NoStop}%
\bibitem [{\citenamefont {Cohen}\ \emph
  {et~al.}(2014{\natexlab{a}})\citenamefont {Cohen}, \citenamefont {Reichman},
  \citenamefont {Millis},\ and\ \citenamefont {Gull}}]{CohenGull2014}%
  \BibitemOpen
  \bibfield  {author} {\bibinfo {author} {\bibfnamefont {G.}~\bibnamefont
  {Cohen}}, \bibinfo {author} {\bibfnamefont {D.~R.}\ \bibnamefont {Reichman}},
  \bibinfo {author} {\bibfnamefont {A.~J.}\ \bibnamefont {Millis}},\ and\
  \bibinfo {author} {\bibfnamefont {E.}~\bibnamefont {Gull}},\ }\bibfield
  {title} {\bibinfo {title} {Green's functions from real-time bold-line monte
  carlo},\ }\href {https://doi.org/10.1103/PhysRevB.89.115139} {\bibfield
  {journal} {\bibinfo  {journal} {Phys. Rev. B}\ }\textbf {\bibinfo {volume}
  {89}},\ \bibinfo {pages} {115139} (\bibinfo {year}
  {2014}{\natexlab{a}})}\BibitemShut {NoStop}%
\bibitem [{\citenamefont {Cohen}\ \emph
  {et~al.}(2014{\natexlab{b}})\citenamefont {Cohen}, \citenamefont {Gull},
  \citenamefont {Reichman},\ and\ \citenamefont {Millis}}]{CohenMillis2014}%
  \BibitemOpen
  \bibfield  {author} {\bibinfo {author} {\bibfnamefont {G.}~\bibnamefont
  {Cohen}}, \bibinfo {author} {\bibfnamefont {E.}~\bibnamefont {Gull}},
  \bibinfo {author} {\bibfnamefont {D.~R.}\ \bibnamefont {Reichman}},\ and\
  \bibinfo {author} {\bibfnamefont {A.~J.}\ \bibnamefont {Millis}},\ }\bibfield
   {title} {\bibinfo {title} {Green's functions from real-time bold-line monte
  carlo calculations: Spectral properties of the nonequilibrium anderson
  impurity model},\ }\href {https://doi.org/10.1103/PhysRevLett.112.146802}
  {\bibfield  {journal} {\bibinfo  {journal} {Phys. Rev. Lett.}\ }\textbf
  {\bibinfo {volume} {112}},\ \bibinfo {pages} {146802} (\bibinfo {year}
  {2014}{\natexlab{b}})}\BibitemShut {NoStop}%
\bibitem [{\citenamefont {Cohen}\ \emph {et~al.}(2015)\citenamefont {Cohen},
  \citenamefont {Gull}, \citenamefont {Reichman},\ and\ \citenamefont
  {Millis}}]{CohenMillis2015}%
  \BibitemOpen
  \bibfield  {author} {\bibinfo {author} {\bibfnamefont {G.}~\bibnamefont
  {Cohen}}, \bibinfo {author} {\bibfnamefont {E.}~\bibnamefont {Gull}},
  \bibinfo {author} {\bibfnamefont {D.~R.}\ \bibnamefont {Reichman}},\ and\
  \bibinfo {author} {\bibfnamefont {A.~J.}\ \bibnamefont {Millis}},\ }\bibfield
   {title} {\bibinfo {title} {Taming the dynamical sign problem in real-time
  evolution of quantum many-body problems},\ }\href
  {https://doi.org/10.1103/PhysRevLett.115.266802} {\bibfield  {journal}
  {\bibinfo  {journal} {Phys. Rev. Lett.}\ }\textbf {\bibinfo {volume} {115}},\
  \bibinfo {pages} {266802} (\bibinfo {year} {2015})}\BibitemShut {NoStop}%
\bibitem [{\citenamefont {Chen}\ \emph
  {et~al.}(2017{\natexlab{a}})\citenamefont {Chen}, \citenamefont {Cohen},\
  and\ \citenamefont {Reichman}}]{ChenReichman2017a}%
  \BibitemOpen
  \bibfield  {author} {\bibinfo {author} {\bibfnamefont {H.-T.}\ \bibnamefont
  {Chen}}, \bibinfo {author} {\bibfnamefont {G.}~\bibnamefont {Cohen}},\ and\
  \bibinfo {author} {\bibfnamefont {D.~R.}\ \bibnamefont {Reichman}},\
  }\bibfield  {title} {\bibinfo {title} {{Inchworm Monte Carlo for exact
  non-adiabatic dynamics. I. Theory and algorithms}},\ }\href
  {https://doi.org/10.1063/1.4974328} {\bibfield  {journal} {\bibinfo
  {journal} {J. Chem. Phys.}\ }\textbf {\bibinfo {volume} {146}},\ \bibinfo
  {pages} {054105} (\bibinfo {year} {2017}{\natexlab{a}})}\BibitemShut
  {NoStop}%
\bibitem [{\citenamefont {Chen}\ \emph
  {et~al.}(2017{\natexlab{b}})\citenamefont {Chen}, \citenamefont {Cohen},\
  and\ \citenamefont {Reichman}}]{ChenReichman2017b}%
  \BibitemOpen
  \bibfield  {author} {\bibinfo {author} {\bibfnamefont {H.-T.}\ \bibnamefont
  {Chen}}, \bibinfo {author} {\bibfnamefont {G.}~\bibnamefont {Cohen}},\ and\
  \bibinfo {author} {\bibfnamefont {D.~R.}\ \bibnamefont {Reichman}},\
  }\bibfield  {title} {\bibinfo {title} {{Inchworm Monte Carlo for exact
  non-adiabatic dynamics. II. Benchmarks and comparison with established
  methods}},\ }\href {https://doi.org/10.1063/1.4974329} {\bibfield  {journal}
  {\bibinfo  {journal} {J. Chem. Phys.}\ }\textbf {\bibinfo {volume} {146}},\
  \bibinfo {pages} {054106} (\bibinfo {year} {2017}{\natexlab{b}})}\BibitemShut
  {NoStop}%
\bibitem [{\citenamefont {Bertrand}\ \emph {et~al.}(2019)\citenamefont
  {Bertrand}, \citenamefont {Florens}, \citenamefont {Parcollet},\ and\
  \citenamefont {Waintal}}]{BertrandWaintal2019}%
  \BibitemOpen
  \bibfield  {author} {\bibinfo {author} {\bibfnamefont {C.}~\bibnamefont
  {Bertrand}}, \bibinfo {author} {\bibfnamefont {S.}~\bibnamefont {Florens}},
  \bibinfo {author} {\bibfnamefont {O.}~\bibnamefont {Parcollet}},\ and\
  \bibinfo {author} {\bibfnamefont {X.}~\bibnamefont {Waintal}},\ }\bibfield
  {title} {\bibinfo {title} {Reconstructing nonequilibrium regimes of quantum
  many-body systems from the analytical structure of perturbative expansions},\
  }\href {https://doi.org/10.1103/PhysRevX.9.041008} {\bibfield  {journal}
  {\bibinfo  {journal} {Phys. Rev. X}\ }\textbf {\bibinfo {volume} {9}},\
  \bibinfo {pages} {041008} (\bibinfo {year} {2019})}\BibitemShut {NoStop}%
\bibitem [{\citenamefont {Erpenbeck}\ \emph {et~al.}(2023)\citenamefont
  {Erpenbeck}, \citenamefont {Gull},\ and\ \citenamefont
  {Cohen}}]{ErpenbeckCohen2023}%
  \BibitemOpen
  \bibfield  {author} {\bibinfo {author} {\bibfnamefont {A.}~\bibnamefont
  {Erpenbeck}}, \bibinfo {author} {\bibfnamefont {E.}~\bibnamefont {Gull}},\
  and\ \bibinfo {author} {\bibfnamefont {G.}~\bibnamefont {Cohen}},\ }\bibfield
   {title} {\bibinfo {title} {Quantum monte carlo method in the steady state},\
  }\href {https://doi.org/10.1103/PhysRevLett.130.186301} {\bibfield  {journal}
  {\bibinfo  {journal} {Phys. Rev. Lett.}\ }\textbf {\bibinfo {volume} {130}},\
  \bibinfo {pages} {186301} (\bibinfo {year} {2023})}\BibitemShut {NoStop}%
\bibitem [{\citenamefont {Feynman}\ and\ \citenamefont
  {Vernon}(1963)}]{FeynmanVernon1963}%
  \BibitemOpen
  \bibfield  {author} {\bibinfo {author} {\bibfnamefont {R.~P.}\ \bibnamefont
  {Feynman}}\ and\ \bibinfo {author} {\bibfnamefont {F.~L.}\ \bibnamefont
  {Vernon}},\ }\bibfield  {title} {\bibinfo {title} {The theory of a general
  quantum system interacting with a linear dissipative system},\ }\href
  {https://doi.org/10.1016/0003-4916(63)90068-X} {\bibfield  {journal}
  {\bibinfo  {journal} {Ann. Phys.}\ }\textbf {\bibinfo {volume} {24}},\
  \bibinfo {pages} {118} (\bibinfo {year} {1963})}\BibitemShut {NoStop}%
\bibitem [{\citenamefont {Negele}\ and\ \citenamefont
  {Orland}(1998)}]{negele1998-quantum}%
  \BibitemOpen
  \bibfield  {author} {\bibinfo {author} {\bibfnamefont {J.~W.}\ \bibnamefont
  {Negele}}\ and\ \bibinfo {author} {\bibfnamefont {H.}~\bibnamefont
  {Orland}},\ }\href@noop {} {\emph {\bibinfo {title} {Quantum Many-Particle
  Systems}}}\ (\bibinfo  {publisher} {Westview Press},\ \bibinfo {year}
  {1998})\BibitemShut {NoStop}%
\bibitem [{\citenamefont {Strathearn}\ \emph {et~al.}(2018)\citenamefont
  {Strathearn}, \citenamefont {Kirton}, \citenamefont {Kilda}, \citenamefont
  {Keeling},\ and\ \citenamefont {Lovett}}]{StrathearnLovett2018}%
  \BibitemOpen
  \bibfield  {author} {\bibinfo {author} {\bibfnamefont {A.}~\bibnamefont
  {Strathearn}}, \bibinfo {author} {\bibfnamefont {P.}~\bibnamefont {Kirton}},
  \bibinfo {author} {\bibfnamefont {D.}~\bibnamefont {Kilda}}, \bibinfo
  {author} {\bibfnamefont {J.}~\bibnamefont {Keeling}},\ and\ \bibinfo {author}
  {\bibfnamefont {B.~W.}\ \bibnamefont {Lovett}},\ }\bibfield  {title}
  {\bibinfo {title} {Efficient non-markovian quantum dynamics using
  time-evolving matrix product operators},\ }\href
  {https://doi.org/10.1038/s41467-018-05617-3} {\bibfield  {journal} {\bibinfo
  {journal} {Nat. Commun.}\ }\textbf {\bibinfo {volume} {9}},\ \bibinfo {pages}
  {3322} (\bibinfo {year} {2018})}\BibitemShut {NoStop}%
\bibitem [{\citenamefont {Thoenniss}\ \emph {et~al.}(2023)\citenamefont
  {Thoenniss}, \citenamefont {Sonner}, \citenamefont {Lerose},\ and\
  \citenamefont {Abanin}}]{ThoennissAbanin2023b}%
  \BibitemOpen
  \bibfield  {author} {\bibinfo {author} {\bibfnamefont {J.}~\bibnamefont
  {Thoenniss}}, \bibinfo {author} {\bibfnamefont {M.}~\bibnamefont {Sonner}},
  \bibinfo {author} {\bibfnamefont {A.}~\bibnamefont {Lerose}},\ and\ \bibinfo
  {author} {\bibfnamefont {D.~A.}\ \bibnamefont {Abanin}},\ }\bibfield  {title}
  {\bibinfo {title} {Efficient method for quantum impurity problems out of
  equilibrium},\ }\href {https://doi.org/10.1103/PhysRevB.107.L201115}
  {\bibfield  {journal} {\bibinfo  {journal} {Phys. Rev. B}\ }\textbf {\bibinfo
  {volume} {107}},\ \bibinfo {pages} {L201115} (\bibinfo {year}
  {2023})}\BibitemShut {NoStop}%
\bibitem [{\citenamefont {Chen}\ \emph
  {et~al.}(2024{\natexlab{a}})\citenamefont {Chen}, \citenamefont {Xu},\ and\
  \citenamefont {Guo}}]{ChenGuo2024a}%
  \BibitemOpen
  \bibfield  {author} {\bibinfo {author} {\bibfnamefont {R.}~\bibnamefont
  {Chen}}, \bibinfo {author} {\bibfnamefont {X.}~\bibnamefont {Xu}},\ and\
  \bibinfo {author} {\bibfnamefont {C.}~\bibnamefont {Guo}},\ }\bibfield
  {title} {\bibinfo {title} {Grassmann time-evolving matrix product operators
  for quantum impurity models},\ }\href
  {https://doi.org/10.1103/PhysRevB.109.045140} {\bibfield  {journal} {\bibinfo
   {journal} {Phys. Rev. B}\ }\textbf {\bibinfo {volume} {109}},\ \bibinfo
  {pages} {045140} (\bibinfo {year} {2024}{\natexlab{a}})}\BibitemShut
  {NoStop}%
\bibitem [{\citenamefont {Chen}\ \emph
  {et~al.}(2024{\natexlab{b}})\citenamefont {Chen}, \citenamefont {Xu},\ and\
  \citenamefont {Guo}}]{ChenGuo2024b}%
  \BibitemOpen
  \bibfield  {author} {\bibinfo {author} {\bibfnamefont {R.}~\bibnamefont
  {Chen}}, \bibinfo {author} {\bibfnamefont {X.}~\bibnamefont {Xu}},\ and\
  \bibinfo {author} {\bibfnamefont {C.}~\bibnamefont {Guo}},\ }\bibfield
  {title} {\bibinfo {title} {Grassmann time-evolving matrix product operators
  for equilibrium quantum impurity problems},\ }\href
  {https://doi.org/10.1088/1367-2630/ad19fa} {\bibfield  {journal} {\bibinfo
  {journal} {New J. Phys.}\ }\textbf {\bibinfo {volume} {26}},\ \bibinfo
  {pages} {013019} (\bibinfo {year} {2024}{\natexlab{b}})}\BibitemShut
  {NoStop}%
\bibitem [{\citenamefont {Chen}\ \emph
  {et~al.}(2024{\natexlab{c}})\citenamefont {Chen}, \citenamefont {Xu},\ and\
  \citenamefont {Guo}}]{ChenGuo2024c}%
  \BibitemOpen
  \bibfield  {author} {\bibinfo {author} {\bibfnamefont {R.}~\bibnamefont
  {Chen}}, \bibinfo {author} {\bibfnamefont {X.}~\bibnamefont {Xu}},\ and\
  \bibinfo {author} {\bibfnamefont {C.}~\bibnamefont {Guo}},\ }\bibfield
  {title} {\bibinfo {title} {Real-time impurity solver using grassmann
  time-evolving matrix product operators},\ }\href
  {https://doi.org/10.1103/PhysRevB.109.165113} {\bibfield  {journal} {\bibinfo
   {journal} {Phys. Rev. B}\ }\textbf {\bibinfo {volume} {109}},\ \bibinfo
  {pages} {165113} (\bibinfo {year} {2024}{\natexlab{c}})}\BibitemShut
  {NoStop}%
\bibitem [{\citenamefont {Guo}\ and\ \citenamefont
  {Chen}(2024{\natexlab{a}})}]{GuoChen2024d}%
  \BibitemOpen
  \bibfield  {author} {\bibinfo {author} {\bibfnamefont {C.}~\bibnamefont
  {Guo}}\ and\ \bibinfo {author} {\bibfnamefont {R.}~\bibnamefont {Chen}},\
  }\bibfield  {title} {\bibinfo {title} {{Efficient construction of the
  Feynman-Vernon influence functional as matrix product states}},\ }\href
  {https://doi.org/10.21468/SciPostPhysCore.7.3.063} {\bibfield  {journal}
  {\bibinfo  {journal} {SciPost Phys. Core}\ }\textbf {\bibinfo {volume} {7}},\
  \bibinfo {pages} {063} (\bibinfo {year} {2024}{\natexlab{a}})}\BibitemShut
  {NoStop}%
\bibitem [{\citenamefont {Chen}\ and\ \citenamefont
  {Guo}(2024)}]{ChenGuo2024g}%
  \BibitemOpen
  \bibfield  {author} {\bibinfo {author} {\bibfnamefont {R.}~\bibnamefont
  {Chen}}\ and\ \bibinfo {author} {\bibfnamefont {C.}~\bibnamefont {Guo}},\
  }\bibfield  {title} {\bibinfo {title} {Solving equilibrium quantum impurity
  problems on the l-shaped kadanoff-baym contour},\ }\href
  {https://doi.org/10.1103/PhysRevB.110.165114} {\bibfield  {journal} {\bibinfo
   {journal} {Phys. Rev. B}\ }\textbf {\bibinfo {volume} {110}},\ \bibinfo
  {pages} {165114} (\bibinfo {year} {2024})}\BibitemShut {NoStop}%
\bibitem [{\citenamefont {Guo}\ and\ \citenamefont
  {Chen}(2024{\natexlab{b}})}]{GuoChen2024e}%
  \BibitemOpen
  \bibfield  {author} {\bibinfo {author} {\bibfnamefont {C.}~\bibnamefont
  {Guo}}\ and\ \bibinfo {author} {\bibfnamefont {R.}~\bibnamefont {Chen}},\
  }\bibfield  {title} {\bibinfo {title} {Infinite grassmann time-evolving
  matrix product operator method in the steady state},\ }\href
  {https://doi.org/10.1103/physrevb.110.045106} {\bibfield  {journal} {\bibinfo
   {journal} {Phys. Rev. B}\ }\textbf {\bibinfo {volume} {110}},\ \bibinfo
  {pages} {045106} (\bibinfo {year} {2024}{\natexlab{b}})}\BibitemShut
  {NoStop}%
\bibitem [{\citenamefont {Guo}\ and\ \citenamefont
  {Chen}(2024{\natexlab{c}})}]{GuoChen2024f}%
  \BibitemOpen
  \bibfield  {author} {\bibinfo {author} {\bibfnamefont {C.}~\bibnamefont
  {Guo}}\ and\ \bibinfo {author} {\bibfnamefont {R.}~\bibnamefont {Chen}},\
  }\bibfield  {title} {\bibinfo {title} {Infinite grassmann time-evolving
  matrix product operator method for zero-temperature equilibrium quantum
  impurity problems},\ }\href {https://doi.org/10.1103/PhysRevB.110.165119}
  {\bibfield  {journal} {\bibinfo  {journal} {Phys. Rev. B}\ }\textbf {\bibinfo
  {volume} {110}},\ \bibinfo {pages} {165119} (\bibinfo {year}
  {2024}{\natexlab{c}})}\BibitemShut {NoStop}%
\bibitem [{\citenamefont {Sun}\ \emph {et~al.}(2024)\citenamefont {Sun},
  \citenamefont {Chen}, \citenamefont {Li},\ and\ \citenamefont
  {Guo}}]{SunGuo2025}%
  \BibitemOpen
  \bibfield  {author} {\bibinfo {author} {\bibfnamefont {Z.}~\bibnamefont
  {Sun}}, \bibinfo {author} {\bibfnamefont {R.}~\bibnamefont {Chen}}, \bibinfo
  {author} {\bibfnamefont {Z.}~\bibnamefont {Li}},\ and\ \bibinfo {author}
  {\bibfnamefont {C.}~\bibnamefont {Guo}},\ }\bibfield  {title} {\bibinfo
  {title} {Infinite grassmann time-evolving matrix product operators for
  non-equilibrium quantum impurity problems},\ }\href
  {https://doi.org/10.48550/arXiv.2412.04702} {\bibfield  {journal} {\bibinfo
  {journal} {arXiv:2412.04702}\ } (\bibinfo {year} {2024})}\BibitemShut
  {NoStop}%
\bibitem [{\citenamefont {J\o{}rgensen}\ and\ \citenamefont
  {Pollock}(2019)}]{JorgensenPollock2019}%
  \BibitemOpen
  \bibfield  {author} {\bibinfo {author} {\bibfnamefont {M.~R.}\ \bibnamefont
  {J\o{}rgensen}}\ and\ \bibinfo {author} {\bibfnamefont {F.~A.}\ \bibnamefont
  {Pollock}},\ }\bibfield  {title} {\bibinfo {title} {Exploiting the causal
  tensor network structure of quantum processes to efficiently simulate
  non-markovian path integrals},\ }\href
  {https://doi.org/10.1103/PhysRevLett.123.240602} {\bibfield  {journal}
  {\bibinfo  {journal} {Phys. Rev. Lett.}\ }\textbf {\bibinfo {volume} {123}},\
  \bibinfo {pages} {240602} (\bibinfo {year} {2019})}\BibitemShut {NoStop}%
\bibitem [{\citenamefont {Popovic}\ \emph {et~al.}(2021)\citenamefont
  {Popovic}, \citenamefont {Mitchison}, \citenamefont {Strathearn},
  \citenamefont {Lovett}, \citenamefont {Goold},\ and\ \citenamefont
  {Eastham}}]{popovic2021-quantum}%
  \BibitemOpen
  \bibfield  {author} {\bibinfo {author} {\bibfnamefont {M.}~\bibnamefont
  {Popovic}}, \bibinfo {author} {\bibfnamefont {M.~T.}\ \bibnamefont
  {Mitchison}}, \bibinfo {author} {\bibfnamefont {A.}~\bibnamefont
  {Strathearn}}, \bibinfo {author} {\bibfnamefont {B.~W.}\ \bibnamefont
  {Lovett}}, \bibinfo {author} {\bibfnamefont {J.}~\bibnamefont {Goold}},\ and\
  \bibinfo {author} {\bibfnamefont {P.~R.}\ \bibnamefont {Eastham}},\
  }\bibfield  {title} {\bibinfo {title} {Quantum heat statistics with
  time-evolving matrix product operators},\ }\href
  {https://doi.org/10.1103/prxquantum.2.020338} {\bibfield  {journal} {\bibinfo
   {journal} {PRX Quantum}\ }\textbf {\bibinfo {volume} {2}},\ \bibinfo {pages}
  {020338} (\bibinfo {year} {2021})}\BibitemShut {NoStop}%
\bibitem [{\citenamefont {Fux}\ \emph {et~al.}(2021)\citenamefont {Fux},
  \citenamefont {Butler}, \citenamefont {Eastham}, \citenamefont {Lovett},\
  and\ \citenamefont {Keeling}}]{fux2021-efficient}%
  \BibitemOpen
  \bibfield  {author} {\bibinfo {author} {\bibfnamefont {G.~E.}\ \bibnamefont
  {Fux}}, \bibinfo {author} {\bibfnamefont {E.~P.}\ \bibnamefont {Butler}},
  \bibinfo {author} {\bibfnamefont {P.~R.}\ \bibnamefont {Eastham}}, \bibinfo
  {author} {\bibfnamefont {B.~W.}\ \bibnamefont {Lovett}},\ and\ \bibinfo
  {author} {\bibfnamefont {J.}~\bibnamefont {Keeling}},\ }\bibfield  {title}
  {\bibinfo {title} {Efficient exploration of hamiltonian parameter space for
  optimal control of non-markovian open quantum systems},\ }\href
  {https://doi.org/10.1103/physrevlett.126.200401} {\bibfield  {journal}
  {\bibinfo  {journal} {Phys. Rev. Lett.}\ }\textbf {\bibinfo {volume} {126}},\
  \bibinfo {pages} {200401} (\bibinfo {year} {2021})}\BibitemShut {NoStop}%
\bibitem [{\citenamefont {Gribben}\ \emph {et~al.}(2021)\citenamefont
  {Gribben}, \citenamefont {Strathearn}, \citenamefont {Fux}, \citenamefont
  {Kirton},\ and\ \citenamefont {Lovett}}]{gribben2021-using}%
  \BibitemOpen
  \bibfield  {author} {\bibinfo {author} {\bibfnamefont {D.}~\bibnamefont
  {Gribben}}, \bibinfo {author} {\bibfnamefont {A.}~\bibnamefont {Strathearn}},
  \bibinfo {author} {\bibfnamefont {G.~E.}\ \bibnamefont {Fux}}, \bibinfo
  {author} {\bibfnamefont {P.}~\bibnamefont {Kirton}},\ and\ \bibinfo {author}
  {\bibfnamefont {B.~W.}\ \bibnamefont {Lovett}},\ }\bibfield  {title}
  {\bibinfo {title} {Using the environment to understand non-markovian open
  quantum systems},\ }\href {https://doi.org/10.22331/q-2022-10-25-847}
  {\bibfield  {journal} {\bibinfo  {journal} {Quantum}\ }\textbf {\bibinfo
  {volume} {6}},\ \bibinfo {pages} {847} (\bibinfo {year} {2021})}\BibitemShut
  {NoStop}%
\bibitem [{\citenamefont {Otterpohl}\ \emph {et~al.}(2022)\citenamefont
  {Otterpohl}, \citenamefont {Nalbach},\ and\ \citenamefont
  {Thorwart}}]{otterpohl2022-hidden}%
  \BibitemOpen
  \bibfield  {author} {\bibinfo {author} {\bibfnamefont {F.}~\bibnamefont
  {Otterpohl}}, \bibinfo {author} {\bibfnamefont {P.}~\bibnamefont {Nalbach}},\
  and\ \bibinfo {author} {\bibfnamefont {M.}~\bibnamefont {Thorwart}},\
  }\bibfield  {title} {\bibinfo {title} {Hidden phase of the spin-boson
  model},\ }\href {https://doi.org/10.1103/physrevlett.129.120406} {\bibfield
  {journal} {\bibinfo  {journal} {Phys. Rev. Lett.}\ }\textbf {\bibinfo
  {volume} {129}},\ \bibinfo {pages} {120406} (\bibinfo {year}
  {2022})}\BibitemShut {NoStop}%
\bibitem [{\citenamefont {Gribben}\ \emph {et~al.}(2022)\citenamefont
  {Gribben}, \citenamefont {Rouse}, \citenamefont {Iles-Smith}, \citenamefont
  {Strathearn}, \citenamefont {Maguire}, \citenamefont {Kirton}, \citenamefont
  {Nazir}, \citenamefont {Gauger},\ and\ \citenamefont
  {Lovett}}]{gribben2022-exact}%
  \BibitemOpen
  \bibfield  {author} {\bibinfo {author} {\bibfnamefont {D.}~\bibnamefont
  {Gribben}}, \bibinfo {author} {\bibfnamefont {D.~M.}\ \bibnamefont {Rouse}},
  \bibinfo {author} {\bibfnamefont {J.}~\bibnamefont {Iles-Smith}}, \bibinfo
  {author} {\bibfnamefont {A.}~\bibnamefont {Strathearn}}, \bibinfo {author}
  {\bibfnamefont {H.}~\bibnamefont {Maguire}}, \bibinfo {author} {\bibfnamefont
  {P.}~\bibnamefont {Kirton}}, \bibinfo {author} {\bibfnamefont
  {A.}~\bibnamefont {Nazir}}, \bibinfo {author} {\bibfnamefont {E.~M.}\
  \bibnamefont {Gauger}},\ and\ \bibinfo {author} {\bibfnamefont {B.~W.}\
  \bibnamefont {Lovett}},\ }\bibfield  {title} {\bibinfo {title} {Exact
  dynamics of nonadditive environments in non-markovian open quantum systems},\
  }\href {https://doi.org/10.1103/prxquantum.3.010321} {\bibfield  {journal}
  {\bibinfo  {journal} {PRX Quantum}\ }\textbf {\bibinfo {volume} {3}},\
  \bibinfo {pages} {010321} (\bibinfo {year} {2022})}\BibitemShut {NoStop}%
\bibitem [{\citenamefont {Ng}\ \emph {et~al.}(2023)\citenamefont {Ng},
  \citenamefont {Park}, \citenamefont {Millis}, \citenamefont {Chan},\ and\
  \citenamefont {Reichman}}]{NgReichman2023}%
  \BibitemOpen
  \bibfield  {author} {\bibinfo {author} {\bibfnamefont {N.}~\bibnamefont
  {Ng}}, \bibinfo {author} {\bibfnamefont {G.}~\bibnamefont {Park}}, \bibinfo
  {author} {\bibfnamefont {A.~J.}\ \bibnamefont {Millis}}, \bibinfo {author}
  {\bibfnamefont {G.~K.-L.}\ \bibnamefont {Chan}},\ and\ \bibinfo {author}
  {\bibfnamefont {D.~R.}\ \bibnamefont {Reichman}},\ }\bibfield  {title}
  {\bibinfo {title} {Real-time evolution of anderson impurity models via tensor
  network influence functionals},\ }\href
  {https://doi.org/10.1103/PhysRevB.107.125103} {\bibfield  {journal} {\bibinfo
   {journal} {Phys. Rev. B}\ }\textbf {\bibinfo {volume} {107}},\ \bibinfo
  {pages} {125103} (\bibinfo {year} {2023})}\BibitemShut {NoStop}%
\bibitem [{\citenamefont {Park}\ \emph {et~al.}(2024)\citenamefont {Park},
  \citenamefont {Ng}, \citenamefont {Reichman},\ and\ \citenamefont
  {Chan}}]{ParkChan2024}%
  \BibitemOpen
  \bibfield  {author} {\bibinfo {author} {\bibfnamefont {G.}~\bibnamefont
  {Park}}, \bibinfo {author} {\bibfnamefont {N.}~\bibnamefont {Ng}}, \bibinfo
  {author} {\bibfnamefont {D.~R.}\ \bibnamefont {Reichman}},\ and\ \bibinfo
  {author} {\bibfnamefont {G.~K.-L.}\ \bibnamefont {Chan}},\ }\bibfield
  {title} {\bibinfo {title} {Tensor network influence functionals in the
  continuous-time limit: Connections to quantum embedding, bath discretization,
  and higher-order time propagation},\ }\href
  {https://doi.org/10.1103/PhysRevB.110.045104} {\bibfield  {journal} {\bibinfo
   {journal} {Phys. Rev. B}\ }\textbf {\bibinfo {volume} {110}},\ \bibinfo
  {pages} {045104} (\bibinfo {year} {2024})}\BibitemShut {NoStop}%
\bibitem [{\citenamefont {Assaad}\ and\ \citenamefont
  {Lang}(2007)}]{assaad2007-diagrammatic}%
  \BibitemOpen
  \bibfield  {author} {\bibinfo {author} {\bibfnamefont {F.~F.}\ \bibnamefont
  {Assaad}}\ and\ \bibinfo {author} {\bibfnamefont {T.~C.}\ \bibnamefont
  {Lang}},\ }\bibfield  {title} {\bibinfo {title} {Diagrammatic determinantal
  quantum monte carlo methods: Projective schemes and applications to the
  hubbard-holstein model},\ }\href {https://doi.org/10.1103/PhysRevB.76.035116}
  {\bibfield  {journal} {\bibinfo  {journal} {Phys. Rev. B}\ }\textbf {\bibinfo
  {volume} {76}},\ \bibinfo {pages} {035116} (\bibinfo {year}
  {2007})}\BibitemShut {NoStop}%
\bibitem [{\citenamefont {Keldysh}(1965)}]{Keldysh1965}%
  \BibitemOpen
  \bibfield  {author} {\bibinfo {author} {\bibfnamefont {L.~V.}\ \bibnamefont
  {Keldysh}},\ }\bibfield  {title} {\bibinfo {title} {Diagram technique for
  non-equilibrium processes},\ }\href
  {https://doi.org/10.1142/9789811279461_0007} {\bibfield  {journal} {\bibinfo
  {journal} {Soviet Physics JETP}\ }\textbf {\bibinfo {volume} {20}},\ \bibinfo
  {pages} {1018} (\bibinfo {year} {1965})}\BibitemShut {NoStop}%
\bibitem [{\citenamefont {Lifshitz}\ and\ \citenamefont
  {Pitaevskii}(1981)}]{LifshitzPitaevskii1981}%
  \BibitemOpen
  \bibfield  {author} {\bibinfo {author} {\bibfnamefont {E.~M.}\ \bibnamefont
  {Lifshitz}}\ and\ \bibinfo {author} {\bibfnamefont {L.~P.}\ \bibnamefont
  {Pitaevskii}},\ }\href@noop {} {\emph {\bibinfo {title} {Course of
  Theoretical Physics Volume 10: Physical Kinetics}}}\ (\bibinfo  {publisher}
  {Elsevier},\ \bibinfo {year} {1981})\BibitemShut {NoStop}%
\bibitem [{\citenamefont {Kadanoff}\ and\ \citenamefont
  {Baym}(1962)}]{kadanoff1962-quantum}%
  \BibitemOpen
  \bibfield  {author} {\bibinfo {author} {\bibfnamefont {L.~P.}\ \bibnamefont
  {Kadanoff}}\ and\ \bibinfo {author} {\bibfnamefont {G.}~\bibnamefont
  {Baym}},\ }\href@noop {} {\emph {\bibinfo {title} {Quantum Statistical
  Mechnics}}}\ (\bibinfo  {publisher} {W. A. Benjamin},\ \bibinfo {address}
  {New York},\ \bibinfo {year} {1962})\BibitemShut {NoStop}%
\bibitem [{\citenamefont {Chen}(2025)}]{Chen2025}%
  \BibitemOpen
  \bibfield  {author} {\bibinfo {author} {\bibfnamefont {R.}~\bibnamefont
  {Chen}},\ }\bibfield  {title} {\bibinfo {title} {Path integral formalism for
  quantum open systems},\ }\href
  {https://doi.org/https://doi.org/10.1016/j.aop.2025.170083} {\bibfield
  {journal} {\bibinfo  {journal} {Ann. Phys.}\ }\textbf {\bibinfo {volume}
  {480}},\ \bibinfo {pages} {170083} (\bibinfo {year} {2025})}\BibitemShut
  {NoStop}%
\bibitem [{\citenamefont {Jordan}\ and\ \citenamefont
  {Wigner}(1928)}]{JordanWigner1928}%
  \BibitemOpen
  \bibfield  {author} {\bibinfo {author} {\bibfnamefont {P.}~\bibnamefont
  {Jordan}}\ and\ \bibinfo {author} {\bibfnamefont {E.}~\bibnamefont
  {Wigner}},\ }\bibfield  {title} {\bibinfo {title} {{{\"U}ber das Paulische
  {\"A}quivalenzverbot}},\ }\href {https://doi.org/10.1007/BF01331938}
  {\bibfield  {journal} {\bibinfo  {journal} {Z. Physik}\ }\textbf {\bibinfo
  {volume} {47}},\ \bibinfo {pages} {631} (\bibinfo {year} {1928})}\BibitemShut
  {NoStop}%
\bibitem [{\citenamefont {Makarov}\ and\ \citenamefont
  {Makri}(1994)}]{makarov1994-path}%
  \BibitemOpen
  \bibfield  {author} {\bibinfo {author} {\bibfnamefont {D.~E.}\ \bibnamefont
  {Makarov}}\ and\ \bibinfo {author} {\bibfnamefont {N.}~\bibnamefont
  {Makri}},\ }\bibfield  {title} {\bibinfo {title} {Path integrals for
  dissipative systems by tensor multiplication. condensed phase quantum
  dynamics for arbitrarily long time},\ }\href
  {https://doi.org/10.1016/0009-2614(94)00275-4} {\bibfield  {journal}
  {\bibinfo  {journal} {Chem. Phys. Lett.}\ }\textbf {\bibinfo {volume}
  {221}},\ \bibinfo {pages} {482} (\bibinfo {year} {1994})}\BibitemShut
  {NoStop}%
\bibitem [{\citenamefont {Makri}(1995)}]{makri1995-numerical}%
  \BibitemOpen
  \bibfield  {author} {\bibinfo {author} {\bibfnamefont {N.}~\bibnamefont
  {Makri}},\ }\bibfield  {title} {\bibinfo {title} {Numerical path integral
  techniques for long time dynamics of quantum dissipative systems},\ }\href
  {https://doi.org/10.1063/1.531046} {\bibfield  {journal} {\bibinfo  {journal}
  {J. Math. Phys.}\ }\textbf {\bibinfo {volume} {36}},\ \bibinfo {pages} {2430}
  (\bibinfo {year} {1995})}\BibitemShut {NoStop}%
\bibitem [{\citenamefont {Schollwöck}(2011)}]{Schollwock2011}%
  \BibitemOpen
  \bibfield  {author} {\bibinfo {author} {\bibfnamefont {U.}~\bibnamefont
  {Schollwöck}},\ }\bibfield  {title} {\bibinfo {title} {The density-matrix
  renormalization group in the age of matrix product states},\ }\href
  {https://doi.org/https://doi.org/10.1016/j.aop.2010.09.012} {\bibfield
  {journal} {\bibinfo  {journal} {Ann. Phys.}\ }\textbf {\bibinfo {volume}
  {326}},\ \bibinfo {pages} {96} (\bibinfo {year} {2011})}\BibitemShut
  {NoStop}%
\bibitem [{\citenamefont {Or{\'u}s}(2014)}]{orus2014-practical}%
  \BibitemOpen
  \bibfield  {author} {\bibinfo {author} {\bibfnamefont {R.}~\bibnamefont
  {Or{\'u}s}},\ }\bibfield  {title} {\bibinfo {title} {A practical introduction
  to tensor networks: Matrix product states and projected entangled pair
  states},\ }\href {https://doi.org/10.1016/j.aop.2014.06.013} {\bibfield
  {journal} {\bibinfo  {journal} {Ann. Phys.}\ }\textbf {\bibinfo {volume}
  {349}},\ \bibinfo {pages} {117} (\bibinfo {year} {2014})}\BibitemShut
  {NoStop}%
\bibitem [{\citenamefont {Sun}\ \emph {et~al.}(2025)\citenamefont {Sun},
  \citenamefont {Chen}, \citenamefont {Li},\ and\ \citenamefont
  {Guo}}]{SunGuo2025b}%
  \BibitemOpen
  \bibfield  {author} {\bibinfo {author} {\bibfnamefont {Z.}~\bibnamefont
  {Sun}}, \bibinfo {author} {\bibfnamefont {R.}~\bibnamefont {Chen}}, \bibinfo
  {author} {\bibfnamefont {Z.}~\bibnamefont {Li}},\ and\ \bibinfo {author}
  {\bibfnamefont {C.}~\bibnamefont {Guo}},\ }\bibfield  {title} {\bibinfo
  {title} {Scalable tensor network algorithm for quantum impurity problems},\
  }\href {https://doi.org/10.48550/arXiv.2507.12722} {\bibfield  {journal}
  {\bibinfo  {journal} {arXiv:2507.12722}\ } (\bibinfo {year}
  {2025})}\BibitemShut {NoStop}%
\bibitem [{\citenamefont {Xu}\ \emph {et~al.}(2024)\citenamefont {Xu},
  \citenamefont {Guo},\ and\ \citenamefont {Chen}}]{XuChen2024}%
  \BibitemOpen
  \bibfield  {author} {\bibinfo {author} {\bibfnamefont {X.}~\bibnamefont
  {Xu}}, \bibinfo {author} {\bibfnamefont {C.}~\bibnamefont {Guo}},\ and\
  \bibinfo {author} {\bibfnamefont {R.}~\bibnamefont {Chen}},\ }\bibfield
  {title} {\bibinfo {title} {Grassmann time-evolving matrix product operators:
  An efficient numerical approach for fermionic path integral simulations},\
  }\href {https://doi.org/10.1063/5.0226167} {\bibfield  {journal} {\bibinfo
  {journal} {J. Chem. Phys.}\ }\textbf {\bibinfo {volume} {161}},\ \bibinfo
  {pages} {151001} (\bibinfo {year} {2024})}\BibitemShut {NoStop}%
\bibitem [{\citenamefont {Chen}(2023)}]{chen2023-heat}%
  \BibitemOpen
  \bibfield  {author} {\bibinfo {author} {\bibfnamefont {R.}~\bibnamefont
  {Chen}},\ }\bibfield  {title} {\bibinfo {title} {Heat current in
  non-markovian open systems},\ }\href
  {https://doi.org/10.1088/1367-2630/acc60a} {\bibfield  {journal} {\bibinfo
  {journal} {New J. Phys.}\ }\textbf {\bibinfo {volume} {25}},\ \bibinfo
  {pages} {033035} (\bibinfo {year} {2023})}\BibitemShut {NoStop}%
\bibitem [{\citenamefont {Mahan}(2000)}]{mahan2000-many}%
  \BibitemOpen
  \bibfield  {author} {\bibinfo {author} {\bibfnamefont {G.~D.}\ \bibnamefont
  {Mahan}},\ }\href@noop {} {\emph {\bibinfo {title} {Many-Particle Physics}}}\
  (\bibinfo  {publisher} {Springer; 3nd edition},\ \bibinfo {year}
  {2000})\BibitemShut {NoStop}%
\bibitem [{\citenamefont {Xu}\ \emph {et~al.}(2022)\citenamefont {Xu},
  \citenamefont {Yan}, \citenamefont {Shi}, \citenamefont {Ankerhold},\ and\
  \citenamefont {Stockburger}}]{XuStockburger2022}%
  \BibitemOpen
  \bibfield  {author} {\bibinfo {author} {\bibfnamefont {M.}~\bibnamefont
  {Xu}}, \bibinfo {author} {\bibfnamefont {Y.}~\bibnamefont {Yan}}, \bibinfo
  {author} {\bibfnamefont {Q.}~\bibnamefont {Shi}}, \bibinfo {author}
  {\bibfnamefont {J.}~\bibnamefont {Ankerhold}},\ and\ \bibinfo {author}
  {\bibfnamefont {J.~T.}\ \bibnamefont {Stockburger}},\ }\bibfield  {title}
  {\bibinfo {title} {Taming quantum noise for efficient low temperature
  simulations of open quantum systems},\ }\href
  {https://doi.org/10.1103/PhysRevLett.129.230601} {\bibfield  {journal}
  {\bibinfo  {journal} {Phys. Rev. Lett.}\ }\textbf {\bibinfo {volume} {129}},\
  \bibinfo {pages} {230601} (\bibinfo {year} {2022})}\BibitemShut {NoStop}%
\bibitem [{\citenamefont {Wolf}\ \emph
  {et~al.}(2014{\natexlab{b}})\citenamefont {Wolf}, \citenamefont {McCulloch},\
  and\ \citenamefont {Schollw\"ock}}]{WolfSchollwock2014}%
  \BibitemOpen
  \bibfield  {author} {\bibinfo {author} {\bibfnamefont {F.~A.}\ \bibnamefont
  {Wolf}}, \bibinfo {author} {\bibfnamefont {I.~P.}\ \bibnamefont
  {McCulloch}},\ and\ \bibinfo {author} {\bibfnamefont {U.}~\bibnamefont
  {Schollw\"ock}},\ }\bibfield  {title} {\bibinfo {title} {Solving
  nonequilibrium dynamical mean-field theory using matrix product states},\
  }\href {https://doi.org/10.1103/PhysRevB.90.235131} {\bibfield  {journal}
  {\bibinfo  {journal} {Phys. Rev. B}\ }\textbf {\bibinfo {volume} {90}},\
  \bibinfo {pages} {235131} (\bibinfo {year} {2014}{\natexlab{b}})}\BibitemShut
  {NoStop}%
\bibitem [{\citenamefont
  {Holstein}(1959{\natexlab{a}})}]{holstein1959-studies}%
  \BibitemOpen
  \bibfield  {author} {\bibinfo {author} {\bibfnamefont {T.}~\bibnamefont
  {Holstein}},\ }\bibfield  {title} {\bibinfo {title} {Studies of polaron
  motion},\ }\href {https://doi.org/10.1016/0003-4916(59)90002-8} {\bibfield
  {journal} {\bibinfo  {journal} {Ann. Phys.}\ }\textbf {\bibinfo {volume}
  {8}},\ \bibinfo {pages} {325} (\bibinfo {year}
  {1959}{\natexlab{a}})}\BibitemShut {NoStop}%
\bibitem [{\citenamefont
  {Holstein}(1959{\natexlab{b}})}]{holstein1959-studies-II}%
  \BibitemOpen
  \bibfield  {author} {\bibinfo {author} {\bibfnamefont {T.}~\bibnamefont
  {Holstein}},\ }\bibfield  {title} {\bibinfo {title} {Studies of polaron
  motion},\ }\href {https://doi.org/10.1016/0003-4916(59)90003-x} {\bibfield
  {journal} {\bibinfo  {journal} {Ann. Phys.}\ }\textbf {\bibinfo {volume}
  {8}},\ \bibinfo {pages} {343–389} (\bibinfo {year}
  {1959}{\natexlab{b}})}\BibitemShut {NoStop}%
\bibitem [{\citenamefont {Chen}\ \emph {et~al.}(2016)\citenamefont {Chen},
  \citenamefont {Cohen}, \citenamefont {Millis},\ and\ \citenamefont
  {Reichman}}]{chen2016-anderson}%
  \BibitemOpen
  \bibfield  {author} {\bibinfo {author} {\bibfnamefont {H.-T.}\ \bibnamefont
  {Chen}}, \bibinfo {author} {\bibfnamefont {G.}~\bibnamefont {Cohen}},
  \bibinfo {author} {\bibfnamefont {A.~J.}\ \bibnamefont {Millis}},\ and\
  \bibinfo {author} {\bibfnamefont {D.~R.}\ \bibnamefont {Reichman}},\
  }\bibfield  {title} {\bibinfo {title} {Anderson-holstein model in two flavors
  of the noncrossing approximation},\ }\href
  {https://doi.org/10.1103/PhysRevB.93.174309} {\bibfield  {journal} {\bibinfo
  {journal} {Phys. Rev. B}\ }\textbf {\bibinfo {volume} {93}},\ \bibinfo
  {pages} {174309} (\bibinfo {year} {2016})}\BibitemShut {NoStop}%
\bibitem [{\citenamefont {Sengupta}\ and\ \citenamefont
  {Georges}(1995)}]{SenguptaGeorges1995}%
  \BibitemOpen
  \bibfield  {author} {\bibinfo {author} {\bibfnamefont {A.~M.}\ \bibnamefont
  {Sengupta}}\ and\ \bibinfo {author} {\bibfnamefont {A.}~\bibnamefont
  {Georges}},\ }\bibfield  {title} {\bibinfo {title} {Non-fermi-liquid behavior
  near a t=0 spin-glass transition},\ }\href
  {https://doi.org/10.1103/PhysRevB.52.10295} {\bibfield  {journal} {\bibinfo
  {journal} {Phys. Rev. B}\ }\textbf {\bibinfo {volume} {52}},\ \bibinfo
  {pages} {10295} (\bibinfo {year} {1995})}\BibitemShut {NoStop}%
\bibitem [{\citenamefont {Si}\ and\ \citenamefont {Smith}(1996)}]{SiSmith1996}%
  \BibitemOpen
  \bibfield  {author} {\bibinfo {author} {\bibfnamefont {Q.}~\bibnamefont
  {Si}}\ and\ \bibinfo {author} {\bibfnamefont {J.~L.}\ \bibnamefont {Smith}},\
  }\bibfield  {title} {\bibinfo {title} {Kosterlitz-thouless transition and
  short range spatial correlations in an extended hubbard model},\ }\href
  {https://doi.org/10.1103/PhysRevLett.77.3391} {\bibfield  {journal} {\bibinfo
   {journal} {Phys. Rev. Lett.}\ }\textbf {\bibinfo {volume} {77}},\ \bibinfo
  {pages} {3391} (\bibinfo {year} {1996})}\BibitemShut {NoStop}%
\bibitem [{\citenamefont {Smith}\ and\ \citenamefont {Si}(2000)}]{SmithSi2000}%
  \BibitemOpen
  \bibfield  {author} {\bibinfo {author} {\bibfnamefont {J.~L.}\ \bibnamefont
  {Smith}}\ and\ \bibinfo {author} {\bibfnamefont {Q.}~\bibnamefont {Si}},\
  }\bibfield  {title} {\bibinfo {title} {Spatial correlations in dynamical
  mean-field theory},\ }\href {https://doi.org/10.1103/PhysRevB.61.5184}
  {\bibfield  {journal} {\bibinfo  {journal} {Phys. Rev. B}\ }\textbf {\bibinfo
  {volume} {61}},\ \bibinfo {pages} {5184} (\bibinfo {year}
  {2000})}\BibitemShut {NoStop}%
\bibitem [{\citenamefont {Chitra}\ and\ \citenamefont
  {Kotliar}(2000)}]{ChitraKotliar2000}%
  \BibitemOpen
  \bibfield  {author} {\bibinfo {author} {\bibfnamefont {R.}~\bibnamefont
  {Chitra}}\ and\ \bibinfo {author} {\bibfnamefont {G.}~\bibnamefont
  {Kotliar}},\ }\bibfield  {title} {\bibinfo {title} {Effect of long range
  coulomb interactions on the mott transition},\ }\href
  {https://doi.org/10.1103/PhysRevLett.84.3678} {\bibfield  {journal} {\bibinfo
   {journal} {Phys. Rev. Lett.}\ }\textbf {\bibinfo {volume} {84}},\ \bibinfo
  {pages} {3678} (\bibinfo {year} {2000})}\BibitemShut {NoStop}%
\bibitem [{\citenamefont {Sun}\ and\ \citenamefont
  {Kotliar}(2002)}]{SunKotliar2002}%
  \BibitemOpen
  \bibfield  {author} {\bibinfo {author} {\bibfnamefont {P.}~\bibnamefont
  {Sun}}\ and\ \bibinfo {author} {\bibfnamefont {G.}~\bibnamefont {Kotliar}},\
  }\bibfield  {title} {\bibinfo {title} {Extended dynamical mean-field theory
  and $\mathrm{GW}$ method},\ }\href
  {https://doi.org/10.1103/PhysRevB.66.085120} {\bibfield  {journal} {\bibinfo
  {journal} {Phys. Rev. B}\ }\textbf {\bibinfo {volume} {66}},\ \bibinfo
  {pages} {085120} (\bibinfo {year} {2002})}\BibitemShut {NoStop}%
\bibitem [{\citenamefont {Huang}\ \emph {et~al.}(2014)\citenamefont {Huang},
  \citenamefont {Ayral}, \citenamefont {Biermann},\ and\ \citenamefont
  {Werner}}]{HuangWerner2014}%
  \BibitemOpen
  \bibfield  {author} {\bibinfo {author} {\bibfnamefont {L.}~\bibnamefont
  {Huang}}, \bibinfo {author} {\bibfnamefont {T.}~\bibnamefont {Ayral}},
  \bibinfo {author} {\bibfnamefont {S.}~\bibnamefont {Biermann}},\ and\
  \bibinfo {author} {\bibfnamefont {P.}~\bibnamefont {Werner}},\ }\bibfield
  {title} {\bibinfo {title} {Extended dynamical mean-field study of the hubbard
  model with long-range interactions},\ }\href
  {https://doi.org/10.1103/PhysRevB.90.195114} {\bibfield  {journal} {\bibinfo
  {journal} {Phys. Rev. B}\ }\textbf {\bibinfo {volume} {90}},\ \bibinfo
  {pages} {195114} (\bibinfo {year} {2014})}\BibitemShut {NoStop}%
\end{thebibliography}%

\end{document}